\makeindex

\documentstyle[12pt,]{tese}

\begin{document}

\begin{center}
{\LARGE  Tese de Doutorado \\ }
\vspace{6cm}
{\LARGE {\bf Geometrodin\^amica Qu\^antica na}}

\vspace{.5cm}

{\LARGE {\bf Interpreta\c{c}\~ao de Bohm-de Broglie}}

\vspace{2cm}
{\Large {\bf Eduardo Sergio Santini} \\ }
\vspace{7cm}
{\Large Centro Brasileiro de Pesquisas F\'{\i}sicas \\
Rio de Janeiro, Maio de 2000  }
\end{center}

\pagenumbering{roman}

\newpage

\centerline{\Large \bf Resumo}
\bigskip
Nesta tese aplicamos a interpreta\c c\~ao de Bohm-de Broglie \`a gravita\c c\~ao 
qu\^antica can\^onica e mostramos que independentemente da regulariza\c c\~ao ou escolha de ordenamento na 
equa\c c\~ao de Wheeler-Dewitt, 
o \'unico efeito qu\^antico relevante que n\~ao quebra a estrutura de espa\c co-tempo \'e uma mudan\c ca 
de assinatura de lorentziana para euclideana. Os outros efeitos qu\^anticos ou s\~ao triviais ou quebram a 
estrutura de espa\c co-tempo. Constru\'{\i}mos uma geometrodin\^amica qu\^antica na vis\~ao de Bohm-de Broglie  
que permite o 
estudo destas estruturas. Por exemplo, mostramos que qualquer solu\c c\~ao real da equa\c c\~ao 
de Wheeler-DeWitt
gera uma geometria compat\'{\i}vel com o limite de gravita\c c\~ao forte da Relatividade Geral e com o 
grupo de Carroll.
Provamos que   a geometrodin\^amica qu\^antica na interpreta\c c\~ao 
de Bohm-de Broglie \'e sempre 
consistente para 
qualquer potencial qu\^antico. Como um passo pr\'evio e introdut\'orio a nossa metodologia, estudamos a 
teoria qu\^antica de campos 
no espa\c co-tempo de Minkowski na interpreta\c c\~ao de Bohm-de Broglie e mostramos um exemplo
concreto onde a invari\^ancia Lorentz \'e quebrada no n\'{\i}vel de processos individuais.

\newpage

\centerline{\Large \bf Abstract}
\bigskip

In this thesis the Bohm-de Broglie interpretation of quantum mechanics is 
applied to canonical quantum gravity. It is shown that, irrespective of any regularization
or choice of factor ordering of the Wheeler-DeWitt equation, the unique
relevant quantum effect which does not break spacetime is the change
of its signature from lorentzian to euclidean. The other
quantum effects are either trivial or break the four-geometry of
spacetime. A Bohm-de Broglie picture of a quantum geometrodynamics is 
constructed,
which allows the investigation of these latter structures.
For instance, it is shown that any real solution of the Wheeler-De Witt
equation yields a generate four-geometry compatible with the strong gravity
limit of General Relativity and the Carroll group. We prove that 
quantum geometrodynamics
 in the 
Bohm-de Broglie interpretation is  consistent for any quantum potential.
As a previous step to introduce our metodology, we study the quantum 
theory of fields in Minkowski spacetime in the Bohm-de Broglie interpretation 
and exhibit a 
concrete example where Lorentz invariance of individual events is broken.

\newpage
\baselineskip=2.0 \normalbaselineskip
{\bf Dedicatoria}

\vspace{3cm}

\hspace{3cm}Odoardo Santini {\it in memoriam}

\hspace{3cm}Ana Mar\'{\i}a Martino de Santini

\hspace{3cm}Juan Jos\'e Santini

\newpage

\vspace{1cm}
{\bf Agrade\c co ao povo do Brasil, que me recebeu de bra\c cos abertos e que \'e
aquele que, na verdade, com o seu trabalho, priva\c c\~oes e sacrif\'{\i}cios, 
paga a nossa pesquisa, fato que nunca devemos esquecer no nosso laborat\'orio.}

\vspace{3cm}
"...O que tenho para dizer \`a Universidade [..] ? Tenho que dizer que se pinte de negro, que se pinte de 
mulato, n\~ao s\'o entre os alunos, mas tamb\'em entre os professores, que se pinte de oper\'arios e de 
camponeses, que se pinte de povo, porque a Universidade n\~ao \'e patrim\^onio de ningu\'em e pertence
ao povo..."

\vspace{1cm}

\hspace{2cm}{\bf Dr. Ernesto Che Guevara}

\newpage

{\bf Agradecimentos}

\vspace{2cm}

Agrade\c co a Nelson Pinto-Neto, orientador, mestre e amigo, pela orienta\c c\~ao
desta tese e pelo ensino constante com a sua atitude na pesquisa. Agrade\c co ao
grupo de cosmologia e gravita\c c\~ao do Lafex e ao pessoal do CBPF/CNPq em geral por estes 
quatro anos de pesquisa e permanente aprendizagem e pela bolsa que permitiu-me realizar
este doutorado. Ao pessoal da secretar\'{\i}a do Lafex e da CFC (Myriam!) muito obrigado.
 "Eu s\'o sei que n\~ao sei nada" mas aqui aprendi algumas coisas: "faz 
sempre a conta, n\~ao interessa de quem \'e o  paper.."(Nelson). "..tudo o que 
vc vai falar ai no quadro tem que saber provar..."(M\'ario Novello). "...sem  mulher n\~ao tem 
f\'{\i}sica nenhuma..."(Leite Lopes)  que eu interpretei como "sem amor n\~ao 
tem coisa nenhuma". A todos eles muito obrigado. Muito obrigado a Helayel pela ajuda e
infinita sensibilidade conosco os posgraduandos.
Agrade\c co a minha familia que sempre me apoiou, seja aqui, l\'a ou em qualquer parte.
\'E a eles que dedico este trabalho.
Aos meus {\bf amigos e amigas} de sempre e de agora...muito obrigado!!!

\baselineskip=1.5 \normalbaselineskip
\tableofcontents

\newpage

{\bf  Nota\c c\~oes e conven\c c\~oes}
\vspace{1cm}

a) assinatura da m\'etrica: $(-,+,+,+)$

b) indices  gregos v\~ao de 0 a  3 e os indices latinos de 1 a 3

c) todas as deltas de Dirac que aparecem s\~ao 3-dimensionais. 
As deltas de Dirac derivadas s\~ao escritas sempre com a 
deriva\c c\~aocom respeito ao primeiro argumento:

\begin{equation}
\delta_i(x,y)\equiv\frac{\partial}{\partial x^i}\delta(x,y)
\end{equation}

\begin{equation}
\delta_i(y,x)\equiv\frac{\partial}{\partial y^i}\delta(y,x)
\end{equation}

d) derivada covariante 4-dimensional de $A^{\beta}$:
\begin{equation}
{\nabla}_{\alpha} A^{\beta} \equiv {\partial}_{\alpha} A^{\beta} +
{\Gamma}^{\beta}_{\alpha\nu} A^{\nu}
\end{equation}
onde
\begin{equation}
{\Gamma}^{\beta}_{\alpha\nu} \equiv \frac{1}{2} g^{\beta\mu}
({\partial}_{\nu}g_{\alpha\mu} + {\partial}_{\alpha}g_{\nu\mu} - 
{\partial}_{\mu}g_{\alpha\nu})
\end{equation}

e ${\partial}_{\alpha} \equiv \frac{\partial}{\partial x^{\alpha}}$

e) derivada covariante 3-dimensional de $A^{j}$:
\begin{equation}
D_{i} A^{j} \equiv {\partial}_{i} A^{j} +
^{3}{\Gamma}^{j}_{ik} A^{k}
\end{equation}
onde
\begin{equation}
^{3}{\Gamma}^{j}_{ik} = \frac{1}{2} h^{jl}
({\partial}_{i}h_{kl} + {\partial}_{k}h_{il} - 
{\partial}_{l}h_{ik})
\end{equation}
onde $h_{ij}$ \'e a m\'etrica 3-dimensional e $h^{ij}$ sua inversa.

f) curvatura 4-dimensional:
\begin{equation} 
R^{\mu}_{\: \nu\alpha\beta}
A^{\nu} \equiv {\nabla}_{\alpha} {\nabla}_{\beta} A^{\mu} -
{\nabla}_{\beta} {\nabla}_{\alpha} A^{\mu}
\end{equation}

g) tensor de  Ricci 4-dimensional:
\begin{equation}
R_{\nu\beta} = R^{\alpha}_{\: \nu\alpha\beta}
\end{equation}

h) equa\c c\~oes de Einstein
\begin{equation}
G_{\mu\nu} \equiv R_{\mu\nu} - \frac{1}{2} R g_{\mu\nu} = 
\frac{\kappa}{2} T_{\mu\nu}
\end{equation}

onde $\kappa=\frac{16 \pi G}{c^4}$

i) curvatura 3-dimensional:
\begin{equation} 
^{3}R^{i}_{\: jkl} A^{j} \equiv {D}_{k} {D}_{l} A^{i} -
{D}_{l} {D}_{k} A^{i}
\end{equation}
sendo o tensor de Ricci 3-dimensional definido de forma an\'aloga

j) simetriza\c c\~ao:
\begin{equation}
A_{(ij)} \equiv \frac{1}{2} (A_{ij} + A_{ji})
\end{equation}

\listoffigures

\pagestyle{headings}

\baselineskip=2.0 \normalbaselineskip

\chapter{Introdu\c{c}\~ao}

\pagenumbering{arabic}   

A mec\^anica qu\^antica \'e aceita na comunidade cient\'{\i}fica mundial como sendo
uma teor\'{\i}a universal e fundamental, aplic\'avel a qualquer sistema
f\'{\i}sico, e da qual pode-se recuperar a f\'{\i}sica cl\'assica. O Universo \'e, claro, 
um sistema f\'{\i}sico valido: existe uma teoria, a Cosmologia Padr\~ao, que 
\'e capaz de descrev\^e-lo em termos f\'{\i}sicos, e fazer previs\~oes 
que podem ser confirmadas ou refutadas pelas observa\c c\~oes. De fato, as
observa\c c\~oes  confirmam at\'e agora  o cen\'ario cosmol\'ogico padr\~ao.
Admitindo-se ent\~ao a universalidade da mec\^anica qu\^antica, concluimos 
que o pr\'oprio  Universo
 deva ser descrito pela teoria  qu\^antica, da qual poder\'{\i}amos recuperar
a Cosmologia Padr\~ao. Mas, a interpreta\c c\~ao de Copenhaguen da 
mec\^anica qu\^antica \cite{bohr,hei,von}\footnote{ Embora estes tr\^es autores
tenham diferentes vis\~oes da teoria qu\^antica, o primeiro deles 
enfatizando a indivisibilidade dos fenomenos qu\^anticos, o segundo a no\c c\~ao
de potencialidade e o terceiro o conceito de estados qu\^anticos, para 
todos eles resulta crucial a existencia de um dom\'{\i}nio cl\'assico. Esta \'e 
a raz\~ao de apresentar suas vis\~oes sob o mesmo 
nome `interpreta\c c\~ao de Copenhaguen'.}, 
ensinada nos cursos de faculdade e  utilizada pela maioria dos f\'{\i}sicos
de todas as \'areas (em especial na apresenta\c c\~ao de von Neumann), n\~ao 
pode ser utilizada numa teoria qu\^antica da Cosmologia.
\'Isto  porque ela imp\~oe a exist\^encia de um dom\'{\i}nio cl\'assico. Na 
formula\c c\~ao de von Neumann, por exemplo, a necessidade de um 
dominio cl\'assico vem da forma como ela  resolve o problema da 
medida (veja  Ref. \cite{omn} para uma boa  discuss\~ao deste ponto). Na medida
impulsiva de um certo observ\'avel, a fun\c c\~ao de onda do sistema medido 
mais o aparelho macrosc\'opico de medida divide-se em muitos ramos que 
praticamente n\~ao se superp\~oem (de modo a se obter uma boa medida),
cada uma delas contendo o sistema observado num autoestado do observ\'avel 
medido, e o ponteiro do aparelho  apontando para o respectivo autovalor.
Por\'em, ao final da medida observamos somente um desses autovalores,
e a medida \'e um processo robusto j\'a que se repetimos  o processo
imediatamente depois, vamos obter o mesmo resultado.
A fun\c c\~ao de onda parece colapsar, os outros ramos desaparecem.
A interpreta\c c\~ao de Copenhaguen asume que este colapso \'e real. 
Mas um colapso real n\~ao pode ser descrito por uma evolu\c c\~ao unit\'aria
de Schr\"{o}dinger. Por isso a interpreta\c c\~ao de Copenhaguen deve
asumir que existe um processo fundamental numa medida o qual deve-se produzir 
fora do mundo qu\^antico, num dom\'{\i}nio cl\'assico. Claro que  se agora   
desejamos quantizar  todo o Universo, n\~ao existe lugar para um mundo 
cl\'assico fora dele, e a interpreta\c c\~ao de Copenhaguen n\~ao pode ser 
aplicada.
Quer dizer que, ao insistir com a interpreta\c c\~ao de Copenhaguen, deve-se
asumir que a teoria qu\^antica n\~ao \'e universal ou, no m\'{\i}nimo, tentar
melhor\'a-la com conceitos adicionais.
Uma possibilidade ser\'{\i}a a de invocar o processo de descoer\^encia \cite{deco}.
De fato , a intera\c c\~ao do sistema qu\^antico observado com o seu ambiente
produz uma diagonaliza\c c\~ao efetiva da matriz densidade reduzida, obtida
do tra\c co com respeito aos graus de liberdade irrelevantes.
A descoer\^encia pode explicar porque a divis\~ao da fun\c c\~ao de onda
\'e dada em termos dos estados do ponteiro, e porque n\~ao vemos 
superposi\c c\~oes de objetos macrosc\'opicos. Desta forma, as propiedades 
cl\'assicas emergem da teoria qu\^antica sem necessidade de serem assumidas.
No contexto da gravita\c c\~ao qu\^antica,  a descoer\^encia tamb\'em pode explicar
o surgimento de uma geometria de fundo cl\'assica num universo 
qu\^antico \cite{2kie}. De fato, esta \'e a primeira quantidade a se  
tornar cl\'assica. Contudo, a descoer\^encia ainda
 n\~ao \'e a resposta completa ao problema da medida \cite{against,zeh}.
 Ela n\~ao consegue explicar o colapso aparente da fun\c c\~ao de onda
 depois que acaba a medida, ou porque apenas um dos elementos 
 da diagonal da 
 matriz densidade sobrevive ap\'os termina o proceso de medida.
A  teoria \'e incapaz de explicar a exist\^encia de fatos, sua unicidade, 
em contraposi\c c\~ao com a multiplicidade dos fen\^omenos poss\'{\i}veis.
Ainda est\~ao em progresso outros desenvolvimetos da teoria  como por exemplo
o de `historias consistentes' \cite{har}, o qual ainda est\'a incompleto.
O  importante papel desempenhado pelos observadores nestas descri\c c\~oes 
ainda n\~ao foi explicado \cite{zur}, e portanto persiste o problema de como
descrever um Universo qu\^antico onde a geometria de fundo ainda n\~ao \'e 
cl\'assica.
Contudo, existem algumas solu\c c\~oes alternativas para este 
dilema cosmol\'ogico qu\^antico os quais, junto com a descoer\^encia, podem resolver
o problema da medida mantendo a universalidade da teoria qu\^antica.
Podemos dizer que a evolu\c c\~ao de Schr\"{o}dinger \'e uma 
aproxima\c c\~ao de uma teoria n\~ao linear mais fundamental que pode 
explicar o colpaso \cite{rim,pen}, ou que o colpaso \'e efetivo mas n\~ao real,
no sentido que os outros ramos desaparecem do observador mas continuam 
existindo. Nesta segunda categoria podemos citar a interpreta\c c\~ao de 
V\'arios Mundos \cite{eve} e a interpreta\c c\~ao de 
Bohm-de Broglie \cite{bohm,hol}. Na primeira, todas as possibilidades
na divis\~ao da fun\c c\~ao de onda s\~ao realizadas. Em cada ramo 
existe um observador com o conhecimento do autovalor correspondente 
a este ramo, mas ele ou ela n\~ao est\'a ciente dos outros observadores e das outras
possibilidades, j\'a que os ramos n\~ao interferem.
Na ultima, \'e suposta a exist\^encia de uma part\'{\i}cula pontual no 
espa\c co de configura\c c\~ao que descreve o sistema observado e o aparelho, 
independentemente de qualquer observa\c c\~ao.
Na divis\~ao, esta particula pontual vai ingressar num dos ramos (isto depende da 
posi\c c\~ao inicial da part\'{\i}cula antes da medida, a 
qual \'e desconhecida),  e os outros ramos estar\~ao vazios.
Pode-se mostrar \cite{hol} que as ondas vazias n\~ao podem interagir 
com outras part\'{\i}culas,  nem  com a part\'{\i}cula pontual que contem o aparelho.
Portanto , nenhum observador pode detetar os outros ramos 
que est\~ao vazios. De novo, temos um colapso efetivo mas n\~ao real (as ondas
vazias continuam existindo), mas agora sem multiplica\c c\~ao de observadores.
Estas interpreta\c c\~oes podem, claro, ser utilizadas em cosmologia qu\^antica.
A evolu\c c\~ao de Schr\"{o}dinger \'e sempre v\'alida, e n\~ao \'e necess\'ario um
dom\'{\i}nio cl\'assico fora do sistema observado.

Nesta tese vamos tratar da aplica\c c\~ao da interpreta\c c\~ao de 
Bohm-de Broglie na cosmologia qu\^antica \cite{vink,sht,val,bola1}. 
Desta forma, o objeto fundamental da gravita\c c\~ao 
qu\^antica, a saber, a geometria das hipersuperf\'{\i}cies espaciais 3-dimensionais,
sup\~oe-se  existir independentemente de qualquer observa\c c\~ao ou medida,
assim como tamb\'em seus momentos can\^onicos conjugados, a curvatura 
extr\'{\i}nseca das hipersuperf\'{\i}cies espaciais.
Sua evolu\c c\~ao, indicada  por algum par\^ametro temporal, \'e ditada 
por uma evolu\c c\~ao qu\^antica  differente da cl\'assica devido
\`a presen\c ca do potencial qu\^antico o qual surge de forma natural 
da  equa\c c\~ao de  Wheeler-DeWitt. Esta interpreta\c c\~ao tem sido aplicada
a muitos modelos de minisuperespa\c co \cite{vink,bola1,kow,hor,bola2,fab}, 
obtidos impondo homogeneidade \`as hipersuperf\'{\i}cies espaciais. Foram discutidas
quest\~oes como o limite cl\'assico, o problema das singularidades, 
o problema da constante cosmol\'ogica e o problema do tempo. Por exemplo, 
em algums desses trabalhos foi mostrado que nos modelos com campos escalares 
ou radia\c c\~ao, que s\~ao bem representativos do conte\'udo de mat\'eria 
no universo primordial, a singularidade pode claramente ser evitada por 
efeitos qu\^anticos. Na descri\c c\~ao segundo a 
interpreta\c c\~ao de Bohm-de Broglie, o potencial qu\^antico   resulta 
importante na vizinhan\c ca da singularidade cl\'assica, produzindo 
uma for\c ca qu\^antica repulsiva, compensando o campo gravitacional, evitando
assim a singularidade e produzindo infla\c c\~ao. O limite cl\'assico (definido
pelo limite para o qual o potencial qu\^antico resulta desprez\'{\i}vel frente
\`a energia cl\'assica) para grandes fatores de escala   s\~ao usualmente 
alcan\c cados, mas para certos modelos com campo escalar ele depende do
estado qu\^antico e das condi\c c\~oes iniciais. De fato, \'e possivel ter
universos cl\'assicos pequenos e universos qu\^anticos grandes \cite{fab}. Com respeito
ao problema do tempo, foi mostrado que, para qualquer escolha da 
fun\c c\~ao lapso, a evolu\c c\~ao qu\^antica das hipersuperf\'{\i}cies homog\^eneas
produz a mesma 4-geometria \cite{bola1}. O que deve agora ser estudado \'e 
se este fato ainda \'e valido na teoria completa, onde n\~ao estamos 
restritos a hipersuperf\'{\i}cies espaciais homog\^eneas.
A quest\~ao \'e: dada uma hipersuperf\'{\i}cie inicial com condi\c c\~oes 
iniciais consistentes,  a evolu\c c\~ao da 3-geometria inicial dada pela 
din\^amica qu\^antica bohmiana vai produzir a mesma 4-geometria para qualquer
escolha das fun\c c\~oes lapso e deslocamento?. E se assim for, qual o tipo de
estrutura espa\c co-temporal formada?. Sabemos que isto \'e verdade 
se a 3-geometria  evolui seguindo a din\^amica da Teoria da Relatividade Geral 
cl\'assica (TRG), produzindo uma 4-geometria n\~ao degenerada, mas isso 
pode ser falso no caso em que a evolu\c c\~ao din\^amica seja a bohmiana.
Nesta tese vamos estudar e responder esta quest\~ao em detalhe.
A id\'eia \'e colocar a din\^amica qu\^antica de Bohm em forma 
hamiltoniana para utilizar resultados poderosos que existem na literatura 
os quais d\~ao a forma mais geral que um hamiltoniano deve ter para 
formar uma 4-geometria n\~ao degenerada  da evolu\c c\~ao  da 3-geometria inicial
\cite{hkt}. 

Nossa conclus\~ao \cite{nossoartigo} \'e que  a evolu\c c\~ao Bohmiana das 3-geometrias
resulta sempre consistente independentemente da escolha das fun\c c\~oes 
lapso e deslocamento, mas somente para estados 
qu\^anticos muito especiais  esta evolu\c c\~ao  produzir\'a uma 4-geometria 
qu\^antica n\~ao
degenerada relevante, que deve ser euclideana. Em geral, a 
evolu\c c\~ao bohmiana das 3-geometrias  produzir\'a uma 
4-geometria 
qu\^antica 
degenerada  onde estar\~ao presentes 
campos de vetores especiais (os autovetores nulos da 4-geometria)\footnote{Por 
exemplo, a 4-geometria do espa\c co Newtoniano \'e degenerada \cite{new},
e o seu unico autovetor nulo \'e a normal das hipersuperf\'{\i}cies de 
simultaneidade, o tempo. Como sabemos, esta estrutura n\~ao representa 
um espa\c co-tempo ja que est\'a quebrada em espa\c co absoluto mais
 tempo absoluto.}. Ent\~ao, se impusermos que o espa\c co-tempo qu\^antico seja
 uma variedade 
quadridimensional com uma 4-geometria n\~ao degenerada definida sobre ele, concluimos que
deva ser euclideano.
Chegamos a estas conclus\~oes sem ter assumido nenhuma regulariza\c c\~ao 
e ordenamento na equa\c c\~ao de Wheeler-DeWitt.
Como sabemos, esta equa\c c\~ao envolve a aplica\c c\~ao do produto de 
operadores locais sobre estados no mesmo ponto do espa\c co, o qual  
n\~ao est\'a 
bem definido \cite{reg}. Portanto, precisamos regulariz\'a-la para resolver
o problema de ordenamento, e ter uma teoria livre de anomalias (para algumas 
propostas
veja as refer\^encias \cite{japa1,japa2,kow2}).  As conclus\~oes desta tese
s\~ao completamente independentes destes problemas. Tamb\'em, no caso geral
onde temos 4-geometrias degeneradas, podemos obter uma imagem da estrutura 
qu\^antica produzida pela din\^amica bohmiana, a qual n\~ao \'e um 
espa\c co-tempo no sentido explicado acima mas algo como 
as 4-geometrias degeneradas compat\'{\i}veis com o grupo de Carroll \cite{poin}.

Esta tese est\'a organizada do seguinte modo: no pr\'oximo cap\'{\i}tulo
exporemos a interpreta\c c\~ao de Bohm-de Broglie da mec\^anica qu\^antica, 
tanto
para uma part\'{\i}cula n\~ao relativ\'{\i}stica quanto para a teoria qu\^antica de campos 
em espa\c co-tempo plano.
No cap\'{\i}tulo 3 estudaremos  algumas carater\'{\i}sticas  desta interpreta\c c\~ao no 
caso da teoria de campos mostrando resultados conceituais importantes, especialmente os concernentes
\`a invari\^ancia relativ\'{\i}stica da teoria. Ao 
mesmo tempo, este cap\'{\i}tulo serve como uma introdu\c c\~ao \`a metodologia 
seguida 
no caso da gravita\c c\~ao qu\^antica.
No cap\'{\i}tulo 4 exporemos a cosmologia qu\^antica can\^onica e no seguinte 
aplicaremos a  interpreta\c c\~ao de Bohm-de Broglie a esta teoria. 
Nele construiremos
a geometrodin\^amica qu\^antica na vis\~ao de Bohm-de Broglie mostrando que, 
independentemente da regulariza\c c\~ao e ordenamento da equa\c c\~ao de 
Wheeler-DeWitt, a evolu\c c\~ao 
bohmiana  das 3-geometrias pode ser obtida a partir de um certo 
hamiltoniano, que resulta, claro, diferente do cl\'assico. Isto ser\'a
utilizado no cap\'{\i}tulo 6 para obter os principais resultados desta 
tese, relativos aos poss\'{\i}veis cen\'arios cosmol\'ogicos na era qu\^antica, 
\`a possibilidade de obter 4-geometrias qu\^anticas 
n\~ao-degeneradas e \`a descri\c c\~ao de outras estruturas qu\^anticas 
diferentes. 
Discutimos tamb\'em o limite cl\'assico destas possibilidades.
Encerramos com as conclus\~oes, discuss\~ao 
e perspectivas para o futuro. No ap\^endice A calculamos explicitamente o colchete
de Poisson relevante para o estudo da invari\^ancia relativ\'{\i}stica concernente 
ao cap\'{\i}tulo 3. No ap\^endice B mostramos um exemplo 
espec\'{\i}fico de quebra de invari\^ancia de Lorentz e calculamos o seu 
respectivo potencial qu\^antico.
No ap\^endice C 
mostramos um exemplo concreto de um midi-superespa\c co que ilustra a 
discuss\~ao do capitulo 6.

\baselineskip=2.0 \normalbaselineskip

\chapter{A Interpreta\c c\~ao de Bohm-de Broglie  }


Como vimos na introdu\c c\~ao, a interpreta\c c\~ao de Bohm-de Broglie, contrariamente
\`a interpreta\c c\~ao de Copenhagen, prescinde da exist\^encia de um mundo cl\'assico
fora do objeto qu\^antico e portanto pode ser usada como interpreta\c c\~ao da 
cosmologia qu\^antica.

Mas existem outras raz~oes para  
estudar esta interpreta\c c\~ao, al\'em da sua aplicabilidade \'a cosmologia qu\^antica. 
De fato, historicamente esta interpreta\c c\~ao surgiu para proporcionar uma 
descri\c c\~ao completa e causal de um fen\^omeno qu\^antico, 
independentemente do ato de 
 observa\c c\~ao.  
Sabemos que nenhum experimento contradiz as previs\~oes da formula\c c\~ao 
ortodoxa e que a concord\^ancia teoria-experimento se d\'a com grande precis\~ao 
(como no caso da Eletrodin\^amica Qu\^antica) \cite{qed}, 
Mas como a mec\^anica 
qu\^antica ortodoxa prediz somente resultados
de experimentos realizados com agregados estat\'{\i}sticos, ela  n\~ao providencia 
uma descri\c c\~ao dos eventos individuais 
da experi\^encia, que parecem acontecer ao acaso e dos quais s\~ao fun\c c\~oes  os fen\^omenos estat\'{\i}sticos.
Resulta ent\~ao um desafio construir uma teoria capaz de descrever os sistemas  
materiais individuais de forma causal,
cada um deles seguindo sua lei de movimento, cujo comportamento em
conjunto reproduza as previs\~oes estat\'{\i}sticas da mec\^anica qu\^antica. 
Assim, os registros no laborat\'orio  poderiam ser explicados como resultado 
de uma sequ\^encia de processos bem definidos ocorridos em sistemas que 
possuem propriedades que existem independentemente do ato da observa\c c\~ao.
Um modo de fazer isto foi constru\'{\i}ido por Louis de Broglie e David Bohm. Al\'em dos 
artigos originais e dos que se seguiram na literatura, j\'a existe hoje o primeiro 
livro  texto de mec\^anica qu\^antica nesta interpreta\c c\~ao \cite{hol}.

Neste cap\'{i}tulo vamos expor as principais carater\'{\i}sticas da 
interpreta\c c\~ao de Bohm-de Broglie da mec\^anica
qu\^antica, que ser\~ao \'uteis no nosso tratamento da gravita\c c\~ao 
qu\^antica e teoria de campos
nos cap\'{\i}tulos seguintes.
Mostraremos primeiro como esta interpreta\c c\~ao se aplica ao caso 
de uma part\'{\i}cula descrita pela equa\c c\~ao de Schr\"{o}dinger e depois
vamos obter, por analogia, a interpreta\c c\~ao causal de uma teoria de
campos no espa\c co-tempo plano.

Comecemos com a interpreta\c c\~ao de Bohm-de Broglie para a  
equa\c c\~ao de Schr\"{o}dinger de uma part\'{\i}cula. Na 
representa\c c\~ao de coordenadas, para uma part\'{\i}cula n\~ao relativ\'{\i}stica
cujo hamiltoniano \'e $
H=p^{2}/2m+V(x),$
a equa\c c\~ao de Schr\"{o}dinger \'e

\begin{equation}
i\hbar \frac{\partial \Psi (x,t)}{\partial t}=\left[ -\frac{\hbar ^{2}}{2m}
\nabla ^{2}+V(x)\right] \Psi (x,t).  \label{bsc}
\end{equation}

Podemos transformar esta equa\c c\~ao diferencial sobre um campo complexo
num par de equa\c c\~oes diferenciais acopladas sobre campos reais, 
escrevendo
$\Psi =A\exp (iS/\hbar )$, onde $A$ e $S$ s\~ao fun\c c\~oes  reais, e
substituindo-a em 
 (\ref{bsc}). Obtemos as seguintes equa\c c\~oes 
 
\begin{equation}
\frac{\partial S}{\partial t}+\frac{(\nabla S)^{2}}{2m}+V-\frac{\hbar ^{2}}{2m}\frac{\nabla ^{2}A}{A}=0,  \label{bqp}
\end{equation}
\begin{equation}
\frac{\partial A^{2}}{\partial t}+\nabla \cdot \biggr(A^{2}\frac{\nabla S}{m}\biggl)=0.
\label{bpr}
\end{equation}

Na interpreta\c c\~ao  de Copenhagen,
a Eq. (\ref{bpr}) \'e uma equa\c c\~ao de continuidade para a densidade de 
probabilidade $A^{2}$ de encontrar a part\'{\i}cula na posi\c c\~ao $x$ e tempo $t$.
Toda a informa\c c\~ao f\'{\i}sica do sistema est\'a contida em   $A^{2}$, e a fase total
$S$ da fun\c c\~ao de onda \'e completamente irrelevante. Nesta interpreta\c c\~ao
nada \'e dito a respeito de $S$ e sua equa\c c\~ao de evolu\c c\~ao (\ref{bqp}).
Mas suponhamos  que o termo 
$\frac{\hbar ^{2}}{2m}\frac{\nabla ^{2}A}{A}$ n\~ao esteja presente na equa\c c\~ao 
(\ref{bqp}). Ent\~ao podemos interpretar (\ref{bqp}) e (\ref{bpr}) como 
 equa\c c\~oes para um conjunto estat\'{\i}stico de part\'{\i}culas cl\'assicas submetidas
ao potencial cl\'assico $V$ satisfazendo a equa\c c\~ao de Hamilton-Jacobi (\ref{bqp}), cuja 
densidade de probabilidade de distribui\c c\~ao no espa\c co,  $A^{2}$,  verifica a 
equa\c c\~ao de continuidade (\ref{bpr}). $\nabla S(x,t) /m$ \'e o campo de
 velocidades 
$v(x,t)$ do conjunto de part\'{\i}culas. Quando o termo 
$\frac{\hbar ^{2}}{2m}\frac{\nabla ^{2}A}{A}$, que chamaremos de potencial qu\^antico,  est\'a presente, podemos ainda
interpretar   (\ref{bqp}) como uma equa\c c\~ao de Hamilton-Jacobi para o
conjunto de part\'{\i}culas. Mas agora, as trajet\'orias n\~ao v\~ao ser as cl\'assicas, devido
\`a presen\c ca do potencial qu\^antico na 
(\ref{bqp}).

A interpreta\c c\~ao de Bohm-de Broglie da mec\^anica qu\^antica esta baseada
nas {\it duas } equa\c c\~oes (\ref{bqp}) e (\ref{bpr}) do modo explicado acima,
n\~ao s\'o na \'ultima como  a interpreta\c c\`ao de Copenhagen.
O ponto de partida \'e que  a posi\c c\~ao $x$ e o momento $p$ est\~ao sempre 
bem definidos, sendo a part\'{\i}cula guiada por um novo campo: o campo qu\^antico. 
Este campo $\Psi $ satisfaz a equa\c c\~ao de Schr\"{o}dinger (\ref{bsc}) a qual 
\'e equivalente \`as duas equa\c c\~oes reais (\ref{bqp}) e (\ref{bpr}). A equa\c c\~ao 
(\ref{bqp})
\'e interpretada como uma equa\c c\~ao tipo Hamilton-Jacobi para a part\'{\i}cula qu\^antica
submetida a um potencial externo, o qual \'e a soma do potencial cl\'assico com 
o novo potencial qu\^antico:

\begin{equation}
Q\equiv -\frac{\hbar ^{2}}{2m}\frac{\nabla ^{2}A}{A}.  
\label{qp}
\end{equation}
O efeito do campo $\Psi $  sobre a trajet\'oria da part\'{\i}cula se d\'a atrav\'es do potencial 
qu\^antico (\ref{qp}). Uma vez obtido $\Psi $ ao resolver a equa\c c\~ao 
de  Schr\"{o}dinger, podemos obter a trajet\'oria da part\'{\i}cula, $x(t),$ integrando
a equa\c c\~ao diferencial $p=m\dot{x}=\nabla S(x,t)$, a qual \'e chamada 
de {\it rela\c c\~ao guia} ou {\it rela\c c\~ao de Bohm} (o ponto acima significa 
derivada temporal). Claro que vamos precisar conhecer a posi\c c\~ao inicial da part\'{\i}cula
para obter a trajet\'oria n\~ao cl\'assica $x(t),$ a partir desta equa\c c\~ao diferencial.
No entanto, n\'os  n\~ao conhecemos a posi\c c\~ao inicial da part\'{\i}cula pois n\~ao sabemos
como medi-la sem perturbar o sistema (esta \'e a variavel escondida da teoria). Para estar de acordo com
todos os experimentos qu\^anticos, \'e preciso postular que, para um conjunto estat\'{\i}stico
de part\'{\i}culas no mesmo campo qu\^antico $ \Psi $, a densidade de probabilidade 
de distribui\c c\~ao das posi\c c\~oes iniciais $x_{0}$ \'e $P(x_{0},t_0)=A^{2}(x_{0},t=t_0)$.
A equa\c c\~ao (\ref{bpr}) garante que $P(x,t)=A^{2}(x,t)$ para todo tempo. Deste modo, as 
previs\~oes estat\'{\i}sticas da teoria qu\^antica na interpreta\c c\~ao de Bohm-de Broglie 
s\~ao exatamente as mesmas que na interpreta\c c\~ao de Copenhaguen.\footnote{ Ja foi mostrado
que sob situa\c c\~oes ca\'oticas t\'{\i}picas, dentro da interpreta\c c\~ao de Bohm-de Broglie,
uma distribui\c c\~ao de probabilidade $P \neq A^2$ deve rapidamente convergir ao valor 
$P=A^2$ \cite{vig,val2}. Neste caso, o postulado da probabilidade inicial n\~ao seria necess\'ario, 
e poderiamos ter situa\c c\~oes, em intervalos de tempo muito curtos, onde 
esta  interpreta\c c\~ao de Bohm-de Broglie modificada poderia diferir da interpreta\c c\~ao de 
Copenhaguen.}

Resulta interessante notar que  o potencial qu\^antico $Q$ depende s\'o da forma de $\Psi $,
n\~ao do seu valor absoluto, como vemos da Eq.(\ref{qp}).
Este fato coloca em evid\^encia o car\'ater n\~ao local e contextual do potencial 
qu\^antico\footnote{A n\~ao localidade de $Q$ resulta evidente ao generalizarmos a 
interpreta\c c\~ao causal para um sistema de muitas part\'{\i}culas.}.
Esta \'e uma carater\'{\i}stica necess\'aria  pois as desigualdades de Bell, junto com os 
experimentos de Aspect, mostram que, em geral, uma teoria qu\^antica deve ser ou n\~ao 
local ou n\~ao ontol\'ogica. Dado que a interpreta\c c\~ao de Bohm-de Broglie \'e 
ontol\'ogica, ent\~ao ela deve ser n\~ao local. O potencial qu\^antico 
n\~ao local e contextual causa os efeitos a qu\^anticos. Ele n\~ao tem paralelo na 
fisica cl\'assica.

A fun\c c\~ao $A$ desempenha um papel duplo na interpreta\c c\~ao de Bohm-de Broglie:
fornece o potencial qu\^antico e tamb\'em a densidade de probabilidade de distribui\c c\~ao
das posi\c c\~oes, mas este \'ultimo    papel \'e secund\'ario. Se tivermos  algum modelo no 
qual a no\c c\~ao de probabilidade n\~ao se aplica, poder\'{\i}amos ainda assim obter informa\c c\~ao 
utilizando
as rela\c c\~oes guia. Neste caso, $A^2$ n\~ao precisa ser normaliz\'avel. A 
interpreta\c c\~ao de Bohm-de Broglie n\~ao \'e {\em em ess\^encia}, uma 
interpreta\c c\~ao probabil\'{\i}stica. Resulta imediata sua aplica\c c\~ao a um 
sistema individual.
O limite cl\'assico pode se obter de uma forma muito simples. S\'o precisamos
achar as condi\c c\~oes segundo as quais  $Q=0$\footnote{Seria interessante estudar
a cone\c c\~ao entre este limite cl\'assico bohmiano  e o fen\^omeno de descoer\^encia.
At\'e onde  sabemos, n\~ao foi feito nehum trabalho nesta dire\c c\~ao, o qual 
poderia iluminar tanto a interpreta\c c\~ao de Bohm-de Broglie quanto  a 
comprens\~ao do fen\^omeno da
descoer\^encia.}.
A quest\~ao de porque numa medida real  n\'os observamos um colapso efetivo da fun\c c\~ao
de onda \'e respondida notando que, numa medi\c c\~ao, a 
fun\c c\~ao
de onda  se divide numa superposi\c c\~ao de ramos que n\~ao se intersectam.
Ent\~ao  a part\'{\i}cula (que na verdade representa o objeto observado mais o aparelho de
medida macrosc\'opico) vai entrar num destes ramos (em qual deles depende das 
condi\c c\~oes iniciais) e ser\'a influenciada somente pelo potencial qu\^antico 
que corresponde
a este ramo particular, que \'e o unico n\~ao desprezivel na regi\~ao onde a part\'{\i}cula 
realmente est\'a. Os outros ramos vazios continuam existindo, mas eles n\~ao 
tem influ\^encia sobre a part\'{\i}cula medida nem sobre qualquer outra \cite{hol}.
Existe um colapso efetivo mas n\~ao real. A equa\c c\~ao de Schr\"{o}dinger \'e
sempre valida. N\~ao \'e necessario que exista um dom\'{\i}nio cl\'assico fora do 
sistema qu\^antico para poder explicar o proceso de medida, nem \'e crucial a 
exist\^encia de observadores ja que esta interpreta\c c\~ao \'e objetiva.

\'E possivel aplicar um racioc\'{\i}nio similar no caso da teoria qu\^antica de campos 
em espa\c co-tempo plano. Como exemplo, a equa\c c\~ao de Schr\"{o}dinger funcional
para um campo  qu\^antico escalar \'e:

\begin{equation}
\label{qsf}
i \hbar \frac{\partial \Psi (\phi ,t)}{\partial t} = 
\frac{1}{2}\int d^3x \biggr\{-\hbar ^2  
\frac{\delta ^2}{\delta \phi ^2} +
(\nabla \phi)^2 + U(\phi) \biggl\} \Psi (\phi ,t) .
\end{equation}
Escrevendo de novo o funcional de onda na forma polar $\Psi = A \exp (iS/\hbar)$, obtemos:

\begin{equation}
\frac{\partial S}{\partial t}+\int d^3x \biggr\{ \frac{1}{2}
\biggr[ \biggr(\frac{\delta S}{\delta \phi}\biggl)^2 +
(\nabla \phi)^2 + U(\phi) \biggl]+ {\cal Q}(\phi)\biggl\} = 0 ,
\end{equation}

\begin{equation}
\label{fp}
\frac{\partial A^2}{\partial t}+\int d^3 x \frac{\delta}{\delta \phi}
\biggr(A^2 \frac{\delta S}{\delta \phi}\biggl) = 0 ,
\end{equation}
onde ${\cal Q}(\phi) = -\hbar ^2 \frac{1}{2A} \frac{\delta ^2 A}{\delta \phi ^2}$ \'e o correspondente potencial qu\^antico (n\~ao regulado).
A primeira equa\c c\~ao \'e interpretada como uma  equa\c c\~ao de Hamilton-Jacobi
que governa a evolu\c c\~ao  de certa configura\c c\~ao inicial de campo no tempo,
a qual vai ser diferente da cl\'assica devido a presen\c ca do potencial qu\^antico.
A rela\c c\~ao guia ser\'a dada  por:

\begin{equation}
\Pi _{\phi} = \dot{\phi} = \frac{\delta S}{\delta \phi}.
\end{equation}
A segunda equa\c c\~ao \'e a equa\c c\~ao de continuidade para a densidade de 
probabilidade
$A^2[\phi(x),t_0]$ de que a configura\c c\~ao de campo  inicial a $t_0$ esteja 
dada por $\phi(x)$.

Um estudo detalhado da interpreta\c c\~ao de Bohm-de Broglie na teoria 
qu\^antica de campos
pode ser encontrado na Ref. \cite{kal} para o caso da eletrodin\^amica 
qu\^antica.

\baselineskip=2.0 \normalbaselineskip

\chapter{Teoria de Campos na Interpreta\c c\~ao de Bohm-de Broglie}


Neste cap\'{\i}tulo vamos estudar 
algumas caracter\'{\i}sticas da interpreta\c c\~ao de Bohm-de Broglie em teoria de 
campos. Al\'em de encontrarmos resultados interessantes para a teoria de campos, 
a saber, a prova da sua consist\^encia geral e a quebra da invari\^ancia 
relativ\'{\i}stica para processos individuais, 
a  metodologia desenvolvida aqui servir\'a como introdu\c c\~ao ao estudo da gravita\c c\~ao 
e cosmologia qu\^anticas
na interpreta\c c\~ao de Bohm-de Broglie, nos cap\'{\i}tulos seguintes.

\section{Teoria de campos parametrizada}
 
Uma caracter\'{\i}stica essencial da geometrodin\^amica \'e a exist\^encia 
dos v\'{\i}nculos 
super-hamiltoniano 
e super-momento, presentes, como sabemos, devido a invari\^ancia 
da TRG sob transforma\c c\~oes
gerais de coordenadas \cite{rack}.
 
 \'E poss\'{\i}vel encontrar (ou simular) uma situa\c c\~ao parecida em 
sistemas 
mec\^anicos com finitos graus de liberdade e tambem em teoria de campos no 
espa\c co-tempo plano, por 
meio de um processo
conhecido como {\it parametriza\c c\~ao} \cite{rack} \cite{lanckzos}.
Isto vai permitir construir a teoria de modo que os estados do campo estejam 
definidos
numa hipersuperf\'{\i}cie espacial geral. Deste modo, resulta manifesta a 
invariancia relativ\'{\i}stica
do formalismo hamiltoniano.
Ademais, esta forma parametrizada de se escrever a a\c c\~ao de campos 
no espa\c co-tempo plano facilitar\'a a implementa\c c\~ao
da interpreta\c c\~ao de Bohm-de Broglie em gravita\c c\~ao 
qu\^antica, onde a a\c c\~ao \'e parametrizada de inicio. De fato, a TRG  j\'a \'e
uma teoria parametrizada 
e at\'e  agora revelou-se imposs\'{\i}vel 
deparametriz\'a-la em geral no sentido de separar os graus de liberdade 
din\^amicos 
(genu\'{\i}nos) dos redundantes (cinem\'aticos).
Na TRG, estamos for\c cados a usar vari\'aveis redundantes  como 
coordenadas can\^onicas e por isso 
aparecem os v\'{\i}nculos.

Concretamente, seja um campo escalar  $\phi(X^{\alpha})$ 
propagando-se
num espa\c co-tempo plano de dimens\~ao $4$ com coordenadas minkowskianas 
$ X^{\alpha} \equiv (T,X^{i})$. Os indices gregos v\~ao de $0$ a $3$ e 
os indices latinos de $1$ a $3$. Consideremos as coordenadas
curvil\'{\i}neas  $ x^{\beta}=(t,x^{i})$ e seja a transforma\c c\~ao:
 
\begin{equation}
X^{\alpha}=X^{\alpha}(x^{\beta})
\end{equation}
Deixando $t$ fixo esta equa\c c\~ao representa  uma hipersuperf\'{\i}cie 
com um sistema de coordenadas espaciais $x^{i}$ definido sobre ela. Para
  diferentes 
valores do par\^ametro $t$ teremos uma fam\'{\i}lia de hipersuperf\'{\i}cies
rotuladas por $t$.

A a\c c\~ao dada em coordenadas minkowskianas \'e:

\begin{equation}
S=\int d^{4}X {\cal L}_{o} \biggr(\phi,\frac{\partial \phi}{\partial 
X^{\alpha}}\biggl)
\end{equation}
onde ${\cal L}_{o}$ representa a densidade lagrangeana em coordenadas 
minkowskianas.
A a\c c\~ao pode ser escrita nas  coordenadas curvil\'{\i}neas  resultando em:

\begin{equation}
S=\int d^{4}x J  {\cal L}_{o} \biggr(\phi, \frac{\partial \phi}
{\partial x^{\beta}}
 \frac{\partial x^{\beta}}{X^{\alpha}}\biggl)=\int d^{4}x {\cal L}\biggr(\phi, \phi_{,i}, 
 \dot{\phi}, X^{\alpha}_{,i}, \dot{X^{\alpha}}\biggl)
\end{equation}
onde $\dot{\phi}\equiv\frac{\partial \phi}{\partial x^{0}}$ e 
$,_{k} \equiv \frac{\partial}{\partial x^{k}}$ e

\begin{equation}
J \equiv \frac{\partial(X^{0}..X^{3})}{\partial(x^{0}..x^{3})}
\end{equation}
\'e o jacobiano da 
transforma\c c\~ao.
Deste modo ${\cal L}$ indica a densidade lagrangeana em coordenadas 
curvilineas. Definindo o momento can\^onico conjugado a $\phi$, $\pi_{\phi}$ na forma
usual:

\begin{equation}
\pi_{\phi}\equiv \frac{\partial {\cal L}}{\partial \dot{\phi}} ,
\end{equation}
obtemos a densidade hamiltoniana

\begin{equation}
{\bf h} = \pi_{\phi} \dot{\phi}-{\cal L} ,
\end{equation}
que \'e poss\'{\i}vel escrever como

\begin{equation}
{\bf h} = \frac{\partial x^{0}}{\partial X^{\alpha}} J T^{\alpha}_{\beta} 
\dot{X^{\beta}} \equiv K_{\beta}\dot{X^{\beta}}
\end{equation}
sendo $T^{\alpha}_{\beta}$ o tensor energia-momento nas coordenadas minkowskianas, dado por

\begin{equation}
\label{tensorme}
T^{\alpha}_{\beta} = \frac{\partial {\cal L}_{o}}{\partial 
\frac{\partial \phi}{\partial X^\alpha}} \frac{\partial \phi}{\partial X^\beta}-{\eta}^{\alpha}_\beta {\cal L}_{o}
\end{equation}
e $K_{\beta}$ foi definido como

\begin{equation}
K_{\beta} \equiv \frac{\partial x^{0}}{\partial X^{\alpha}} J T^{\alpha}_{\beta}
\end{equation}

A densidade hamiltoniana ${\bf h}$ resulta ter uma depend\^encia linear  
nas `velocidades cinematicas' $\dot{X^{\beta}}$, j\'a que $K_{\beta}$ 
independe delas. A densidade lagrangeana ser\'a ent\~ao dada por:

\begin{equation}
{\cal L}=\pi_{\phi} \dot{\phi}-K_{\beta}\dot{X^{\beta}}
\end{equation}
Podemos definir  os momentos `cinem\'aticos' como

\begin{equation}
\Pi_{\alpha}\equiv\frac{\partial {\cal L}}{\partial \dot{X^{\alpha}}} = 
-K_{\alpha} ,
\end{equation}
o que produz na verdade o v\'{\i}nculo

\begin{equation}
\Pi_{\alpha}+K_{\alpha}=0 ,
\end{equation}
ou seja,

\begin{equation}
\label{vice}
{\cal H}_{\alpha} \equiv \Pi_{\alpha} + \frac{\partial x^{0}}{\partial X^{\beta}} J T^{\beta}_{\alpha} = 0 .
\end{equation}
Deste modo \'e possivel  escrever a a\c c\~ao numa forma linear tanto nas 
velocidades din\^amicas 
$\dot{\phi}$ quanto nas velocidades cinem\'aticas
 $\dot{X^{\beta}}$, a saber

\begin{equation}
\label{alin}
 S=\int d^{4}x(\pi_{\phi} \dot{\phi}+\Pi_{\beta}\dot{X^{\beta}}) .
\end{equation}

Para que possamos variar livremente a a\c c\~ao sem nos preocuparmos 
com o v\'{\i}nculo (\ref{vice}), devemos acrecentar \`a mesma o 
termo $N^{\alpha}{\cal H}_{\alpha}$ sendo  $N^{\alpha}$ multiplicadores 
de Lagrange. Assim

\begin{equation}
\label{Talin}
 S=\int d^{4}x(\pi_{\phi} \dot{\phi}+\Pi_{\beta}\dot{X^{\beta}}-N^{\alpha}{\cal H}_{\alpha}) .
\end{equation}

Os v\'{\i}nculos (\ref{vice}) podem ser projetados nas dire\c c\~oes normal e paralela \'a 
hipersuperf\'{\i}cies $t=cte$

\begin{equation}
{\cal H}\equiv{\cal H}_{\alpha}n^{\alpha}
\end{equation}
\begin{equation}
{\cal H}_{i}\equiv{\cal H}_{\alpha}X^{\alpha}_{i}
\end{equation}
onde $X^{\alpha}_{i}$ s\~ao as componentes dos vetores  tangentes \`a hipersuperficie 
na base $\frac{\partial}{\partial X^{\alpha}}$, 
$\frac{\partial}{\partial x^i}=\frac{\partial X^{\alpha}}{\partial x^i}\frac{\partial}{\partial X^{\alpha}}$ e
o vetor normal \'e definido por

\begin{equation}
\eta_{\alpha \beta}n^{\alpha}n^{\beta}=\epsilon=\mp 1
\end{equation}

\begin{equation}
n_{\alpha} X^{\alpha}_{i}=0
\end{equation}
($-$ para assinatura hiperb\'olica e  $+$ para euclideana).

Portanto  a forma geral dos v\'{\i}nculos ser\'a dada pela  soma de uma parte cinem\^atica e de uma
parte din\^amica ou de campo:

\begin{equation}\label{vinculop0}
{\cal H} \equiv \Pi_{\alpha}n^{\alpha} + 
\frac{\partial x^{0}}{\partial X^{\beta}} J T^{\beta}_{\alpha}n^{\alpha} = 0
\end{equation}

\begin{equation}\label{vinculopi}
{\cal H}_{i} \equiv \Pi_{\alpha}X^{\alpha}_{i} + \frac{\partial x^{0}}{\partial X^{\beta}
} J T^{\beta}_{\alpha}X^{\alpha}_{i} = 0
\end{equation}
O v\'{\i}nculo ${\cal H}$ \'e chamado de super-hamiltoniano e o v\'{\i}nculo  
${\cal H}_{i}$ de super-momento. Expandindo $N^{\alpha}$ na base $(n^{\alpha}, X^{\alpha}_{i})$, 
$N^{\alpha} = Nn^{\alpha} + N^{i} X^{\alpha}_{i}$, teremos, para a a\c c\~ao,

\begin{equation}
\label{ali}
 S=\int d^{4}x(\pi_{\phi} \dot{\phi} + \Pi_{\beta}\dot{X^{\beta}} - N{\cal H} 
 - N^{i}{\cal H}_{i})
\end{equation}

Nesta a\c c\~ao, as vari\'aveis can\^onicas $\phi, \pi_{\phi}, X^{\alpha}, \Pi_{\alpha}$ 
s\~ao variadas, como j\'a dissemos, independentemente. As equa\c c\~oes de Hamilton que resultam v\~ao
determinar a evolu\c c\~ao dessas vari\'aveis can\^onicas com o tempo $t$.
 Ao variar com respeito aos multiplicadores 
de Lagrange $N$ e $N^i$
obtemos os v\'{\i}nculos

\begin{equation}\label{VV}
{\cal H}\approx 0, \hspace{0.5cm} {\cal H}_{i}\approx 0
\end{equation}

Utilizamos, nestas \'ultimas equa\c c\~oes, a nota\c c\~ao e terminologia de Dirac: 
os v\'{\i}nculos s\~ao fracamente
iguais a zero, indicando com isso que os par\^enteses de Poisson de uma quantidade 
$A(\phi,\pi_{\phi},X^{\alpha}, \Pi_{\alpha})$ com um v\'{\i}nculo fracamente zero, n\~ao \'e zero 
necessariamente.
Para que a teoria resulte consistente, os v\'{\i}nculos  devem ser preservados no 
tempo $t$, o qual significa que seus par\^enteses de Poisson com a hamiltoniana devem se anular, 
quer dizer, devem ser fracamente zero. 
A hamiltoniana est\'a dada por 

\begin{equation}
H = \int d^{3}x(N{\cal H} + N^{i}{\cal H}_{i}) \, ,
\end{equation}
Ent\~ao, devido \'a arbitrariedade dos multiplicadores $N$ e $N^i$, ser\'a 
$\dot{{\cal H}}\approx 0$ e
$\dot{{\cal H}_i}\approx 0$ somente si os colchetes de Poisson dos v\'{\i}nculos 
avaliados em dois pontos
$x$ e $y$ da hipersuperf\'{\i}cie 
$\{ {\cal H} (x), {\cal H} (y)\}$, 
$\{{\cal H}_i(x),{\cal H}(y)\}$ e $\{{\cal H}_i(x),{\cal H}_j(y)\}$ s\~ao fracamente zero.
Este c\'alculo foi feito por Dirac (com $\epsilon=-1$), que mostrou que este colchetes 
se escrevem como uma certa combina\c c\~ao linear dos v\'{\i}nculos originais 
(\'e dizer que n\~ao aparecem novos v\'{\i}nculos) e  
 satisfazem a seguinte \'algebra (chamada de `\'algebra de Dirac')\footnote{Rigurosamente 
nao \'e uma \'algebra ja que as constantes de estrutura dependem da 
m\'etrica\cite{hkt}}\cite{rack}
 \cite{dirac}:

\begin{equation}
\label{algebra1c}
\{ {\cal H} (x), {\cal H} (y)\}={\cal H}^i(x) {\partial}_i \delta^3(x,y)- 
{\cal H}^i(y){\partial}_i \delta^3(y,x)
\end{equation}
\begin{equation}
\label{algebra2c} 
\{{\cal H}_i(x),{\cal H}(y)\}={\cal H}(x) {\partial}_i \delta^3(x,y) 
\end{equation}
\begin{equation}
\label{algebra3c} 
\{{\cal H}_i(x),{\cal H}_j(y)\}={\cal H}_i(x) {\partial}_j \delta^3(x,y)- 
{\cal H}_j(y){\partial}_i \delta^3(y,x) 
\end{equation}
onde os \'{\i}ndices dos super-momentos sobem com a m\'etrica $h_{ij}$ induzida na hipersuperf\'{\i}cie $t=cte$, 
a qual est\'a dada por $h_{ij}=\eta_{\alpha \beta}X^{\alpha}_{,i}X^{\alpha}_{,j}$.
Dirac obteve este resultado com os v\'{\i}nculos dados na forma (\ref{vice}). 

Neste ponto resulta apropriado colocar um resultado, que ser\'a  de import\^ancia 
fundamental no nosso estudo, e que foi obtido por Claudio Teitelboim \cite{tei1}. Ele obteve esta \'algebra 
(mas com  a assinatura aparecendo explicitamente como vamos ver a seguir) de uma forma bem geral 
que independe da forma dos v\'{\i}nculos e sem ter assumido necessariamente um espa\c co-tempo 
de Minkowski .
 Nesse trabalho s\~ao estudadas as deforma\c c\~oes 
de uma hipersuperf\'{\i}cie embutida num espa\c co-tempo riemanniano. Intuitivamente, uma 
hipersuperf\'{\i}cie rotulada  pode ser deformada em geral segundo duas opera\c c\~oes: 
deixar ela fixa no espa\c co-tempo 
no qual esta embutida e simplesmente re-rotular seus pontos, ou bem manter os r\'otulos e deforma-la.
A primeira opera\c c\~ao representa uma deforma\c c\~ao $\delta t N^i$, tangencial 
\`a hipersuperf\'{\i}cie, sendo
governada por $\bar{{\cal H}_{i}}$. A segunda opera\c c\~ao representa uma deforma\c c\~ao
 $\delta t N$, ortogonal 
\`a hipersuperf\'{\i}cie e est\'a governada por $\bar{{\cal H}}$.
Qualquer funcional $F$ das vari\'aveis can\^onicas (campos e vari\'aveis cinem\^aticas) definidos na 
hipersuperf\'{\i}cie v\~ao mudar quando esta \'e  deformada, de acordo com o hamiltoniano dado por
\begin{equation}
\label{hgm}
\bar{H} = \int d^3x (N\bar{{\cal H}} + N^i\bar{{\cal H}}_i) \, ,
\end{equation}
de modo que

\begin{equation}
\delta F = \int d^3x \{ F, \delta N \bar{{\cal H}} + \delta N^i\bar{{\cal H}}_i \}\, .
\end{equation}  

Teitelboim utiliza  um argumento puramente geom\'etrico, baseado na `independ\^encia de caminho'
da evolu\c c\~ao din\^amica: a mudan\c ca nas vari\'aveis can\^onicas durante a evolu\c c\~ao 
desde uma dada hipersuperf\'{\i}cie inicial at\'e uma dada hipersuperf\'{\i}cie final deve 
ser independente da sequ\^encia particular de hipersuperf\'{\i}cies intermedi\'arias, utilizadas na 
avalia\c c\~ao desta mudan\c ca. Assumindo ent\~ao que as 3-geometr\'{\i}as est\~ao imersas numa variedade
4-dimensional n\~ao degenerada, juntamente com a consistencia da teoria, ele mostrou que os v\'{\i}nculos 
$\bar{{\cal H}} \approx 0$ e $\bar{{\cal H}}_i
\approx 0$ devem satisfazer a seguinte \'algebra

\begin{eqnarray}
\{ \bar{{\cal H}} (x), \bar{{\cal H}} (x')\}&=&-\epsilon[\bar{{\cal 
H}}^i(x) {\partial}_i \delta^3(x',x)
-  \bar{{\cal H}}^i(x') {\partial}_i \delta^3(x',x)] \, ,
\label{algebra1m} \\
\{\bar{{\cal H}}_i(x),\bar{{\cal H}}(x')\} &=& \bar{{\cal H}}(x)  
{\partial}_i \delta^3(x,x')  \, ,
\label{algebra2m} \\
\{\bar{{\cal H}}_i(x),\bar{{\cal H}}_j(x')\} &=& \bar{{\cal H}}_i(x)  
{\partial}_j \delta^3(x,x')- 
\bar{{\cal H}}_j(x') {\partial}_i \delta^3(x',x)  \, ,
\label{algebra3m}  
\end{eqnarray} 
onde os \'{\i}ndices dos super-momentos sobem com a m\'etrica $h_{ij}$ induzida na hipersuperf\'{\i}icie $t=cte$, 
a qual est\'a dada agora por $h_{ij}=g_{\alpha \beta}X^{\alpha}_{,i}X^{\alpha}_{,j}$, 
sendo $g_{\alpha \beta}$ a m\'etrica do espa\c co de fundo onde as hipersuperf\'{\i}cies est\~ao embutidas.
A constante 
$\epsilon$ na  Eq.(\ref{algebra1m}) pode ser $\pm 1$ dependendo se a 4-geometria 
na qual 
 as 3-geometrias est\~ao imersas, \'e euclideana
($\epsilon = 1$) ou hiperb\'olica ($\epsilon = -1$).
Esta an\'alise 
se aplica tanto 
para uma teoria de campos num fundo riemanniano j\'a dado, quanto para o caso em que o fundo \'e gerado
pela evolu\c c\~ao, como na TRG.
No primeiro caso  a estrutura da \'algebra dos v\'{\i}nculos
imp\~oe condi\c c\~oes para que a invari\^ancia de Lorentz n\~ao seja quebrada. No caso da TRG a \'algebra  fornece as condi\c c\~oes para 
a exist\^encia de um espa\c co-tempo: condi\c c\~oes de imersibilidade que asseguram que a 
evolu\c c\~ao das 
3-geometrias pode ser interpretada como o movimento de um `corte' 3-dimensional num espa\c co-tempo 
4-dimensional com assinatura lorentziana.
Este resultado aplicado ao caso da teoria de campos parametrizada em espa\c co-tempo plano  que estamos estudando,
implica que os v\'{\i}nculos da teoria devem satisfazer justamente a \'algebra  dada por (\ref{algebra1c})
 (\ref{algebra2c}) (\ref{algebra3c}).
 
Vamos considerar o caso de um campo escalar em espa\c co-tempo plano, cujo lagrangeano est\'a dado por

\begin{equation}
{\cal L}_{o} =-\frac{1}{2}\biggr(\eta^{\alpha \beta}\frac{\partial \phi}{\partial X^\alpha}
\frac{\partial \phi}{\partial X^\beta} + U(\phi)\biggl) \, ,
\end{equation}
onde $\eta^{\alpha \beta}=\eta_{\alpha \beta}=diag(-1,1,1,1)$
Calculando com este lagrangeano  o tensor momento-energia, Eq.(\ref{tensorme}),
e substituindo nas (\ref{vinculop0}) e (\ref{vinculopi})  vamos obter 
os v\'{\i}nculos super-hamiltoniano e super-momento na forma

\begin{equation}
\label{shc}
{\cal H}=\frac{1}{\nu}(\Pi_{\alpha}\nu^{\alpha} + \frac{1}{2}\pi_{\phi}^{2} + 
\frac{1}{2} \nu^2 (h^{ij}\phi_{,i}\phi_{,j} +  U(\phi))) = 0 \, ,
\end{equation}

\begin{equation}
\label{supmc}
{\cal H}_i= \Pi_ {\alpha} X^{\alpha}_{i} + \pi_\phi \phi_{,i} = 0   \, ,
\end{equation}
onde o  vetor normal \`a hipersuperf\'{\i}icie foi escrito na forma (veja\cite{lovelock} cap.7)
$n^{\alpha}=\frac{\nu^{\alpha}}{\nu}$, sendo

\begin{equation}
\label{nu}
\nu_{\alpha}\equiv -\frac{1}{3!}\epsilon_{\alpha \alpha1 \alpha2 \alpha 3}
\frac{\partial ( X^{\alpha 1} X^{\alpha 2} X^{\alpha 3})}{\partial ( x^{1} x^{2} x^{3})}
\end{equation}
e $\nu$ \'e a norma de $\nu^{\alpha}$
\begin{equation}
\nu= \sqrt{-\nu^\alpha \nu_{\alpha}}
\end{equation}

Pode se mostrar que resulta $-\nu^\alpha \nu_{\alpha}=h$ onde $h\equiv det(h_{ij})$ \'e o 
determinante da m\'etrica induzida na 
hipersuperf\'{\i}cie.

Os v\'{\i}nculos satisfazem, como vimos, a \'algebra de Dirac.
Na se\c c\~ao seguinte vamos quantizar este modelo e interpretar segundo Bohm-de Broglie, mas 
pasando a uma vis\~ao hamiltoniana da mesma.

\section{Teoria de campos parametrizadas na 
interpreta\c c\~ao de Bohm-de Broglie}

Nesta se\c c\~ao vamos estudar a interpreta\c c\~ao de 
Bohm-de Broglie da teoria de campos parametrizada, desenvolvida na se\c c\~ao 
anterior. 
Primeiramente  quantizaremos seguindo a prescri\c c\~ao de Dirac. As 
coordenadas $\phi^A \equiv (X^{0}, X^{1}, X^{2}, X^{3}, \phi)$ e 
os momentos $\pi_A \equiv (\Pi_{0}, \Pi_{1}, \Pi_{2}, \Pi_{3}, \pi_\phi)$
se tornam operadores, satisfazendo as rela\c c\~oes de comuta\c c\~ao

\begin{equation}
[\phi^{A}(x),\phi^{B}(y)] = 0 , [\pi_{A}(x),\pi^{B}(y)] = 0
\end{equation} 
\begin{equation}
[\phi^{A}(x),\pi_{B}(y)] = i \hbar \delta^A_B \delta(x,y)
\end{equation} 
e $x, y$ s\~ao dois pontos da hipersuperf\'{\i}cie.
Os v\'{\i}nculos atuam aniquilando o estado, produzindo 
condi\c c\~oes sobre os estados poss\'{\i}veis:

\begin{equation}
\label{smoc}
\hat{{\cal H}}_i \mid \Psi  \! > = 0 
\end{equation}
\begin{equation}
\label{wdwc}
\hat{{\cal H}} \mid \Psi  \! > = 0 
\end{equation}

Na representa\c c\~ao  $\phi^{A}(x)$ (`de coordenadas') o estado do campo 
escalar est\'a 
descrito pelo funcional $\Psi[\phi^{A}(x)]$ 
e o operador
momento \'e uma derivada funcional: 
$\pi_{A}(x)=-i\hbar\frac{\delta}{\delta \phi^{A}(x)}$.
Substituindo na Eq. (\ref{smoc}) e levando  em conta o super-momento Eq.(\ref{supmc}) temos

\begin{equation}
\label{suminv}
 -i \hbar X^{\alpha}_{i} \frac{\delta \Psi}{\delta X^{\alpha}(x)} -
 i \hbar  \phi_{,i} \frac{\delta \Psi}{\delta \phi(x)} = 0
\end{equation}
Esta equa\c c\~ao implica que $\Psi$ \'e um invariante sob 
transforma\c c\~oes de coordenadas espaciais 
na hipersuperf\'{\i}cie.

Substituindo na Eq. (\ref {wdwc}) o super-hamiltoniano, dado na se\c c\~ao anterior na Eq. (\ref{shc}), temos

\begin{equation}
\label{supHpsi}
{\cal H}(x)\Psi=\frac{1}{\nu}\biggr(-i\hbar \nu^{\alpha}\frac{\delta \Psi}{\delta X^{\alpha}(x) } -
 (\hbar)^2\frac{1}{2}\frac{\delta^2 \Psi}{\delta \phi(x)^2} + 
 \frac{1}{2} \nu^2 \biggr( h^{ij}(x)\phi(x)_{,i}\phi(x)_{,j} +  U(\phi(x))\biggl)\Psi \biggl) = 0
\end{equation}

Para interpretar segundo Bohm-de Broglie fazemos como \'e usual,  
escrevemos o funcional de onda em forma polar 
 $\Psi=A e^{\frac{i}{\hbar}S}$. Substituindo na (\ref{suminv}) vamos obter duas equa\c c\~oes que 
 indicam que tanto $S$ quanto $A$ s\~ao  invariantes sob transforma\c c\~oes gerais 
 de coordenadas espaciais
 
 \begin{equation}
\label{suminvS}
 X^{\alpha}_{i}\frac{\delta S}{\delta X^{\alpha}(x)} + \phi_{,i} \frac{\delta S}{\delta \phi(x)} = 0
\end{equation}
\begin{equation}
\label{suminvA}
 X^{\alpha}_{i}\frac{\delta A}{\delta X^\alpha(x)} + \phi_{,i} \frac{\delta A}{\delta \phi(x)} = 0
\end{equation}

Substituindo a forma polar da $\Psi$ na Eq.(\ref{supHpsi}) vamos obter duas equa\c c\~oes 
que depender\~ao
do ordenamento escolhido. Por\'em, a equa\c c\~ao que sai da parte real, depois de dividir pela 
amplitude $A$, ser\'a

\begin{equation}
\label{hjc}
\frac{1}{\nu}\biggr( \nu^{\alpha}\frac{\delta S}{\delta X^{\alpha}(x) } + \frac{1}{2}\biggr(\frac{\delta S}{\delta \phi }\biggl)^2
+\frac{1}{2} \nu^2 (h^{ij}(x)\phi(x)_{,i}\phi(x)_{,j} +  U(\phi(x)))\biggl) +{\cal Q} = 0
\end{equation}
Esta \'e uma equa\c c\~ao tipo Hamilton-Jacobi modificada pelo potencial 
qu\^antico,  dado pelo \'ultimo termo. Vemos que somente este termo vai depender da 
regulariza\c c\~ao e ordenamento, ja que os outros termos desta equa\c c\~ao est\~ao 
bem definidos. Segundo a forma n\~ao regulada dada na Eq (\ref{supHpsi}), ${\cal Q}$ resulta:

\begin{equation}
{\cal Q}=-\frac{1}{\nu}\frac{\hbar^2}{2A}\frac{\delta^2 A}{\delta \phi(x)^2}
\end{equation}
a outra equa\c c\~ao, que sai reordenando a  parte imaginaria, \'e

\begin{equation}
\nu^{\alpha}\frac{\delta A^2}{\delta X^{\alpha}}+
\frac{\delta(A^2 \frac{\delta S}{\delta \phi})}{\delta \phi}=0
\end{equation}

Notamos que na interpreta\c c\~ao de Bohm-de Broglie as vari\'aveis can\^onicas existem independentemente
da observa\c c\~ao, e, como vimos no cap\'{\i}tulo 2, a evolu\c c\~ao das coordenadas 
can\^onicas $\phi$ e $X^{\alpha}$ \'e obtida das 
 rela\c c\~oes guia de Bohm, dadas por:

\begin{equation}
\label{grx}
\Pi_{\alpha} = \frac{\delta S(\phi, X^{\alpha})}{\delta X^{\alpha}}
\end{equation} 

\begin{equation}
\label{grc}
\pi_\phi=\frac{\delta S(\phi, X^{\alpha})}{\delta \phi}
\end{equation}

Dados os valores iniciais do campo $\phi(t_0, x^i )$ e das vari\'aveis cinem\^aticas $ X^{\alpha}(t_0, x^i)$ 
numa hipersuperf\'{\i}cie inicial $x^{0}=t_0=cte$, podemos integrar estas equa\c c\~oes de primeiro ordem 
e obter assim as trajet\'orias bohmianas, isto \'e,  os valores do campo $\phi(t,x^i)$ e das $X^{\alpha}(t,x^i)$ 
para todo valor do par\^ametro $t$. 
A evolu\c c\~ao desses
campos ser\'a differente  da cl\'assica devido \`a presen\c ca do potencial qu\^antico na Eq. de 
Hamilton-Jacobi da teoria de Bohm-de Broglie, Eq. (\ref{hjc}). Como sabemos, o limite cl\'assico \'e obtido 
exigindo-se que 
${\cal Q}=0$. Neste caso o funcional $S$ \'e solu\c c\~ao da equa\c c\~ao de Hamilton -Jacobi 
cl\'assica, e sabemos que ao integrar as equa\c c\~oes (\ref{grx}) e (\ref{grc}), as solu\c c\~oes que se obtem
representam
um campo $\phi$ evoluindo num espaco-tempo de Minkowski. Isto segue do fato de que os v\'{\i}nculos da teoria 
cl\'assica
satisfazem a 
algebra de Dirac (\ref{algebra1m}) (\ref{algebra2m}) (\ref{algebra3m}) com $\epsilon=-1$. 
Mas, se o potencial qu\^antico n\~ao \'e zero, ent\~ao $S$ \'e solu\c c\~ao da equa\c c\~ao 
de Hamilton-Jacobi {\it modificada} (\ref{hjc}) e, portanto, n\~ao podemos assegurar que a solu\c c\~ao obtida 
para $\phi^A$
represente todavia um campo num espa\c co tempo de Minkowski. Os efeitos qu\^anticos poderiam quebrar 
a invari\^ancia de Lorentz e modificar assim a causalidade einsteniana da relatividade especial. Ent\~ao
perguntamos: qual  tipo de estrutura  corresponder\'a \`a este caso?. Para encarar esta quest\~ao vamos
re-escrever a teoria de Bohm-de Broglie, que est\'a formulada usualmente em termos da equa\c c\~ao de
Hamilton-Jacobi, numa forma Hamiltoniana.

As rela\c c\~oes de Bohm (\ref{grx}) (\ref{grc})
permitem escrever  (\ref{hjc}) na forma:

\begin{equation}
\label{vshcn}
\frac{1}{\nu}\biggr( \nu^{\alpha}\Pi_{\alpha} + \frac{1}{2} \pi_\phi{^2}
+\frac{1}{2} \nu^2 \biggr(h^{ij}(x)\phi(x)_{,i}\phi(x)_{,j} +  U(\phi(x))\bigg)\biggl) + {\cal Q} = 0 
\end{equation}

O potencial qu\^antico ${\cal Q}$ resulta ser uma densidade escalar de peso 1. Isto pode ser visto considerando a 
expres\~ao para ${\cal Q}$ que se obtem da equa\c c\~ao de Hamilton-Jacobi modificada, Eq.(\ref{hjc})

\begin{equation}
{\cal Q}= -\frac{1}{\nu}\biggr( \nu^{\alpha}\frac{\delta S}{\delta X^{\alpha}(x) } + \frac{1}{2}(\frac{\delta S}{\delta \phi })^2
+\frac{1}{2} \nu^2 \biggr(h^{ij}(x)\phi(x)_{,i}\phi(x)_{,j} +  U(\phi(x))\biggl)\biggl)
\end{equation}
Lembramos que $\nu=\sqrt{h}$ \'e uma densidade escalar de peso 1, e que 
a fase $S$ \'e um invariante perante transforma\c c\~oes gerais de coordenadas na hipersuperf\'{\i}cie 
(isto resulta do v\'{\i}nculo super-momento aplicado a $\Psi$, Eq.(\ref{suminvS})). Ent\~ao  
$\frac{\delta S}{\delta X^{\alpha}}$ \'e uma densidade vetorial, que estando 
contra\'{\i}da com o vetor normal,
resulta em uma densidade escalar de peso 1. Para o segundo termo usamos o mesmo racioc\'{\i}nio e 
o terceiro \'e obviamente uma densidade de peso 1. Assim ${\cal Q}$ \'e uma  soma de densidades 
escalares de peso 1, e portanto 
ele  tamb\'em \'e.

Podemos escrever a Eq.(\ref{vshcn}) da seguinte forma

\begin{equation}
{\cal H} +  {\cal Q} = 0  
\end{equation}
onde ${\cal H}$ \'e o super-hamiltoniano cl\'assico dado por (\ref{shc} ).
Ent\~ao, o super-hamiltoniano qu\^antico 
ou de Bohm vai ser:

\begin{equation}
\label{hqc}
{\cal H}_Q \equiv {\cal H} + {\cal Q}
\end{equation}
O hamiltoniano
que gera as trajetorias bohmianas, uma vez satisfeitas inicialmente as 
rela\c c\~oes  guia (\ref{grx}) e (\ref{grc}) ser\'a: 

\begin{equation}
\label{hqca}
H_Q = \int d^3x\biggr[N {\cal H}_Q + N^i{\cal H}_i\biggl] \, .
\end{equation}
Isto pode-se ver notando que as rela\c c\~oes guia s\~ao consistentes com os v\'{\i}nculos 
${\cal H}_Q \approx 0$ 
e ${\cal H}_i \approx 0$, pois $S$ satisfaz (\ref{suminvS}) e (\ref{hjc}). Ademais elas s\~ao 
conservadas na evolu\c c\~ao gerada pelo hamiltoniano (\ref{hqca}). Vamos mostrar isto. Escrevemos
primeiramente as rela\c c\~oes de Bohm (\ref{grx}) (\ref{grc})  de uma forma adaptada ao 
formalismo hamiltoniano, a saber

\begin{equation} 
\label{grxv}
\Phi_\alpha \equiv \Pi_\alpha - \frac{\delta S}{\delta X^{\alpha}} \approx 0 \, ,
\end{equation} 

\begin{equation}
\label{grcv}
\Phi_\phi \equiv \pi_\phi - \frac{\delta S}{\delta \phi} \approx 0 \, .
\end{equation}

A conserva\c c\~ao no tempo das rela\c c\~oes de Bohm  significa que $\dot{\Phi_{\phi}} \equiv \{\Phi_{\phi}, H_Q \} = 0$ e $\dot{\Phi_{\alpha}} \equiv \{\Phi_{\alpha}, H_Q\}= 0$. 
Isto por  sua vez equivale a provar que os par\^enteses de Poisson com 
os v\'{\i}nculos ${\cal H}_Q$ e ${\cal H}_i$ se anulam.
Calculemos ent\~ao  $\{\Phi_\phi, {\cal H}_Q \}$, $\{\Phi_\alpha, {\cal H}_Q \}$, $\{\Phi_\phi, {\cal H}_i \}$ e
$\{\Phi_\alpha, {\cal H}_i \}$. Para simplificar a nota\c c\~ao, definimos 
$W \equiv h^{ij}(x)\phi(x)_{,i}\phi(x)_{,j} +  U(\phi(x))$, de modo que o hamiltoniano qu\^antico se escreve

\begin{equation}
{\cal H}_Q \equiv {\cal H} + {\cal Q}=\frac{1}{\nu}\biggr( \nu^{\alpha}\Pi_{\alpha} +
 \frac{1}{2} \pi_\phi{^2}
+\frac{1}{2} \nu^2 W \biggl) +{\cal Q}  
\end{equation}
Calculando temos
\begin{eqnarray}
\{\Phi_{\phi}(y), {\cal H}_Q(x) \}=\biggr\{ \Pi_\alpha - \frac{\delta S}{\delta X^{\alpha}}, 
\frac{1}{\nu}\biggr( \nu^{\alpha}\Pi_{\alpha} + \frac{1}{2} \pi_\phi{^2}
+\frac{1}{2} \nu^2 W \biggl) +{\cal Q} \biggl\} = \nonumber \\ 
-\frac{\delta}{\delta \phi(y)} \biggr\{\frac{1}{\nu}\biggr( \nu^{\alpha}\frac{\delta S}{\delta X^{\alpha}(x) } + 
\frac{1}{2}\biggr(\frac{\delta S}{\delta \phi }\biggl)^2
+\frac{1}{2} \nu^2 W \biggl) +{\cal Q}\bigg\}-\frac{1}{\nu}\frac{\delta^2 S}{\delta \phi^2}\Phi_{\phi}
\end{eqnarray}
onde o primeiro termo do lado direito desta equa\c c\~ao respresenta a derivada funcional 
com rela\c c\~ao a $\phi(y)$, do  
lado esquerdo da  da equa\c c\~ao de Hamilton-Jacobi modificada, Eq (\ref{hjc}). Por tanto \'e identicamente 
zero. O segundo termo do  lado direito resulta ser fracamente zero em virtude da rela\c c\~ao de Bohm
(\ref{grcv}). Temos ent\~ao que

\begin{equation}
\{\Phi_\phi(y), {\cal H}_Q(x) \}=-\frac{1}{\nu}\frac{\delta^2 S}{\delta \phi(y)^2}\Phi_{\phi}(x)\approx 0
\end{equation}

Para o par\^enteses $\{\Phi_\alpha(y), {\cal H}_Q(x)\}$ temos

\begin{eqnarray}
\{\Phi_\alpha(y), {\cal H}_Q(x)\}=-\frac{\delta }{\delta X^{\alpha}(y)}\biggr\{\frac{1}{\nu}\biggr( \nu^{\alpha}\frac{\delta S}{\delta X^{\alpha}(x) } 
+ \frac{1}{2}\biggr(\frac{\delta S}{\delta \phi(x)}\biggl)^2 
+\frac{1}{2} \nu^2 W \biggl) +{\cal Q} \biggl\} \nonumber \\ -\frac{1}{\nu}\frac{\delta \nu^\beta}{\delta X^\alpha(y)}\Phi_\beta-
\frac{\delta \nu^{-1}}{\delta X^{\alpha}(y)}\nu^{\beta}\Phi_{\beta}- 
\frac{1}{2}\frac{\delta \nu^{-1}}{\delta X^{\alpha}(y)}\biggr(\Phi_{\phi}^2+2\frac{\delta S}{\delta \phi} \Phi_{\phi}\biggl)
-\frac{1}{\nu}\frac{\delta^2 S}{\delta\phi(x) \delta X^{\alpha}(y)}\Phi_{\phi}\approx 0
\end{eqnarray}
onde o primeiro termo do  lado direito desta equa\c c\~ao respresenta a derivada funcional 
com rela\c c\~ao a $\ X^{\alpha}(y)$, do lado esquerdo   da equa\c c\~ao de 
Hamilton-Jacobi modificada, Eq (\ref{hjc}), sendo, portanto, identicamente zero. Os outros termos s\~ao fracamente zero em 
virtude das rela\c c\~oes de Bohm (\ref{grxv}) (\ref{grcv}).

Para calcular os par\^enteses de Poisson que envolvem o v\'{\i}nculo super-momento usamos o fato 
de que este \'e 
gerador de transforma\c c\~oes espaciais de coordenadas. Temos que, sendo $S$ um invariante, ent\~ao $\Phi_\alpha$ 
\'e uma densidade vetorial e $\Phi_\phi$ uma densidade escalar.
Portanto temos

\begin{equation}
\{\Phi_\phi(y), {\cal H}_i(x) \}=-\Phi_{\phi}(x) \partial_{i}\delta(y,x)\approx 0 \, ,
\end{equation}

\begin{equation}
\{\Phi_\alpha(y), {\cal H}_i(x) \}=\Phi_i(x)\partial_{\alpha}\delta(y,x)-\Phi_\alpha(y)\partial_{i}\delta(y,x)\approx 0 \, .
\end{equation}

Combinando estes resultados obtemos

\begin{equation}
\dot{\Phi}_{\phi}=\{\Phi_\phi, H_Q\} \approx 0 \, ,
\end{equation}

\begin{equation}
\dot{\Phi}_{\alpha,}=\{\Phi_\alpha,, H_Q\} \approx 0 \, .
\end{equation}
Ou seja, as rela\c c\~oes guia de Bohm s\~ao conservadas.

Dado que o potencial qu\^antico n\~ao depende dos momentos, temos que as 
defini\c c\~oes dos momentos em termos das velocidades continuam sendo as mesmas do caso cl\'assico:

\begin{equation}
\dot{\phi}=\{\phi,H_Q\}=\{\phi,H\} \, ,
\end{equation}

\begin{equation}
\dot{X^{\alpha}}=\{X^{\alpha},H_Q\}=\{X^{\alpha},H\} \, .
\end{equation}

Expressamos a teoria de Bohm-de Broglie em forma hamiltoniana, e
estamos interesados em estudar que tipo de estrutura vai corresponder \`a evolu\c c\~ao bohmiana
gerada pelo  hamiltoniano (\ref{hqca}). Os v\'{\i}nculos ${\cal H}_i\approx0$ e ${\cal H}_Q\approx0$  devem
se manter no tempo para que a teoria resulte consistente. A consist\^encia da teoria equivale a que os 
v\'{\i}nculos tenham par\^enteses de Poisson fracamente zero entre eles.
No contexto do trabalho de Teitelboim explicado antes, vamos analizar
a \'algebra satisfeita pelos v\'{\i}nculos ${\cal H}_i\approx0$ e ${\cal H}_Q\approx0$.
O par\^enteses de Poisson 
$\{{\cal H}_i (x),{\cal H}_j (y)\}$ satisfaz a Eq.
(\ref{algebra3c}) ja que o ${\cal H}_i$ de $H_Q$ definido na  Eq.
(\ref{hqca}) \'e o mesmo que na teoria cl\'assica.
Da mesma forma, $\{{\cal H}_i (x),{\cal H}_Q (y)\}$ satisfaz a Eq. (\ref{algebra2c}) pois 
${\cal H}_i$ \'e o gerador de transforma\c c\~oes espaciais de coordenadas,
e como ${\cal H}_Q$ \'e uma 
densidade  escalar de peso 1 pois  $Q$ \'e uma densidade escalar 
de peso 1, 
ent\~ao ele deve satisfazer esta rela\c c\~ao de colchetes de Poisson 
com ${\cal H}_i$. O que resta ser verificado \'e se 
o colchete de Poisson
$\{{\cal H}_Q (x),{\cal H}_Q (y)\}$ fecha e se \'e como na Eq. 
(\ref{algebra1c}).
Vamos ver que efetivamente este colchete resulta fracamente zero 
independentemente 
do potencial qu\^antico. Este fato significa que a teoria \'e consistente 
para qualquer ${\cal Q}$ e, portanto,
para qualquer estado. Assim,

\begin{equation}
\label{cp1}
\{{\cal H}_Q (x),{\cal H}_Q (y)\}=\{{\cal H}(x),{\cal H} (y)\} + 
\{{\cal H} (x),Q (y)\} + \{ Q (x),{\cal H} (y)\} \, .
\end{equation}
Da equa\c c\~ao (\ref{hjc}) podemos escrever o potencial qu\^antico como:

\begin{equation}
{\cal Q}= -\frac{1}{\nu}\biggr( \nu^{\alpha}\frac{\delta S}{\delta X^{\alpha}(x) } + \frac{1}{2}(\frac{\delta S}{\delta \phi })^2
+\frac{1}{2} \nu^2 \biggr(h^{ij}(x)\phi(x)_{,i}\phi(x)_{,j} +  U(\phi(x))\biggl)\biggl)
\end{equation}

Substituindo esta \'ultima na Eq.(\ref{cp1}) e usando as rela\c c\~oes guia de 
Bohm dadas por (\ref{grxv}) (\ref{grcv})
encontramos que
\begin{eqnarray}
\{{\cal H}_Q (x),{\cal H}_Q (y)\}=\frac{1}{\nu(x)\nu(y)}\biggr(\biggr(\frac{\delta S}{\delta \phi(y)}
\frac{\delta^2 S}{\delta \phi(x) \delta \phi(y)}+ \nonumber \\
\nu^{\alpha}(y)\frac{\delta^2 S}{\delta X^{\alpha}(y) \delta \phi(x)}\biggl)\Phi_{\phi}(x)-
\biggr(\frac{\delta S}{\delta \phi(x)}
\frac{\delta^2 S}{\delta \phi(y) \delta \phi(x)}+ \nonumber \\
\nu^{\alpha}(x)\frac{\delta^2 S}{\delta X^{\alpha}(x) \delta \phi(y)}\biggl)\Phi_{\phi}(y) +
\nu^{\alpha}(y)\frac{\delta \nu^{\beta}(x)}{\delta X^{\alpha}(y)}\Phi_{\beta}(y) - 
\nu^{\alpha}(x)\frac{\delta \nu^{\beta}(y)}{\delta X^{\alpha}(x)}\Phi_{\beta}(y)\biggl) \approx 0
\end{eqnarray}
O lado direito desta equa\c c\~ao \'e fracamente zero em virtude das   
rela\c c\~oes de Bohm (\ref{grxv}) e (\ref{grcv}).

Vemos, portanto, que a interpreta\c c\~ao de Bohm-de Broglie de um 
campo escalar  num
 fundo de Minkowski,  \'e uma 
teoria consistente. Mas, a \'algebra dos v\'{\i}nculos  n\~ao fecha necessariamente 
segundo a \'algebra de Dirac. Isto vai depender da forma do $Q$. Se o potencial 
qu\^antico quebra a \'algebra de Dirac
ent\~ao, de acordo com o trabalho de Teitelboim sintetizado na se\c c\~ao anterior,
 a estrutura do espa\c co-tempo de fundo vai ser modificada, n\~ao 
 ser\'a mais  Minkowski. Isto significa que a invari\^ancia de Lorentz 
ser\'a quebrada. Uma situa\c c\~ao  an\'aloga vai ser mostrada no caso da 
geometrodin\^amica qu\^antica, onde o potencial qu\^antico vai determinar  
a estrutura qu\^antica do Universo. 
A seguir mostraremos que j\'a o estado de v\'acuo do campo escalar livre produz um 
potencial qu\^antico que quebra a \'algebra de Dirac. O c\'alculo deste potencial qu\^antico
 \'e presentado no 
apendice A, onde estudamos um campo escalar livre  num espa\c co-tempo de Minkowski.
 Ali \'e mostrado que o  potencial qu\^antico para o estado de v\'acuo \'e

\begin{equation}
Q=-\frac{1}{2}\int d^3X  \biggr( \int d^3Y \frac{d^3k}{(2\pi)^3} \omega_k \cos k.(X-Y ) \phi(Y)\biggl)^2 + \frac{1}{2}\int d^3 X \int d^3k \omega_k 
\end{equation}
que escrevemos como

\begin{equation}
Q=\int d^3X f( X^i, \phi) \, ,
\end{equation}
onde $f$ \'e uma fun\c c\~ao de $X^i$
e um funcional de $\phi$ dado por

\begin{equation}
f\equiv -\frac{1}{2}\biggr(\int d^3Y \frac{d^3k}{(2\pi)^3} \omega_k \cos\{k.(X-Y)\}
 \phi(Y)\biggl)^2 + \frac{1}{2}\int d^3k \omega_k  \, .
\end{equation}
Escreveremos  $f$ como

\begin{equation}
f=-B^2+E_0\, ,
\end{equation}
onde

\begin{equation}
B\equiv\sqrt{\frac{1}{2}}\int d^3Y \frac{d^3k}{(2\pi)^3} \omega_k \cos\{k.(X-Y)\}
 \phi(Y)\, ,
\end{equation}
\begin{equation}
E_0\equiv \frac{1}{2}\int d^3 X \int d^3k \omega_k 
\end{equation}
Pasando \`as coordenadas da hipersuperf\'{\i}cie $x^i$ temos

\begin{equation}
Q=\int d^3x J f(X^i(x^j),\phi) \, ,
\end{equation}
sendo $J$ o jacobiano $ J=\frac{1}{3!}\epsilon_{ijk}\epsilon^{abc}
\frac{\partial X^i}{\partial x^a}\frac{\partial X^j}{\partial x^b}\frac{\partial X^k}{\partial x^c}$.
Assim, o ${\cal Q}$ (a densidade) que entra no super-hamiltoniano qu\^antico 
Eq. (\ref{hqc}) ser\'a:

\begin{equation}
{\cal Q}= J f(X^i(x^j),\phi)
\end{equation}

Calculemos o par\^enteses de Poisson $\{{\cal H}_Q (x),{\cal H}_Q (y)\}$. Temos

\begin{eqnarray}
\{{\cal H}_Q (x),{\cal H}_Q (y)\}=\{{\cal H}(x),{\cal H} (y)\} + 
\{{\cal H} (x),Q (y)\} + \{ Q (x),{\cal H} (y)\}= \nonumber \\ 
{\cal H}^i(x) {\partial}_i \delta^3(x,y)- 
{\cal H}^i(y){\partial}_i \delta^3(y,x)+ \{{\cal H} (x),Q (y)\} + \{ Q (x),{\cal H} (y)\}
\end{eqnarray}
onde os dois primeiros termos  do lado direito s\~ao exatamente aqueles 
que aparecem na \'algebra de Dirac Eq.(\ref{algebra1c}). Portanto, para que a \'algebra de 
Dirac se mantenha, \'e necess\'ario que $ \{{\cal H} (x),Q (y)\} + \{ Q (x),{\cal H} (y)\}$ seja fortemente zero.
Entretanto,
 
\begin{eqnarray}
 \{{\cal H} (x),Q (y)\} + \{ Q (x),{\cal H} (y)\}= +2\frac{\nu^\alpha(y)}{\nu(y)}f(y)\epsilon_{\alpha j k}\epsilon^{abc}
\frac{\partial X^j}{\partial y_b}\frac{\partial X^k}{\partial y_c}\frac{\partial \delta(y,x)}{\partial y_a}+ \nonumber \\
2\frac{J(y)}{\nu(x)}\pi_{\phi}B(y)\int\frac{d^3k}{(2\pi)^3} \omega_k \cos{k.(X(y)-x)} - 
2\frac{\nu^\alpha(x)}{\nu(x)}f(x)\epsilon_{\alpha j k}\epsilon^{abc}
\frac{\partial X^j}{\partial x_b}\frac{\partial X^k}{\partial x_c}\frac{\partial \delta(x,y)}{\partial x_a}+ \nonumber \\
2\frac{J(x)}{\nu(y)}\pi_{\phi}(y)B(x)\int\frac{d^3k}{(2\pi)^3} \omega_k \cos{k.(X(x)-y)}  
\end{eqnarray}
e o lado direito desta equa\c c\~ao \'e evidentemente diferente de zero.
Assim a \'algebra de Dirac n\~ao \'e satisfeita neste exemplo particular. 
Isto est\'a nos dizendo, no 
contexto geometrodin\^amico
do trabalho de Teitelboim j\'a discutido, que as trajet\'orias bohmianas est\~ao gerando uma 
estrutura que n\~ao corresponde a um campo relativ\'{\i}stico no  espa\c co-tempo de Minkowski. 
Em outras palavras, mostramos a quebra
 da invari\^ancia
de Lorentz da teoria em termos da quebra da \'algebra de Dirac dos v\'{\i}nculos.  Ressaltamos que a 
invari\^ancia relativ\'{\i}stica \'e perdida a n\'{\i}vel de eventos individuais. As propriedades
observ\'aveis do campo s\~ao b\'asicamente estat\'{\i}sticas e est\~ao contidas nos valores esperados 
dos operadores

\begin{equation}
  <  \Psi \mid \hat{A} \mid \Psi  > = \int \Psi*[\phi](\hat{A}\Psi)[\phi]D\phi
\end{equation}
os quais continuam sendo invariantes. Enquanto a invari\^ancia dos eventos 
individuais pode  em 
geral ser quebrada, como vimos explicitamente  no exemplo acima, a relatividade 
especial vai ser verificada 
nos experimentos
que testem e confirmem o formalismo probabil\'{\i}stico. A invari\^ancia 
Lorentz \'e um efeito estat\'{\i}stico \cite{bohm87} \cite{hol}. Salientamos que
os resultados obtidos neste cap\'{\i}tulo n\~ao dependem de se ter 
assumido dimens\~ao 4 e s\~ao v\'alidos para um espa\c co-tempo de Minkowski 
de qualquer dimens\~ao $n \geq 2$.

\baselineskip=2.0 \normalbaselineskip

\chapter{Qu\^antiza\c c\~ao Can\^onica da Gravita\c c\~ao}


N\~ao existe ainda uma teoria qu\^antica da gravita\c c\~ao estabelecida.
A TRG conduz a sua propria inaplicabilidade: 
suaves condi\c c\~oes iniciais podem evoluir em singularidades na forma 
de buracos negros, talvez num grande colapso final (`big crunch'), ou evoluiram de uma 
singularidade inicial cosmol\'ogica chamada de `big bang'. 
Nessas configura\c c\~oes  a teoria perde o seu poder preditivo. Este \'e o 
conte\'udo dos `teoremas das
singularidades'  \cite{singul}. Assim, \'e nosso trabalho procurar outra 
teor\'{\i}a que 
possa descrever 
 o Universo, mesmo num hipot\'etico 
instante de cria\c c\~ao. Neste caso, temos que admitir que as suaves 
condi\c c\~oes 
iniciais assumidas nos teoremas de singularidades n\~ao s\~ao mais v\'alidas
sob situa\c c\~oes extremas de 
densidades de energia 
e curvatura muito altas (escala de Planck). Podemos dizer que a TRG
 e outras teorias que descrevem 
os campos de mat\'eria devam ser modificadas sob estas condi\c{c}\~oes 
extremas.

Para realizar esta modifica\c{c}\~ao, poder\'{\i}amos pensar (por analogia 
com a eletrodin\^amica qu\^antica) que os efeitos qu\^anticos da 
gravita\c{c}\~ao come\c{c}am, nestas condi\c{c}\~oes, a ser importantes. 
Este \'e um procedimento natural pois historicamente  teorias que desenvolveram
singularidades foram resolvidas por meio da quantiza\c c\~ao, como a
 eletrodin\^amica.
Al\'em disso, um universo de campos quantizados (eletro-fraca e cromodin\^amica 
qu\^anticas) em intera\c c\~ao com um campo gravitacional fundamental 
cl\'assico resulta inconsistente \cite{wal}\cite{dew10}.
Ademais, as constantes fundamentais 
$G$ (constante de Newton ), 
$\hbar$ (constante de Planck) e $c$ (velocidade da luz) indicam  a escala 
 onde  ser\'a relevante uma teoria qu\^antica da gravita\c c\~ao:

\begin{eqnarray}
L_{pl} &=& \sqrt{\frac{\hbar G}{c^3}} \sim 10^{-33}cm \nonumber \\
T_{pl} &=& \sqrt{\frac{\hbar G}{c^5}} \sim 10^{-45}s \nonumber \\  
M_{pl} &=& \sqrt{\frac{\hbar c}{G}}   \sim 10^{-5}g \nonumber \\
{\rho}_{pl} &=& \frac{c^5}{\hbar G^2} \sim 10^{94}g/cm^3
\end{eqnarray}
 que s\~ao respetivamente, o comprimento de Planck, o tempo de Planck, a massa de 
Planck e a densidade de  Planck.

Outra raz\~ao que justifica  quantizar a gravita\c c\~ao  \'e que 
uma teoria qu\^antica da gravidade ao ser aplicada na cosmologia pode 
constituir uma teoria de condi\c c\~oes iniciais  para o Universo. Uma teoria
de condi\c c\~oes iniciais do Universo \'e necess\'aria em Cosmologia desde que desejemos
explicar porque o Universo em que vivimos hoje tem as marcadas propriedades 
de isotropia e homogeneidade, com pequenas desvios deste estado altamente sim\'etrico que s\~ao 
amplificados pela intera\c c\~ao gravitacional. As solu\c c\~oes das 
equa\c c\~oes de Einstein com estas simetrias s\~ao de medida nula, ent\~ao porque 
n\~ao vivemos num universo altamente n\~ao homogeneo e anisotr\'opico? A cosmologia qu\^antica 
\'e uma teoria 
que poderia explicar isto \cite{hh} \cite{vilenkin}. A id\'eia de infla\c c\~ao \cite{gut}\cite{kol}  
ajuda  nesta quest\~ao mas 
n\~ao a resolve porque tamb\'em precisa  de condi\c c\~oes iniciais para acontecer.
A cosmologia qu\^antica \'e o estudo da aplica\c c\~ao
de  teorias de gravita\c c\~ao qu\^antica  ao  problema  espec\'{\i}fico da 
origem e evolu\c c\~ao do Universo  \cite{Halliwell90}\cite{DeWitt67}\cite{nelson95}.

Quantizaremos a  TRG. Utilizaremos o 
conhecido formalismo 
can\^onico
que est\'a baseado na vers\~ao hamiltoniana da TRG. A id\'eia aqui 
\'e obter uma equa\c c\~ao funcional qu\^antica para o funcional do universo, 
a qual \'e analoga \`a equa\c c\~ao de Schr\"odinger. Este formalismo n\~ao 
\'e muito popular entre as outras teorias de campos (para uma compara\c c\~ao
deste com os  formalismos mais usuais veja por exemplo \cite{kie1,gut2,fou}). 
Para construir o hamiltoniano da TRG, o espa\c co-tempo 
\'e folheado por uma familia de hipersuperf\'{\i}cies tipo-espa\c co, uma para cada 
valor de $t=cte$, coordenada que define uma dire\c c\~ao tipo-tempo. Isto implica
que estamos  restringindo o espa\c co-tempo a variedades 4-dimensionais do tipo 
$M^{4} = R \bigotimes M^{3}$. Ent\~ao, al\'em de destruir a covari\^ancia 
manifesta da teoria, este procedimento n\~ao permite considerar espa\c cos de 
outra topologia diferente desta, como por exemplo espa\c cos 
com rota\c c\~ao e com curvas de tipo-tempo fechadas como espa\c cos de Goedel 
\cite{goedel} \cite{hawking3}. Quer dizer que quest\~oes acerca de da exist\^encia
de curvas tipo-tempo fechadas n\~ao podem ser respondidas neste formalismo. 

\section{Formalismo hamiltoniano cl\'assico}

Comecemos ent\~ao com a separa\c c\~ao 3+1 do espa\c co-tempo \cite{adm} \cite{mtw} \cite{nelsonF}.
Primeiramente escolhemos um par\^ametro que descrever\'a a evolu\c c\~ao din\^amica 
do sistema, seja por exemplo $x^0$, que desempenhar\'a o papel de tempo $x^0=t$. O estado 
f\'{\i}sico do sistema num determinado instante $t$ significa dar a configura\c c\~ao das 
vari\'aveis can\^onicas numa certa 
hipersuperf\'{\i}cie 3-dimensional tipo-espa\c co. Uma determinada hipersuperf\'{\i}cie
$t=cte$ pode ser descrita pelas equa\c c\~oes  param\'etricas

\begin{equation}
X^\alpha=X^\alpha(x^i) \hspace{1cm}\alpha=0,1,2,3 \hspace{1cm}i=1,2,3
\end{equation}
Seus tr\^es vetores tangentes $\frac{\partial}{\partial x^i}$ tem componentes $\frac{\partial X^\alpha}{\partial x^i}$ na base
 $\frac{\partial}{\partial X^\alpha}$ e o vetor normal unit\'ario a esta hipersuperf\'{\i}cie 
 ${\bf n}=n^{\alpha}\frac{\partial}{\partial X^\alpha}$ est\'a definido por
\begin{equation} 
{\bf g}({\bf n},{\bf n})=-1 \hspace{.5cm} {\bf g}({\bf n},\frac{\partial}{\partial x^i})=0 \, ,
\end{equation}
onde ${\bf g}$ \'e o tensor m\'etrico. No
 sistema de coordenadas de base $\frac{\partial}{\partial X^\alpha}$ ficam

\begin{equation}
\tilde{g}_{\alpha \beta}n^{\alpha}n^{\beta}=- 1 \, ,
\end{equation}

\begin{equation}
\tilde{g}_{\alpha \beta}X^{\alpha}_{i}n^{\alpha} = 0 \, ,
\end{equation}
Como no cap\'{\i}tulo precedente.
Para cada valor do par\^ametro $t=cte$ vamos ter uma  hipersuperf\'{\i}cie diferente
 e o conjunto de todas elas, que  forma o espa\c co-tempo, pode ser descrito por 
 
\begin{equation}
 X^\alpha=X^\alpha(x^i,t) \, .
\end{equation}

O vetor deforma\c c\~ao, que conecta dois 
pontos com o mesmo r\'otulo $x^i$ em duas hipersuperf\'{\i}icies pr\'oximas, 
\'e dado por $\frac{\partial}{\partial t}$ que se escreve 
$\frac{\partial}{\partial t}=\frac{\partial X^\alpha}{\partial t}\frac{\partial}{\partial X^\alpha}$ 
e chamamos $N^\alpha\equiv \frac{\partial X^\alpha}{\partial t}$ 
suas componentes na base $\frac{\partial}{\partial X^\alpha}$.  Este vetor pode ser escrito na base dada pelo vetor 
normal unitario ${\bf n}$, e os vetores tangentes \'a hipersuperf\'{\i}cie 
$\frac{\partial}{\partial x^i}$ como sendo

\begin{equation}
\label{deform}
\frac{\partial}{\partial t}=N{\bf n}+N^i\frac{\partial}{\partial x^i} \, ,
\end{equation}
de modo que para as componentes na base $\frac{\partial}{\partial X^\alpha}$ temos

\begin{equation}
N^\alpha=N n^\alpha + N^i\frac{\partial X^\alpha}{\partial x^i} \, ,
\end{equation}
onde $N$ \'e conhecida como a fun\c c\~ao lapso e $N^i$ fun\c c\~ao deslocamento. Notemos 
que estas quantidades aparecen devido ao fato de as linhas coordenadas do tempo $t$, das quais
$\frac{\partial}{\partial t}$ \'e tangente, n\~ao serem necessariamente ortogonais \`a  hipersuperf\'{\i}cie.
No sistema de coordenadas $x^{\mu}=(t,x^i)$, temos para a m\'etrica espacial

\begin{equation}
g_{ij}\equiv{\bf g}(\frac{\partial}{\partial x^i},\frac{\partial}{\partial x^j}) \, ,
\end{equation}
que em termos da m\'etrica na base antiga \'e

\begin{equation}
g_{ij}=\tilde{g}_{\mu \nu}X^{\mu}_i X^{\nu}_j \, .
\end{equation}
Assim, $g_{ij}$ \'e a m\'etrica induzida na hipersuperf\'{\i}cie (ou m\'etrica intr\'{\i}nseca), 
que chamaremos de $h_{ij}$ ( $h_{ij}\equiv g_{ij}$) e que
sobe e desce os \'{\i}ndices na hipersuperf\'{\i}cie. 
Em particular, $N_j \equiv N^{i}h_{ij}$. Tamb\'em

\begin{equation}
g_{0i}\equiv{\bf g}(\frac{\partial}{\partial t}, \frac{\partial}{\partial x^i})=
{\bf g}(N{\bf n}+N^i\frac{\partial}{\partial x^i},\frac{\partial}{\partial x^i})=h_{ij}N^j=N_{i} \, ,
\end{equation}

\begin{equation}
g_{00}\equiv{\bf g}(\frac{\partial}{\partial t}, \frac{\partial}{\partial t})=
{\bf g}(N{\bf n}+N^i\frac{\partial}{\partial x^i},N{\bf n}+N^j\frac{\partial}{\partial x^j})=-N^2+ N^i N^j h_{ij}=-N^2 + N^i N_i\, ,
\end{equation}
Calculando a inversa $g^{\mu\nu}$ temos

\begin{equation}
\label{31}
g^{00} = -\frac{1}{N^2}; \; g^{0i} = \frac{N^{i}}{N^2};
\; g^{ij} = h^{ij} - \frac{N{i} N^{j}}{N^2}
\, .
\end{equation}
Isto permite escrever o intervalo $ds^2=g_{\mu \nu} dx^\mu dx^\nu$ na forma conhecida 
como {\it ADM} ou  3+1 \cite{adm}

\begin{eqnarray}
\label{32}
ds^2 &=& g_{\mu\nu}dx^{\mu}dx^{\nu} \nonumber \\
&=& (N_iN^i-N^2)dt^2 + 2 N_i dx^i dt + h_{ij} dx^i dx^j =
\nonumber \\ &=& N^2 dt^2 + h_{ij} (N^i dt + dx^i) (N^j dt + dx^j)
\end{eqnarray}
 
Examinando a equa\c c\~ao (\ref{32}), podemos ver que 
$N(t,x^k)$ mede a taxa de varia\c c\~ao com rela\c c\~ao ao tempo coordenado
$t$, do tempo pr\'oprio do observador com 4-velocidade $n^{\mu}(t,x^{k})$ 
no ponto $(t,x^k)$. Por isso o nome  fun\c c\~ao  lapso. 
Tamb\'em, $N^{i}(t,x^k)$
nos d\'a a taxa de varia\c c\~ao com rela\c c\~ao ao tempo coordenado $t$,
do deslocamento  dos pontos com o mesmo r\'otulo 
$x^i$ ao pasar da hipersuperf\'{\i}cie $t=cte$ \`a hipersuperf\'{\i}cie $t+dt=cte$.
Por isso o nome fun\c c\~ao deslocamento. A interpreta\c c\~ao geom\'etrica
pode-se ver ao considerar o vetor deforma\c c\~ao na forma (\ref{deform}).

Uma hipersuperf\'{\i}cie caraterizada pela m\'etrica $h_{ij}(x^a)$ tem uma curvatura
intr\'{\i}nseca associada a sua 3-geometria, que pode ser calculada da forma usual. 
Mas esta hipersuperf\'{\i}cie pode estar curvada com rela\c c\~ao
\`a variedade quadridimensional na qual est\'a imersa, de maneira arbitr\'aria.
Ent\~ao, para descrever  univocamente  o tipo de folhea\c c\~ao  a que foi 
 submetido o espa\c co-tempo
 quadridimensional, \'e preciso caracterizar a curvatura das hipersuperf\'{\i}cies em rela\c c\~ao ao 
 espa\c co-tempo quadridimensional. O vetor normal \`a hipersuperf\'{\i}cie serve para 
 realizar esta tarefa por meio de sua varia\c c\~ao ao ser 
 transportado paralelamente ao longo da hipersuperf\'{\i}cie. Esta varia\c c\~ao est\'a 
 dada pela derivada covariante. Deste modo definimos o tensor 
 curvatura extr\'{\i}nseca

\begin{equation}
\nonumber
K_{\mu\nu} \equiv - h^{\alpha}_{\mu} \, h^{\beta}_{\nu} \,
{\nabla}_{(\alpha} {n}_{\beta)}
\, ,
\end{equation}
sendo
 $h^{\alpha}_{\mu}\equiv\delta^{\alpha}_{\mu}+n^\alpha n_\mu$  o projetor 
sobre a hipersuperf\'{\i}cie da qual $n^\mu$ \'e normal, e 
$\nabla_{\alpha} n_{\beta}\equiv n_{\alpha,\beta}-\Gamma^{\epsilon}_{\alpha \beta}n_{\epsilon}$ 
\'e a derivada covariante de $n^\mu$.
As componentes que ser\~ao relevantes s\~ao:

\begin{eqnarray}
\label{33}
K_{ij} &=& - N {\Gamma}^{0}_{ij} \nonumber \\
&=& \frac{1}{2N}(D_{(i}N_{j)}-{\partial}_th_{ij}) ,
\end{eqnarray}
onde $D_i$ \'e a derivada covariante intr\'{\i}nseca \`a hipersuperf\'{\i}cie 
3-dimensional de m\'etrica $h_{ij}(x)$, de modo que 

\begin{equation}
D_{i} N_{j} \equiv {\partial}_{i} N_{j} - ^{3}{\Gamma}^{a}_{ij}N_{a} \, ,
\end{equation}
sendo $^{3}\Gamma^{a}_{ij}$ a conex\~ao intr\'{\i}nseca a esta hipersuperf\'{\i}cie. 
Usando (\ref{31}), (\ref{32}) e (\ref{33}), obtemos para o escalar de
 curvatura \cite{nelsonF}:
\begin{equation}
\label{34}
R=R^{(3)} +K^{ki}K_{ki}+K^2-\frac{2}{N}{\partial}_{t}K+\frac{2N^i}{N}
{\partial}_iK-\frac{2}{N}D_k({\partial}^kN) \, ,
\end{equation}
onde $R^{(3)}$ \'e o escalar de curvatura das hipersuperf\'{\i}cies construido com os $^{3}\Gamma^{a}_{ij}$.

A densidade  lagrangeana de Einstein-Hilbert pode se escrever como
\begin{eqnarray}
\label{35}
{\cal L}_{E} &=& \frac{1}{\kappa}\sqrt{-g}R =\frac{1}{\kappa} Nh^{1/2}R \nonumber \\
&=& \frac{1}{\kappa}(Nh^{1/2}(R^{(3)}+K_{ij}K^{ij}-K^2)-2{\partial}_{t}(h^{1/2}K)+
\nonumber \\
& & + 2{\partial}_i(h^{1/2}KN^i-h^{1/2}h^{ki}{\partial}_kN)).
\end{eqnarray}
Aqui, $\kappa=\frac{16 \pi G}{c^{4}}$. O termo dado pela derivada temporal total vai ser eliminado j\'a que conduz a 
inconsist\^encias na formula\c c\~ao em integrais de  trajetoria da 
teoria qu\^antica \cite{haw1},  e al\'em disso exige impor novas condi\c c\~oes nos bordos
ao fazer a varia\c c\~ao da a\c c\~ao \cite{nelson95}. O termo dado pela derivada 
espacial total nao \'e importante j\'a que
n\~ao muda as equa\c c\~oes de movimento. Estes termos podem ser importantes no caso de
espa\c cos-tempos abertos ao se obter as corretas equa\c c\~oes de Hamilton, e  a 
energia total gravitacional (caso exista) \cite{energia}. S\~ao importantes no estudo de
buracos negros qu\^anticos. Na cosmologia qu\^antica de um universo fechado, estes termos s\~ao zero
e portanto podem ser descartados da lagrangeana. Existem argumentos, baseados em integrais 
de trajet\'oria, contra a existencia de universos abertos. Al\'em disso os universos fechados
s\~ao conceitualmente e tecnicamente mais simples. Iremos concentrar nosso estudo nestes
universos. Assim a densidade lagrangeana gravitacional se reduz a   
   
\begin{equation}
\label{36}
{\cal L}[N,N^i,h_{ij}]=\frac{1}{\kappa}Nh^{1/2}(R^{(3)}+K^{ij}K_{ij}-K^2) .
\end{equation}
Para o  lagrangeano temos

\begin{equation}
\label{36b}
L = \int {\cal L} d^3x \, ,
\end{equation}
Variando com rela\c c\~ao a $N$, $N^i$ e $h_{ij}$ obtemos as equa\c c\~oes de 
Einstein projetadas $G_{\mu\nu}
{n}^{\mu}{n}^{\nu}=0$, $G_{\mu\nu}
{n}^{\mu} h^{\nu}_{\alpha} = 0$ and $G_{\mu\nu}
h^{\mu}_{\beta} h^{\nu}_{\alpha} = 0$, respectivamente.

Vamos construir o hamiltoniano da TRG. Os momentos 
can\^onicos conjugados a $N$ e $N^i$  s\~ao v\'{\i}nculos, j\'a que a densidade 
lagrangeana (\ref{36}) n\~ao depende de $\partial_0 N$ e $\partial_{0}N^{i}$:

\begin{equation} 
\label{v365}
\Pi _{i } = \frac{\delta L}{\delta (\partial_{0}N^{i})} \approx 0
\end{equation}

\begin{equation} 
\label{v365b}
\Pi  = \frac{\delta L}{\delta (\partial_0N)} \approx 0
\end{equation}

A TRG \'e portanto uma teoria com v\'{\i}nculos e por isso ser\'a estudada
com o formalismo de Dirac \cite{dirac,sun}. Os v\'{\i}nculos 
(\ref{v365}) (\ref{v365b}) s\~ao chamados `prim\'arios'.
Os momentos canonicamente conjugados a  $h^{ij}$ est\~ao dados por:

\begin{equation} 
\label{37}
\Pi _{ij} = \frac{\delta L}{\delta (\partial _t h^{ij})} = 
- h^{1/2}(K_{ij}-h_{ij}K) \, .
\end{equation}
A densidade hamiltoniana ser\'a: 
 ${\cal H}_{c} = \Pi _{ij} \; \partial _t h^{ij} - {\cal L}$, e o hamiltoniano
 se escreve como
 
\begin{eqnarray}
\label{38}
H_c &=& \int d^3x {\cal H}_c \\
\nonumber
&=& \int d^3x(N{\cal H}+N_j{\cal H}^j) ,
\end{eqnarray}
onde
\begin{eqnarray}
\label{39}
{\cal H} &=&\kappa G_{ijkl}\Pi ^{ij}\Pi ^{kl}-\frac{1}{\kappa} h^{1/2}R^{(3)}   ,\\
\label{supermomentum}
{\cal H}^j &=& -2D_i\Pi ^{ij}  ,
\end{eqnarray}
e
\begin{equation}
\label{301}
G_{ijkl}=\frac{1}{2}h^{-1/2}(h_{ik}h_{jl}+h_{il}h_{jk}-h_{ij}h_{kl}) \, ,
\end{equation}
esta chamada de superm\'etrica ou m\'etrica de DeWitt. Sua inversa 
 $G^{ijkl}$ est\'a dada por

\begin{equation}
\label{300}
G^{ijkl}=\frac{1}{2}h^{1/2}(h^{ik}h^{jl}+h^{il}h^{jk}-2h^{ij}h^{kl}),
\end{equation}
de modo que ($G_{ijkl}G^{ijab}=\delta ^{ab}_{kl}$) .

A densidade hamiltoniana total se obtem acrecentando os v\'{\i}nculos 
(\ref{v365}) e (\ref{v365b}) por meio de multiplicadores de Lagrange $\lambda^\mu$:

\begin{equation}
\label{302}
{\cal H}_T=N{\cal H}+N_j{\cal H}^j+\lambda ^{\mu}\Pi _{\mu}
\end{equation}

Para ter uma teoria consistente, os v\'{\i}nculos prim\'arios devem se conservar 
no tempo: $\dot{\Pi }_{\mu}=\{\Pi_{\mu},{\cal H}_c\}=0$. Daqui temos que
as quantidades (\ref{39}) e (\ref{supermomentum})  devem ser fracamente zero:

\begin{equation}
{\cal H} = \kappa G_{ijkl}\Pi ^{ij}\Pi ^{kl}-\frac{1}{\kappa}h^{1/2}R^{(3)} \approx 0 
\end{equation}
\begin{equation}
\label{hi}
{\cal H}^j = -2D_i\Pi ^{ij} \approx 0.
\end{equation}

Estes s\~ao chamados de v\'{\i}nculo super-hamiltoniano e v\'{\i}nculo super-momento
respectivamente, e s\~ao v\'{\i}nculos secund\'arios. Sua conserva\c c\~ao no tempo
n\~ao conduz a novos v\'{\i}nculos. J\'a que $N$ e $N^{i}$ n\~ao t\^em din\^amica
e aparecem no hamiltoniano total multiplicando v\'{\i}nculos secund\'arios, ent\~ao 
podem ser interpretados como multiplicadores de Lagrange destes v\'{\i}nculos , podendo
ser eliminados do espa\c co de fase da teoria \cite{sun}. Portanto o hamiltoniano da 
TRG  no v\'acuo ser\'a:

\begin{equation}
\label{303}
H_{GR} = \int d^3x(N{\cal H}+N_j{\cal H}^j) \, .
\end{equation}
Estes v\'{\i}nculos secund\'arios s\~ao de primeira classe: eles possuem par\^enteses de 
Poisson nulos entre eles. A conjectura de Dirac (provada em \cite{costa})
estabelece que todos os v\'{\i}nculos de primeira classe s\~ao geradores de 
transforma\c c\~oes de gauge. Para os v\'{\i}nculos super-hamiltoniano e 
super-momento pode-se mostrar que:

\begin{equation}
\label{304}
\delta h_{ij} (x) = \{ h_{ij}(x), \int d^3 y \xi ^k(y) {\cal H}_k (y)\} =
D_j \xi _i(x) + D_i \xi _j(x) = {\bf \pounds _{\xi}} h_{ij}
\, ,
\end{equation}
\begin{equation}
\label{305}
\delta h_{ij} (x) = \{ h_{ij}(x), \int d^3 y \zeta (y) {\cal H} (y)\} =
- 2 \zeta (x) K_{ij}(x) =  \zeta (x) {\bf \pounds _{n}} h_{ij}
\, ,
\end{equation}
onde 
${\bf \pounds _{\xi}}$ \'e a derivada de Lie ao longo do vetor 
infinitesimal tipo-espa\c co ${\bf \xi}$ e ${\bf \pounds _{n}}$ \'e a 
derivada de Lie ao longo da dire\c c\~ao ortogonal \`a hipersuperf\'{\i}cie 
tipo-espa\c co com m\'etrica $h_{ij}$. a fun\c c\~ao $\zeta (x)$ \'e 
infinitesimal. Temos um resultado an\'alogo para os momentos $\Pi _{ij}$.
Ent\~ao o v\'{\i}nculo super-momento \'e o gerador das transforma\c c\~oes espacias 
de coordenadas e o v\'{\i}nculo super-hamiltoniano \'e o gerador de 
reparametriza\c c\~oes no tempo, as quais s\~ao as transforma\c c\~oes de calibre
da teoria. Podemos ver tamb\'em da (\ref{305}) que o v\'{\i}nculo super-hamiltoniano 
determina a din\^amica da teoria.
Variando $H_{GR}$ com rela\c c\~ao a $N$ e $N^{i}$  obtem-se os v\'{\i}nculos 
${\cal H} = 0$ e ${\cal H}^{i} = 0$, que s\~ao as equa\c c\~oes de Einstein 
$G_{\mu\nu}
{n}^{\mu}{n}^{\nu}=0$ e $G_{\mu\nu}
{n}^{\mu} h^{\nu}_{\alpha} = 0$, respectivamente. A equa\c c\~ao de 
evolu\c c\~ao para $h^{ij}$ d\'a a defini\c c\~ao de $\Pi _{ij}$ mostrada em
(\ref{37}), a qual combinada com a equa\c c\~ao de evolu\c c\~ao para $\Pi _{ij}$
permite obter e equa\c c\~ao de Einstein 
$G_{\mu\nu} h^{\mu}_{\beta} h^{\nu}_{\alpha} = 0$. 
Vemos que as equa\c c\~oes de Einstein no v\'acuo
s\~ao obtidas partindo de um espa\c co de fases dado por todas as m\'etricas $h^{ij}(x)$ possiveis das 
hipersuperf\'{\i}cies e seus momentos canonicamente conjugados $\Pi_{ij}(x)$, o que 
significa que o espa\c co de configura\c c\~ao da teoria \'e composto por todas as $h^{ij}(x)$  possiveis.
Um espa\c co-tempo que seja solu\c c\~ao particular das equa\c c\~oes de Einstein, pode ser vissualizado 
como uma trajetoria no espa\c co de todas as  $h^{ij}(x)$.
O hamiltoniano (\ref{303}) resulta igual a zero, como no caso mostrado no 
cap\'itulo anterior, fato caracter\'{\i}stico das teorias 
invariantes por reparametriza\c c\~ao temporal.

A seguir vamos quantizar a TRG com um campo 
escalar minimamente acoplado 
e com um potencial arbitr\'ario.
Todos os resultados que encontramos ser\~ao essencialmente os mesmos para 
qualquer campo que esteja acoplado com a 
m\'etrica e n\~ao com as suas derivadas. 
O hamiltoniano neste caso ser\'a dado por:

\begin{equation}
\label{hgr}
H = \int d^3x(N{\cal H}+N^j{\cal H}_j) 
\end{equation}
onde
\begin{eqnarray}
\label{h0}
{\cal H} &=& \kappa G_{ijkl}\Pi ^{ij}\Pi ^{kl} + 
\frac{1}{2}h^{-1/2}\Pi ^2 _{\phi}+\nonumber\\
& & + h^{1/2}\biggr[-{\kappa}^{-1}(R^{(3)} - 2\Lambda)+
\frac{1}{2}h^{ij}\partial _i \phi\partial _j \phi+U(\phi)\biggl]\\
\label{hii}
{\cal H}_j &=& -2D_i\Pi ^i_j + \Pi _{\phi} \partial _j \phi .
\end{eqnarray}
 Nestas equa\c c\~oes $h_{ij}$ \'e a m\'etrica das 3-hipersuperf\'{\i}cies  espaciais 
fechadas, e $\Pi ^{ij}$ s\~ao os momentos canonicamente conjugados, dados 
como vimos, por:

\begin{equation} 
\label{ph}
\Pi ^{ij} = - h^{1/2}(K^{ij}-h^{ij}K) =
G^{ijkl}({\dot{h}}_{kl} -  D _k N_l - D _l N_k ),
\end{equation}
O  momento can\^onico do campo escalar \'e

\begin{equation}
\label{pf}
\Pi _{\phi} = \frac{h^{1/2}}{N}\biggr(\dot{\phi}-N^i \partial _i \phi \biggl).
\end{equation}
A 4-m\'etrica cl\'assica 
\begin{equation}
\label{4g}
ds^{2}=-(N^{2}-N^{i}N_{i})dt^{2}+2N_{i}dx^{i}dt+h_{ij}dx^{i}dx^{j}
\end{equation}
e o campo escalar, que s\~ao solu\c c\~oes das equa\c c\~oes de Einstein, 
podem ser obtidas  por meio das equa\c c\~oes de Hamilton

\begin{equation}
\label{hh}
{\dot{h}}_{ij} = \{h_{ij},H\},
\end{equation}
\begin{equation}
\label{hp}
{\dot{\Pi}}^{ij} = \{\Pi ^{ij},H\},
\end{equation}
\begin{equation}
\label{hf}
{\dot{\phi}} = \{\phi,H\},
\end{equation}
\begin{equation}
\label{hpf}
{\dot{\Pi _{\phi}}}= \{\Pi _{\phi},H\},
\end{equation}
para alguma escolha de $N$ e $N^i$, e impondo condi\c c\~oes iniciais 
compat\'{\i}veis com os v\'inculos

\begin{equation}
\label{hh0}
{\cal H} \approx 0 ,
\end{equation}
\begin{equation}
\label{hhi}
{\cal H}_i \approx 0.
\end{equation}
Uma caracter\'{\i}stica do hamiltoniano da TRG \'e que as 4-m\'etricas (\ref{4g}) 
constru\'{\i}das dessa forma
, com as mesmas condi\c c\~oes iniciais, descrevem a mesma 4-geometria para 
qualquer escolha de $N$ e $N^i$.

Sejam $x,x'$ dois pontos na hipersuperf\'{\i}cie  $t=cte$. Os v\'{\i}nculos satisfazem 
tamb\'em a seguinte \'algebra

\begin{eqnarray}\label{algebra}
\{ {\cal H} (x), {\cal H} (x')\}&=&{\cal H}^i(x) {\partial}_i \delta^3(x,x')- 
{\cal H}^i(x'){\partial}_i \delta^3(x',x) \, ,\nonumber \\ 
\{{\cal H}_i(x),{\cal H}(x')\}&=&{\cal H}(x) {\partial}_i \delta^3(x,x') \, ,\\ 
\{{\cal H}_i(x),{\cal H}_j(x')\}&=&{\cal H}_i(x) {\partial}_j \delta^3(x,x')- 
{\cal H}_j(x'){\partial}_i \delta^3(x',x) \, .\nonumber  
\end{eqnarray} 
Como j\'a dissemos no cap\'{\i}tulo 3, esta \'algebra foi obtida primeiramente por Dirac no caso de teoria de 
campos parametrizadas em espa\c co-tempo de Minkowski. No caso da TRG, a \'algebra (\ref{algebra}) foi obtida
por DeWitt calculando diretamente de (\ref{h0})(\ref{hii})\cite{DeWitt67}. Os super-momentos que aparecem
nos lados direitos de (\ref{algebra}) t\^em o \'{\i}ndice contravariante subidos por meio da 
m\'etrica espacial $h_{ij}$ como no cap\'{\i}tulo precedente.
Como j\'a vimos, Teitelboim provou
que esta \'algebra \'e independente da forma dos v\'{\i}nculos  no caso da teoria de campos e que ela \'e,
na verdade, um padr\~ao que segue a din\^amica das deforma\c c\~oes de uma hipersuperf\'{\i}cie 
num espa\c co riemanniano de m\'etrica n\~ao degenerada. Tanto a TRG quanto a teoria de campos 
parametrizadas, apresentam este 
 padr\~ao  por tratarem de espa\c cos-tempos com m\'etrica n\~ao degenerada \cite{tei1}.

\section{Quantiza\c c\~ao can\^onica}

Agora vamos quantizar este sistema f\'{\i}sico com v\'{\i}nculos seguindo o 
formalismo de Dirac.
Os v\'{\i}nculos se transformam em condi\c c\~oes sobre os estados 
poss\'{\i}veis do sistema qu\^antico, 
resultando nas seguintes equa\c c\~oes qu\^anticas

\begin{eqnarray}
\label{smo}
\hat{{\cal H}}_i \mid \Psi  \! > &=& 0 \, ,\\
\label{wdw}
\hat{{\cal H}} \mid \Psi  \! > &=& 0 \, .
\end{eqnarray}
Na representa\c c\~ao da m\'etrica e do campo, a primeira equa\c c\~ao \'e

\begin{equation}
\label{smo2}
-2 h_{li}D_j\frac{\delta \Psi(h_{ij},\phi)}{\delta h_{lj}} + 
\frac{\delta \Psi(h_{ij},\phi)}{\delta \phi} \partial _i \phi = 0 ,
\end{equation}
que implica ser o funcional de onda $\Psi$  um invariante sob 
transforma\c c\~oes  de coordenadas espaciais.

A segunda \'e a equa\c c\~ao de  Wheeler-DeWitt  \cite{whe}\cite{DeWitt67}. 
Se a escrevemos de maneira n\~ao regulada em representa\c c\~ao de 
coordenadas temos

\begin{equation}
\label{wdw2}
\biggr\{-\hbar ^2\biggr[\kappa G_{ijkl}\frac{\delta}{\delta h_{ij}} 
\frac{\delta}{\delta h_{kl}}
 + \frac{1}{2}h^{-1/2} \frac{\delta ^2}{\delta \phi ^2}\biggl] + 
V\biggl\}\Psi(h_{ij},\phi) = 0 ,
\end{equation}
onde $V$ \'e o potencial cl\'assico dado por

\begin{equation}
\label{v}
V = h^{1/2}\biggr[-{\kappa}^{-1}(R^{(3)} - 2\Lambda)+
\frac{1}{2}h^{ij}\partial _i \phi\partial _j \phi+
U(\phi)\biggl] .
\end{equation}
Esta equa\c c\~ao envolve produtos de operadores locais no mesmo ponto do 
espa\c co;
 ent\~ao deve ser regularizada. Depois de ter feito isso 
devemos encontrar  um ordenamento que deixe a teoria livre de anomalias, 
no sentido que os comutadores dos v\'{\i}nculos 
(sendo agora operadores) fechem na mesma \'algebra que os par\^enteses de 
Poisson cl\'assicos
(\ref{algebra}). Portanto, a equa\c c\~ao  (\ref{wdw2}) \'e apenas 
uma equa\c c\~ao formal, que  deve ser usada com cuidado
\cite{japa1,japa2,kow2}.

\baselineskip=2.0 \normalbaselineskip

\chapter{Geometrodin\^amica Qu\^antica na Vis\~ao de Bohm-de Broglie}

\section{A perspectiva de Hamilton-Jacobi}

Vamos ver agora como  a interpreta\c c\~ao de Bohm-de Broglie se aplica \`as
 solu\c c\~oes das 
Eqs. (\ref{smo}) e (\ref{wdw}) na representa\c c\~ao da m\'etrica e do campo.
Primeiro escreveremos o funcional de onda em forma polar
$\Psi = A\exp (iS/\hbar )$, onde $A$ e $S$ s\~ao  funcionais de
$h_{ij}$ e $\phi$. Substituindo na Eq. (\ref{smo2}), obtemos duas 
equa\c c\~oes, as quais estabelecem que $A$ e $S$ s\~ao invariantes
sob  transforma\c c\~oes gerais de coordenadas espaciais:

\begin{equation}
\label{smos}
-2 h_{li}D_j\frac{\delta S(h_{ij},\phi)}{\delta h_{lj}} + 
\frac{\delta S(h_{ij},\phi)}{\delta \phi} \partial _i \phi = 0 ,
\end{equation}

\begin{equation}
\label{smoa}
-2 h_{li}D_j\frac{\delta A(h_{ij},\phi)}{\delta h_{lj}} + 
\frac{\delta A(h_{ij},\phi)}{\delta \phi} \partial _i \phi = 0 .
\end{equation}

Exatamente como no cap\'{\i}tulo 3, as duas equa\c c\~oes que obtemos para $A$ e $S$ ao 
substituirmos
$\Psi = A\exp (iS/\hbar )$ na Eq. (\ref{wdw}) depender\~ao do ordenamento 
escolhido. No entanto, seja como for, uma das equa\c c\~oes ser\'a

\begin{equation}
\label{hj}
\kappa G_{ijkl}\frac{\delta S}{\delta h_{ij}} 
\frac{\delta S}{\delta h_{kl}}
 + \frac{1}{2}h^{-1/2} \biggr(\frac{\delta S}{\delta \phi}\biggl)^2
+V+{\cal Q}=0 ,
\end{equation}
onde  $V$ \'e o potencial cl\'assico dado na  Eq. (\ref{v}).
Ao contr\'ario dos outros termos na  Eq. (\ref{hj}), os quais j\'a est\~ao 
bem definidos, a forma precisa de ${\cal Q}$ depende da regulariza\c c\~ao e 
do ordenamento escolhido na equa\c c\~ao de Wheeler-DeWitt. 
Na forma n\~ao regulada dada na Eq. (\ref{wdw2}), ${\cal Q}$ \'e dado por

\begin{equation}
\label{qp1}
{\cal Q} = -{\hbar ^2}\frac{1}{A}\biggr(\kappa G_{ijkl}\frac{\delta ^2 A}
{\delta h_{ij} \delta h_{kl}} + \frac{h^{-1/2}}{2} \frac{\delta ^2 A}
{\delta \phi ^2}\biggl) .
\end{equation}
A outra equa\c c\~ao que aparece neste caso \'e

\begin{equation}
\label{pr}
\kappa G_{ijkl}\frac{\delta}{\delta h_{ij}}\biggr(A^2
\frac{\delta S}{\delta h_{kl}}\biggl)+\frac{1}{2}h^{-1/2} 
\frac{\delta}{\delta \phi}\biggr(A^2
\frac{\delta S}{\delta \phi}\biggl) = 0 .
\end{equation}

Vamos implementar agora a interpreta\c c\~ao de Bohm-de Broglie da 
gravita\c c\~ao qu\^antica can\^onica.
Primeiramente, notamos que
as Eqs. (\ref{smos}) e (\ref{hj}), que s\~ao validas independentemente do 
ordenamento da equa\c c\~ao 
Wheeler-DeWitt, s\~ao como as equa\c c\~oes de  Hamilton-Jacobi para a TRG, 
acrescentadas de  um termo extra
 ${\cal Q}$ no caso da  Eq. (\ref{hj}), o qual chamaremos de potencial qu\^antico. 
Por analogia com os casos de uma part\'{\i}ula n\~ao relativ\'{\i}stica e de uma  
teoria de campos em espa\c co-tempo plano, postularemos
que  a 3-m\'etrica das hipersuperf\'{\i}cies espaciais, o campo escalar, e os 
momentos can\^onicamente conjugados sempre existem,
 independentemente da observa\c c\~ao, e que a evolu\c c\~ao da 3-m\'etrica 
 e do campo escalar pode ser obtida das rela\c c\~oes
 guia
\begin{equation}
\label{grh}
\Pi ^{ij} = \frac{\delta S(h_{ab},\phi)}{\delta h_{ij}} ,
\end{equation}
\begin{equation}
\label{grf}
\Pi _{\phi} = \frac{\delta S(h_{ij},\phi)}{\delta \phi} ,
\end{equation}
com $\Pi ^{ij}$ e $\Pi _{\phi}$ dados pelas Eqs. (\ref{ph}) e (\ref{pf}), 
respectivamente. Como antes, estas s\~ao
equa\c c\~oes diferenciais de primeira ordem que podem ser integradas para 
produzir a  3-m\'etrica e o campo escalar para todo
valor do par\^ametro $t$ . Estas solu\c c\~oes dependem dos valores iniciais  
da 3-m\'etrica e do campo escalar em 
alguma hipersuperf\'{\i}cie inicial.  A evolu\c c\~ao desses campos ser\'a 
diferente, claro, da 
cl\'assica devido \`a presen\c ca do potencial qu\^antico ${\cal Q}$ na Eq. (\ref{hj}).  
O limite cl\'assico \'e, outra vez, conceitualmente simples: est\'a dado 
pela regi\~ao onde 
o potencial qu\^antico ${\cal Q}$ torna-se desprezivel com respeito \`a energia 
cl\'assica.

A \'  unica e importante diferen\c ca com os casos anteriores de uma part\'{\i}cula n\~ao relativ\'{\i}stica 
e uma teoria de 
campos em espa\c co-tempo plano \'e o fato de que as equivalentes das 
 Eqs. (\ref{bpr}) e (\ref{fp}) para gravita\c c\~ao qu\^antica can\'onica, 
 que no ordenamento 
 ing\^enuo \'e a Eq. (\ref{pr}), n\~ao pode ser interpretada como uma equa\c c\~ao 
 de continuidade
  para uma densidade de probabilidade
 $A^2$ por causa da natureza hiperb\'olica da m\'etrica de  DeWitt  $G_{ijkl}$.
 Contudo, embora sem uma no\c c\~ao de probabilidade, a qual neste caso poderia
significar a densidade de probabilidadede de distrbui\c c\~ao para valores 
iniciais da 3-m\'etrica e o 
 campo escalar numa hipersuperf\'{\i}cie inicial, podemos extrair muita 
 informa\c c\~ao da Eq. (\ref{hj}) qualquer que seja  
 o potencial qu\^antico ${\cal Q}$, como veremos agora. Depois de obter estes 
 resultados vamos voltar \`a quest\~ao da
probabilidade no \'ultimo cap\'{\i}tulo.

Primeiro notamos que, qualquer que seja a forma do potencial qu\^antico
 ${\cal Q}$, ele deve ser uma densidade escalar de peso 1. Isto sai da equa\c c\~ao 
 de Hamilton-Jacobi 
 (\ref{hj}). Desta  equa\c c\~ao podemos expresar ${\cal Q}$ como 

\begin{equation}
\label{pothj}
{\cal Q} = -\kappa G_{ijkl}\frac{\delta S}{\delta h_{ij}} 
\frac{\delta S}{\delta h_{kl}}
- \frac{1}{2}h^{-1/2} \biggr(\frac{\delta S}{\delta \phi}\biggl)^2 - V .
\end{equation}
Sendo   $S$ \'e um invariante perante transforma\c c\~oes gerais 
de coordenadas (veja Eq. (\ref{smos})), ent\~ao 
$\delta S / \delta h_{ij}$ e $\delta S /\delta \phi$ devem ser
 densidades tensoriais, de ordem dois e  escalar, 
ambas de peso 1, respectivamente. 
Quando o seu produto \'e contra\'{\i}do com $G_{ijkl}$ e multiplicado por  
$h^{-1/2}$, respectivamente, eles formam uma 
densidade escalar de peso 1, como tamb\'em \'e $V$. Portanto, ${\cal Q}$ \'e uma densidade escalar de peso 1.
Alem disso, ${\cal Q}$ s\'o deve depender de  $h_{ij}$ e $\phi$ j\'a que ele \'e oriundo
 do funcional de onda que  por sua vez s\'o depende  destas vari\'aveis.
Por outro lado pode ser n\~ao local (mostramos um exemplo no ap\^endice B),
isto \'e,  dependente de integrais dos campos sobre todo o espa\c co.

Agora vamos investigar o seguinte problema importante. Das rela\c c\~oes guia
(\ref{grh}) e  (\ref{grf}) obtemos as seguintes equa\c c\~oes diferenciais
 em derivadas parciais:

\begin{equation} 
\label{hdot}
{\dot{h}}_{ij} =  
2NG_{ijkl}\frac{\delta S}{\delta h_{kl}} + D _i N_j + D _j N_i \, ,
\end{equation}
e
\begin{equation}
\label{fdot}
\dot{\phi}=Nh^{-1/2}\frac{\delta S}{\delta \phi} + N^i \partial _i \phi  \, ,
\end{equation}
A quest\~ao \'e, dadas certas condi\c c\~oes iniciais para a 3-m\'etrica e para o 
campo escalar,  que  tipo de estrutura  vamos obter ao integrar estas 
equa\c c\~oes no par\^ametro $t$? Ser\'a que esta estrutura forma uma 
geometria 4-dimensional com um campo escalar para qualquer escolha das 
fun\c c\~oes  lapso e 
deslocamento? Fica claro que se o funcional  $S$ for solu\c c\~ao da 
equa\c c\~ao
de  Hamilton-Jacobi cl\'assica, que n\~ao contem o termo ${\cal Q}$, ent\~ao a 
resposta seria afirmativa j\'a  que estar\'{\i}amos no caso  da TRG.
Mas $S$ \'e uma solu\c c\~ao da equa\c c\~ao de  Hamilton-Jacobi 
{\it modificada}  
(\ref{hj}), e ent\~ao n\~ao podemos garantir que isso  continuar\'a 
sendo verdadeiro.
Podemos obter uma estrutura completamente diferente devido aos efeitos 
qu\^anticos 
dados pelo termo  correspondente ao  potencial qu\^antico na  Eq. (\ref{hj}). 

Para responder a esta quest\~ao, vamos mudar da vis\~ao de 
Hamilton-Jacobi 
da geometrodin\^amica qu\^antica para uma vis\~ao hamiltoniana.
Vamos fazer  isto porque existem resultados poderosos  na geometrodin\^amica 
que foram obtidos
no formalismo hamiltoniano \cite{hkt} \cite{tei1}.
Vamos construir um formalismo hamiltoniano consistente com as rela\c c\~oes 
guia (\ref{grh}) e (\ref{grf}), exatamente como fizemos no cap\'{\i}tulo 3.
Ele produz as trajet\'orias bohmianas  (\ref{hdot}) e (\ref{fdot}) sempre 
que as rela\c c\~oes guia sejam satisfeitas inicialmente. Uma vez que tenhamos o 
hamiltoniano, podemos fazer uso de resultados bem conhecidos da literatura 
e assim obter resultados poderosos
acerca da vis\~ao Bohm-de Broglie da geometrodin\^amica qu\^antica. 

\section{A perspectiva hamiltoniana}

Examinando as Eqs. (\ref{smos}) e (\ref{hj}), podemos facilmente esperar que 
o hamiltoniano
que gera as trajetorias bohmianas, uma vez satisfeitas inicialmente as 
rela\c c\~oes  guia 
 (\ref{grh}) e (\ref{grf}), deva ser 

\begin{equation}
\label{hq}
H_Q = \int d^3x\biggr[N {\cal H}_Q + N^i{\cal H}_i\biggl] 
\end{equation}
onde definimos
\begin{equation}
\label{hq0}
{\cal H}_Q \equiv {\cal H} + {\cal Q} .
\end{equation}
As quantidades  ${\cal H}$ e ${\cal H}_i$ s\~ao os v\'{\i}nculos  super-hamiltoniano
 e  super-momento da TRG
dados pelas Eqs. (\ref{h0}) e (\ref{hii}).
Vamos mostrar isto. De fato, as rela\c c\~oes guia  (\ref{grh}) e (\ref{grf}) s\~ao
 consistentes 
com os v\'{\i}nculos
${\cal H}_Q \approx 0$ e ${\cal H}_i \approx 0$ 
j\'a que $S$ satisfaz (\ref{smos}) e (\ref{hj}). Al\'em disso, s\~ao conservados 
na evolu\c c\~ao hamiltoniana
dada por (\ref{hq}), analogamente ao que acontece no cap\'{\i}tulo 3. Vejamos
isto com algum detalhe.

Escrevamos Eqs. (\ref{grh}) e (\ref{grf}) de uma forma levemente diferente, 
definindo

\begin{equation}
\label{ch}
\Phi^{ij} \equiv \Pi ^{ij} - \frac{\delta S(h_{ab},\phi)}{\delta h_{ij}} 
\approx 0 ,
\end{equation}
e
\begin{equation}
\label{cf}
\Phi _{\phi} \equiv \Pi _{\phi} - \frac{\delta S(h_{ij},\phi)}{\delta \phi}
\approx 0 .
\end{equation}
Calculemos agora  $\{\Phi^{ij} (x),{\cal H}_Q (x')\}, \{ \Phi_{\phi} (x),{\cal H}_Q (x')\}, \{\Phi^{ab} (x),{\cal H}_i (x')\}$ e 
$\{ \Phi _{\phi} (x),{\cal H}_i (x')\}$, e verifiquemos se as rela\c c\~oes guia
(\ref{ch}) e (\ref{cf}), visualizadas agora como v\'{\i}nculos, s\~ao conservadas 
pelo hamiltoniano
$H_Q$.

\begin{eqnarray}
\{ {\cal H}_Q (x),\Phi^{ij} (x')\}&=&\kappa \frac{\delta G_{{abcd}}}{\delta h_{ij}'}\Pi^{ab} \Pi^{cd} +  
\frac{1}{2}\frac{\delta h^{-1/2}}{\delta h_{ij}'}\Pi_{\phi}^2 +  
\frac{\delta (V+{\cal Q})}{\delta h_{ij}'} \nonumber \\
& & + 2\kappa G_{abcd} \Pi^{ab} 
\frac{\delta^2 S}{\delta h_{cd} \delta h_{ij}'} +  h^{-\frac{1}{2}}
\Pi_{\phi} \frac{\delta^2 S}{\delta \phi \delta h_{ij}'} \nonumber \\ 
&=& \kappa\frac{\delta G_{abcd}}{\delta h_{ij}'}\biggr(\Phi^{ab} \Phi^{cd} +
2\Phi^{ab} \frac{\delta S}{\delta h_{cd}}\biggl)+\frac{1}{2}\frac{\delta h^{-\frac{1}{2}}}{\delta h_{ij}'}\biggr(\Phi _{\phi}^2+  
2 \Phi _{\phi} \frac{\delta S}{\delta \phi}\biggl) \nonumber \\ 
& & + 
2\kappa G_{abcd} \Phi^{ab} \frac{\delta^2 S}{\delta h_{cd} \delta h_{ij}'}+ 
 h^{-\frac{1}{2}} \Phi _{\phi} \frac{\delta S}{\delta \phi \delta h_{ij}'} \nonumber \\
& & +
 \frac{\delta }{\delta h_{ij}'}\biggr[\kappa G_{abcd} \frac{\delta S}{\delta h_{ab}}\frac{\delta S}{\delta h_{cd}} +
 \frac{1}{2}h^{-\frac{1}{2}}\biggr(\frac{\delta S}{\delta \phi}
\biggl)^2 +V+{\cal Q}\biggl]
\end{eqnarray}
onde a linha significa avaliar em $x'$.
O \'ultimo termo \'e zero por causa da Eq.(\ref{hj}), e assim obtemos

\begin{eqnarray}
\{ {\cal H}_Q (x),\Phi ^{ij} (x')\}&=&\biggr\{\kappa \biggr[-\frac{1}{2}
G_{abcd} h^{ij} \nonumber \\
& & + 
\frac{1}{2} h^{-\frac{1}{2}}(4\delta^{ij}_{ac} h_{bd}- 
\delta^{ij}_{ab}h_{cd}-\delta^{ij}_{cd}h_{ab})\biggl]\biggr(\Phi^{ab} \Phi ^{cd}+2\Phi^{ab} \frac{\delta S}{\delta h_{cd}}\biggl) \nonumber \\
& & -
\frac{1}{4}h^{-\frac{1}{2}} h^{ij} \biggr(\Phi _{\phi}^2 + 
2 \Phi _{\phi} \frac{\delta S}{\delta \phi}\biggl)\biggl\}\delta^3(x,x') \nonumber \\
& &
+ 2 \kappa G_{abcd} \frac{\delta^2 S}{\delta h_{cd} \delta h_{ij}'}\Phi^{ab} + 
h^{-\frac{1}{2}} \frac{\delta^2 S}{\delta \phi \delta h_{ij}'}\Phi _{\phi} \approx 0 .
\end{eqnarray}
Da mesma maneira podemos provar que

\begin{eqnarray}
{\{\cal H}_Q (x),\Phi _{\phi} (x')\}=2\kappa G_{abcd} \frac{\delta^2 S}{\delta h_{ab} \delta \phi'}\Phi^{cd}+
h^{-\frac{1}{2}}\frac{\delta^2 S}{\delta \phi \delta \phi'}\Phi_\phi\approx 0 ,
\end{eqnarray}
onde utilizamos o fato de a derivada funcional
 da Eq. (\ref{hj}) com respeito a $\phi$ \'e zero. 
 
Para os par\^enteses de Poisson que envolvem o v\'{\i}nculo super-momento, sendo  $S$ 
 um invariante j\'a que verifica a Eq. (\ref{smos}), ent\~ao $\Phi^{ij}$
e $\Phi_{\phi}$ s\~ao uma densidade tensorial de segundo ordem e uma densidade 
escalar, respectivamente, ambas de peso 1. 
Dado que  ${\cal H}_i$ \'e o gerador de transforma\c c\~oes espaciais de 
coordenadas, temos 

\begin{equation}
\{ {\cal H}_i(x),\Phi ^{ab} (x')\}=-2\delta^{ab}_{ci}\Phi^{cj}(x')
\partial _j \delta ^3(x,x')+
\Phi^{ab}(x) \partial _i \delta ^3(x,x')\approx 0 ,
\end{equation}

e
 
\begin{equation}
{\{\cal H}_i(x),\Phi _{\phi} (x')\}=\Phi _{\phi}
\partial _i \delta ^3(x,x')\approx 0
\end{equation}
Combinando estes resultados obtemos

\begin{equation}
\label{cch}
{\dot{\Phi}}^{ij} = \{\Phi ^{ij},H_Q\} \approx  0 ,
\end{equation}
e
\begin{equation}
\label{ccf}
{\dot{\Phi}}_{\phi} = \{\Phi _{\phi},H_Q\} \approx 0 .
\end{equation}
Ademais, os par\^enteses de Poisson entre 
 (\ref{ch}) e (\ref{cf})
s\~ao zero. Finalmente, as defini\c c\~oes dos momentos em termos das velocidades continuam 
sendo as mesmas do caso cl\'assico j\'a que o potencial qu\^antico ${\cal Q}$
n\~ao depende dos momentos:
\begin{equation}
\label{hhq}
{\dot{h}}_{ij} =  \{h_{ij},H_Q\} =  \{h_{ij},H\},
\end{equation}
e
\begin{equation}
\label{hfq}
{\dot{\phi}} = \{\phi,H_Q\} = \{\phi,H\}.
\end{equation}  
recuperando as Eqs.(\ref{hdot}) e 
(\ref{fdot}).

Temos agora um hamiltoniano
$H_Q$, que gera as trajet\'orias bohmianas uma vez que as rela\c c\~oes guia 
(\ref{grh}) e (\ref{grf}) s\~ao impostas inicialmente. No que segue, podemos 
investigar se a evolu\c c\~ao dos campos
dada por  $H_Q$ forma uma 4-geometria semelhante ao caso da geometrodin\^amica 
cl\'assica.
Primeiro lembremos o resultado obtido por 
 Claudio Teitelboim \cite{tei1} que usamos no cap\'{\i}tulo 3.
Neste trabalho, ele mostra que se as configura\c c\~oes de 3-geometrias e 
campos definidas numa hipersuperficie
s\~ao evolu\'{\i}das por um certo hamiltoniano da forma
\begin{equation}
\label{hg}
\bar{H} = \int d^3x(N\bar{{\cal H}} + N^i\bar{{\cal H}}_i) \, ,
\end{equation}
e se esta evolu\c c\~ao pode ser vista como o `movimento' de um corte  
3-dimensional num espa\c co-tempo 4-dimensional
 (as 3-geometrias podem ser embutidas numa 4-geometria), ent\~ao os v\'{\i}nculos
$\bar{{\cal H}} \approx 0$ e $\bar{{\cal H}}_i
\approx 0$ devem satisfazer a \'algebra de Dirac

\begin{eqnarray}
\{ \bar{{\cal H}} (x), \bar{{\cal H}} (x')\}&=&-\epsilon[\bar{{\cal 
H}}^i(x) {\partial}_i \delta^3(x',x)
-  \bar{{\cal H}}^i(x') {\partial}_i \delta^3(x',x)] \, ,
\label{algebra1} \\
\{\bar{{\cal H}}_i(x),\bar{{\cal H}}(x')\} &=& \bar{{\cal H}}(x)  
{\partial}_i \delta^3(x,x') \, , 
\label{algebra2} \\
\{\bar{{\cal H}}_i(x),\bar{{\cal H}}_j(x')\} &=& \bar{{\cal H}}_i(x)  
{\partial}_j \delta^3(x,x')- 
\bar{{\cal H}}_j(x') {\partial}_i \delta^3(x',x) \, .
\label{algebra3}  
\end{eqnarray} 
A constante 
 $\epsilon$ na  (\ref{algebra1}) pode ser $\pm 1$ dependendo se a 4-geometria 
 na qual 
 as 3-geometrias est\~ao embutidas  \'e euclideana
($\epsilon = 1$) ou hiperb\'olica ($\epsilon = -1$).
Estas s\~ao as condi\c c\~oes para a exist\^encia de um espa\c c\~o-tempo de m\'etrica
n\~ao degenerada.
A \'algebra acima \'e a mesma que
(\ref{algebra}) da TRG se escolhermos $\epsilon = -1$.

O hamiltoniano (\ref{hq}) \'e diferente do hamiltoniano da TRG
somente devido a presen\c ca do potencial qu\^antico
${\cal Q}$ em ${\cal H}_Q$. 
O colchete de Poisson 
$\{{\cal H}_i (x),{\cal H}_j (x')\}$ satisfaz a Eq.
(\ref{algebra3}) j\'a que  ${\cal H}_i$ de $H_Q$ definido na  Eq.
(\ref{hq}) \'e o mesmo que na TRG.
Da mesma forma 
$\{{\cal H}_i (x),{\cal H}_Q (x')\}$ satisfaz Eq. (\ref{algebra2}) pois 
 ${\cal H}_i$ \'e o gerador de transforma\c c\~oes espaciais de coordenadas,
 e como ${\cal H}_Q$ \'e uma 
 densidade  escalar de peso 1 (lembrar que ${\cal Q}$ tamb\'em \'e uma densidade escalar 
 de peso 1), 
 ent\~ao ele deve satisfazer esta rela\c c\~ao de colchetes de Poisson 
 com ${\cal H}_i$. O que resta ser verificado, como no cap\'{\i}tulo 3, \'e se 
 o colchete de Poisson
$\{{\cal H}_Q (x),{\cal H}_Q (x')\}$ fecha como na Eq. (\ref{algebra1}).

Agora lembramos um  resultado importante estabelecido por Hojman, 
Kucha$\check{\mbox{r}}$,  
e Teitelboim  \cite{hkt}. Neste trabalho mostra-se que caso
um super-hamiltoniano geral  $\bar{{\cal H}}$  satisfa\c ca a Eq.
(\ref{algebra1}), seja uma densidade escalar de peso 1 cujos graus de 
liberdade geom\'etricos 
sejam  dados  somente pela 3-m\'etrica $h_{ij}$ e seus momentos canonicamente 
conjugados, e contenham 
somente pot\^encias pares e nehum termo n\~ao-local nos momentos (estes dois 
\'ultimos requisitos, juntamente com os anteriores,
s\~ao satisfeitos por  ${\cal H}_Q$ porque ele \'e quadr\'atico nos momentos 
e o potencial qu\^antico nao contem nehum
termo n\~ao-local nos momentos), ent\~ao 
$\bar{{\cal H}}$ deve ter a seguinte forma:

\begin{equation}
\label{h0g}
\bar{{\cal H}} = \kappa G_{ijkl}\Pi ^{ij}\Pi ^{kl} + 
\frac{1}{2}h^{-1/2}\pi ^2 _{\phi} + V_G ,
\end{equation}
onde

\begin{equation}
\label{vg}
V_G \equiv -\epsilon h^{1/2}\biggl[-{\kappa}^{-1}(R^{(3)} - 2\bar{\Lambda})+
\frac{1}{2}h^{ij}\partial _i \phi\partial _j \phi+\bar{U}(\phi)\biggr] .
\end{equation}

Com este resultado  podemos agora estabelecer dois poss\'{\i}veis cen\'arios 
para a geometrodin\^amica qu\^antica 
na interpreta\c c\~ao de Bohm-de Broglie, dependendo da forma do potencial 
qu\^antico. Eles ser\~ao apresentados no pr\'oximo cap\'{\i}tulo

\section{Consist\^encia da teoria}

Finalmente, vamos mostrar a que  os par\^enteses de Poisson  
$\{{\cal H}_Q (x),{\cal H}_Q (x')\}$ s\~ao sempre fracamete iguais a zero 
independentemente
 do potencial qu\^antico e, portanto,  a geometrodin\^amica qu\^antica 
na vis\~ao de Bohm-de Broglie \'e  consistente. Temos

\begin{eqnarray}
\{ {\cal H}_Q (x), {\cal H}_Q (x')\} = \{{\cal H}(x), {\cal H}(x')\}  
-2 \kappa G_{abcd}(x)\Pi^{cd}(x)\frac{\delta {\cal Q}(x')}{\delta h_{ab}(x)} \nonumber \\
+2 \kappa G_{abcd}(x')\Pi^{cd}(x')\frac{\delta {\cal Q}(x)}{\delta h_{ab}(x')}- 
h^{-1/2}(x)\Pi_{\phi}(x)\frac{\delta {\cal Q}(x')}{\delta \phi(x)} + 
h^{-1/2}(x')\Pi_{\phi}(x')\frac{\delta {\cal Q}(x)}{\delta \phi(x')}\, .
\end{eqnarray}
Nesta \'ultima equa\c c\~ao vamos substituir o potencial qu\^antico 
que sai da equa\c c\~ao de Hamilton Jacobi
modificada Eq.(\ref{hj}):

\begin{equation}
\label{hjerrata}
{\cal Q} = -\kappa G_{ijkl}\frac{\delta S}{\delta h_{ij}} 
\frac{\delta S}{\delta h_{kl}}
- \frac{1}{2}h^{-1/2} \biggr(\frac{\delta S}{\delta \phi}\biggl)^2 - V .
\end{equation}
Ent\~ao,

\begin{eqnarray}
\{ {\cal H}_Q (x), {\cal H}_Q (x')\} = \{{\cal H}(x), {\cal H}(x')\} 
+2 \kappa G_{abcd}(x)\Pi^{cd}(x)\frac{\delta V(x')}{\delta h_{ab}(x)} + h^{-1/2}(x)\Pi \frac{\delta V(x')}{\delta \phi(x)} \nonumber \\
- \kappa G_{abcd}(x')\Pi^{cd}(x')\frac{\delta V(x)}{\delta h_{ab}(x')}-h^{-1/2}(x')\Pi \frac{\delta V(x)}{\delta \phi(x')} \nonumber \\
+4\kappa^2 G_{abcd}(x)\Pi^{cd}(x)\frac{\delta^2 S}{\delta h_{ab}(x) h_{ij}(x')}\frac{\delta S}{\delta h_{kl}(x')}G_{ijkl}(x') \nonumber \\
+2\kappa G_{abcd}(x)\Pi^{cd}(x)h^{-1/2}(x')\frac{\delta S}{\delta \phi(x')}\frac{\delta^2 S}{\delta h_{ab}(x) \delta\phi(x')} \nonumber \\
+2\kappa h^{-1/2}(x)\Pi_{\phi}(x) G_{ijkl}(x')\frac{\delta^2 S}{\delta \phi(x) \delta h_{ij}(x')}\frac{\delta S}{\delta h_{kl}(x')}  \nonumber \\
+h^{-1/2}(x)h^{-1/2}(x')\Pi_{\phi}(x)\frac{\delta S}{\delta \phi(x')}\frac{\delta^2 S}{\delta \phi (x)\phi(x')}
-(x \longleftrightarrow x')\, ,
\end{eqnarray}
onde $(x \longleftrightarrow x')$ significa a mesma express\~ao, mas com $x$ e $x'$ trocados entre sim. 
No lado direito desta equa\c c\~ao  foram cancelados os termos proporcionais a $\delta^3(x',x)$ 
(estes termos v\^em das derivadas funcionais 
$\frac{\delta G_{ijkl}(x')}{\delta h_{ab}(x)}$ e $ \frac{\delta h^{-1/2}(x')}{\delta h_{ab}(x)}$) 
com os  que vem do termo $-(\leftrightarrow)$. Os quatro termos que seguem depois de 
$\{{\cal H}(x), {\cal H}(x')\}$ v\~ao produzir exatamente $-\{{\cal H}(x), {\cal H}(x')\}$ e portanto se cancelam.
Substituindo agora os momentos expressos segundo as rela\c c\~oes guia de Bohm,

\begin{equation}
\Pi ^{ij}(x)=  \Phi^{ij}(x) + \frac{\delta S}{\delta h_{ij}(x)} \, ,
\end{equation}

\begin{equation}
\Pi_{\phi}(x)=\Phi_{\phi}(x)+\frac{\delta S}{\delta \phi(x)} \, ,
\end{equation}
\'e simples de ver, utilizando as propriedades de simetria da $G_{ijkl}$, que todos os termos que 
n\~ao s\~ao fracamente zero se cancelam em pares. Finalmente temos

\begin{eqnarray}
\{ {\cal H}_Q (x), {\cal H}_Q (x')\} = 
+4\kappa^2 G_{abcd}(x)\Phi^{cd}(x)\frac{\delta^2 S}{\delta h_{ab}(x) h_{ij}(x')}\frac{\delta S}{\delta h_{kl}(x')}G_{ijkl}(x') \nonumber \\
+2\kappa G_{abcd}(x)\Phi^{cd}(x)h^{-1/2}(x')\frac{\delta S}{\delta \phi(x')}\frac{\delta^2 S}{\delta h_{ab}(x) \delta\phi(x')} \nonumber \\
+2\kappa h^{-1/2}(x)\Phi_{\phi}(x) G_{ijkl}(x')\frac{\delta^2 S}{\delta \phi(x) \delta h_{ij}(x')}\frac{\delta S}{\delta h_{kl}(x')}  \nonumber \\
+h^{-1/2}(x)h^{-1/2}(x')\Phi_{\phi}(x)\frac{\delta S}{\delta \phi(x')}\frac{\delta^2 S}{\delta \phi (x)\phi(x')}
-(x \longleftrightarrow x')\, .
\end{eqnarray} 
O  lado direito desta equa\c c\~ao \'e fracamente zero devido as rela\c c\~oes de Bohm e, portanto

\begin{equation}
\{ {\cal H}_Q (x), {\cal H}_Q (x')\} \approx 0
\end{equation}
Isto prova a consist\^encia. 
Note que foi muito importante utilizar as rela\c c\~oes guia para fechar 
a \'algebra.
Isto significa que a evolu\c c\~ao hamiltoniana com o potencial qu\^antico 
(\ref{hjerrata}) \'e consistente s\'o quando  restrita \`as trajet\'orias bohmianas.
Para outras trajetorias \'e inconsistente. Assim, quando nos restringimos 
\`a trajetorias bohmianas, uma \'algebra que n\~ao fecha em geral pode 
fechar, como vimos acima.
Isto \'e um ponto importante na interpreta\c c\~ao de  Bohm-de Broglie da 
cosmologia qu\^antica, que algumas vezes n\~ao \'e notado.
Resta saber em que casos a \'algebra formada \'e a \'algebra de Dirac.

\baselineskip=2.0 \normalbaselineskip

\chapter{Cen\'arios Poss\'{\i}veis}


O potencial qu\^antico vai determinar se os colchetes de Poisson dos 
v\'{\i}nculos fecham  segundo 
a \'algebra de Dirac ou segundo uma outra diferente. Vimos que
o colchete de Poisson relevante para esta determina\c c\~ao ser\'a aquele do 
super-hamiltoniano com ele mesmo, $\{{\cal H}_Q (x),{\cal H}_Q (x')\}$, j\'a que os 
outros continuam a fechar da mesma forma  que no caso cl\'assico. Deste modo, o potencial 
qu\^antico permite definir os seguintes cen\'arios cosmologicos:

\section{A geometrodin\^amica qu\^antica 
gera uma 4-geometria n\~ao degenerada}

Neste caso, o colchete de Poisson
 $\{{\cal H}_Q (x),{\cal H}_Q (x')\}$ 
deve satisfazer a Eq. (\ref{algebra1}). Ent\~ao, segundo os resultados de \cite{hkt},
 ${\cal Q}$ deve ser tal 
que $V+{\cal Q}=V_G$ com  $V_G$ dado por 
(\ref{vg}) obtendo-se:
\begin{equation}
\label{q4}
{\cal Q} = -h^{1/2}\biggr[(\epsilon + 1)\biggr(-{\kappa}^{-1} R^{(3)}+
\frac{1}{2}h^{ij}\partial _i \phi\partial _j \phi\biggl)+
\frac{2}{\kappa}(\epsilon\bar{\Lambda} + \Lambda)+
\epsilon\bar{U}(\phi) + U(\phi)\biggl] .
\end{equation}
Temos ent\~ao duas possibilidades:

\subsection{ O espa\c co tempo formado \'e hiperb\'olico ($\epsilon = -1$)}

Neste caso ${\cal Q}$ \'e

\begin{equation}
\label{q4a}
{\cal Q} = -h^{1/2}\biggr[\frac{2}{\kappa}(-\bar{\Lambda} + \Lambda)
-\bar{U}(\phi) + U(\phi)\biggl] .
\end{equation}
Ent\~ao
 ${\cal Q}$ \'e tipo um potencial cl\'assico. Seu efeito \'e renormalizar a constante 
 cosmologica e o potencial cl\'assico do campo escalar. A Geometrodin\^amica 
  qu\^antica \'e indistinguivel da cl\'assica.
 N\~ao \'e preciso tomar o limite cl\'assico  ${\cal Q}=0$ porque $V_G=V+{\cal Q}$ ja 
 descreve o Universo cl\'assico em 
 que vivemos.

\subsection{ O espa\c co-tempo formado \'e euclideano ($\epsilon = 1$)}

Aqui temos que ${\cal Q}$ resulta em

\begin{equation}
\label{q4b}
{\cal Q} = -h^{1/2}\biggr[2\biggr(-{\kappa}^{-1} R^{(3)}+
\frac{1}{2}h^{ij}\partial _i \phi\partial _j \phi\biggl)+
\frac{2}{\kappa}(\bar{\Lambda} + \Lambda)+
\bar{U}(\phi) + U(\phi)\biggl] .
\end{equation}
Agora ${\cal Q}$ n\~ao s\'o renormaliza a constante cosmologica e o potencial 
cl\'assico do campo escalar
como tamb\'em muda a assinatura do espa\c co-tempo. O potencial total 
 $V_G=V+{\cal Q}$ pode descrever
 alguma era do universo primordial quando tinha assinatura 
 euclideana, mas n\~ao a era atual 
 em que \'e hiperb\'olico. 
A transi\c c\~ao entre esas duas fases deve acontecer numa hipersuperf\'{\i}cie 
onde  
  ${\cal Q}=0$, o qual \'e o limite cl\'assico. 
  
Destas considera\c c\~oes podemos concluir que se existe um espa\c co-tempo 
qu\^antico com
caracter\'{\i}sticas diferentes do cl\'assico que observamos, ent\~ao este deve ser 
euclideano.
Em outras palavras, o \'unico efeito qu\^antico relevante que  mantem a 
natureza n\~ao-degenerada
da 4-geometria \'e a mudan\c ca da assinatura da geometria para euclideana.
Os outros efeitos qu\^anticos ou s\~ao irrelevantes por serem indistingu\'{\i}veis de efeitos cla\'ssicos 
ou quebram  a 
estrutura de espa\c co-tempo.
Estes resultados apontam na dire\c c\~ao da Ref.
\cite{haw}.
\vspace{1.0cm}

\section{A geometrodin\^amica qu\^antica gera
uma 4-geometria  degenerada}

Neste caso, o colchete $\{{\cal H}_Q (x),{\cal H}_Q (x')\}$ 
n\~ao satisfaz a Eq. (\ref{algebra1}) mas \'e fracamente zero numa 
outra forma. Vamos examinar alguns exemplos.

\subsection{Solu\c c\~oes reais da equa\c c\~ao de  Wheeler-DeWitt}

No caso de solu\c c\~oes reais da equa\c c\~ao de Wheeler-DeWitt,  
que \'e uma equa\c c\~ao real, a fase  $S$ \'e nula. Ent\~ao, da Eq. (\ref{hj}), podemos
ver que ${\cal Q}=-V$. 
Portanto , o super-hamiltoniano qu\^antico
(\ref{hq0}) conter\'a apenas termos cin\'eticos, dando

\begin{equation}
\label{car}
\{{\cal H}_Q (x),{\cal H}_Q (x')\} = 0.
\end{equation}
Esta \'e uma igualdade forte. Este caso tem rela\c c\~ao com o limite da gravita\c c\~ao forte da TRG
 \cite{tei2,hen,san1}. 
 Se tomarmos o limite de constante gravitacional  $G$ grande (ou velocidade da 
 luz $c$ pequena, onde chegamos ao grupo de  Carroll \cite{poin}), 
ent\~ao  o potencial no v\'{\i}nculo  super-hamiltoniano da TRG pode ser 
desprezado e resulta um  super-hamiltoniano  contendo um termo cin\'etico apenas.
A interpreta\c c\~ao de  Bohm-de Broglie est\'a nos dizendo que qualquer 
solu\c c\~ao real
da equa\c c\~ao de Wheeler-DeWitt produz uma geometrodin\^amica qu\^antica 
que satisfaz 
precisamente este 
limite de gravita\c c\~ao forte .
O limite cl\'assico ${\cal Q}=0$ neste caso implica tambem em $V=0$.
Ser\'{\i}a interessante investigar esta estrutura em profundidade.

\subsection{Potenciais qu\^anticos n\~ao locais}

Um potencial qu\^antico  n\~ao-local qualquer quebra o espa\c co-tempo.
Tomemos como exemplo um potencial da forma

\begin{equation}
\label{non}
{\cal Q}=\gamma V ,
\end{equation}
onde  $\gamma$ \'e uma fun\c c\~ao do funcional  $S$ 
(aqui aparece a n\~ao-localidade). 
No ap\^endice B mostraremos um funcional solu\c c\~ao de um modelo de 
midisuperespa\c co que produz este tipo de potencial qu\^antico.

Calculemos $\{{\cal H}_Q (x),{\cal H}_Q (x')\}$:

\begin{eqnarray}
\{ {\cal H}_Q (x), {\cal H}_Q (x')\}&=&\{{\cal H}(x) + Q(x),{\cal H}(x') + Q(x')\} \nonumber \\
 &=&  
\{{\cal H}(x), {\cal H}(x')\}+\{T(x), {\cal Q}(x')\}+\{{\cal Q}(x),T(x')\} \nonumber
\end{eqnarray}
onde $T$ \'e o termo cin\'etico do super-hamiltoniano qu\^antico.
Desenvolvendo os dois \'ultimos termos temos

\begin{eqnarray}
\{ {\cal H}_Q (x), {\cal H}_Q (x')\} &=& \{{\cal H}(x), {\cal H}(x')\} + 
\gamma \{{\cal H}(x), {\cal H}(x')\} \nonumber \\
& & -  
 \frac{d \gamma }{d S}V(x') \biggr[2 \kappa G_{klij}(x)\Pi^{ij}(x)\frac{\delta S}{\delta h_{kl}(x)}+ 
h^{-\frac{1}{2}}\Pi_{\phi}(x)\frac{\delta S}{\delta \phi(x)}\biggl] \nonumber \\
& & + \frac{d \gamma }{d S}V(x)\biggr[2 \kappa G_{klij}(x')\Pi^{ij}(x')\frac{\delta S}{\delta h_{kl}(x')} +  
h^{-\frac{1}{2}}\Pi_{\phi}(x')\frac{\delta S}{\delta \phi(x')}\biggl] 
\nonumber  \\ 
&=& 
(1+\gamma)\{{\cal H}(x), {\cal H}(x')\} \nonumber \\
& & - 
\frac{d \gamma }{d S}V(x')\biggr[2{\cal H}_Q (x) - 2 \kappa G_{klij}(x) \Pi^{ij}(x)\biggr(\Pi^{kl}(x)  
-  \frac{\delta S}{\delta h_{kl}(x)}\biggl) \nonumber \\
& &  -   h^{-\frac{1}{2}}\Pi_{\phi}(x)\biggr(\Pi_{\phi}(x) - \frac{\delta S}{\delta \phi(x)}\biggl)\biggl] \nonumber \\
& &+\frac{d \gamma }{d S}V(x)\biggr[2{\cal H}_Q (x') - 2 \kappa G_{klij}(x')\Pi^{ij}(x')\biggr(\Pi^{kl}(x') - 
\frac{\delta S}{\delta h_{kl}(x')}\biggl) \nonumber \\
& & - h^{-\frac{1}{2}}\Pi_{\phi}(x')\biggr(\Pi_{\phi}(x')-
\frac{\delta S}{\delta \phi(x')}\biggl)\biggl]
\end{eqnarray}
Fazendo uso da \'algebra (\ref{algebra}) e das defini\c c\~oes
(\ref{ch}) e (\ref{cf}) temos:

\begin{eqnarray}
\label{algnao}
\{ {\cal H}_Q (x), {\cal H}_Q (x')\}&=& (1+\gamma)[{\cal H}^i(x) 
{\partial}_i \delta^3(x,x') - {\cal H}^i(x') {\partial}_i \delta^3(x',x)] \nonumber \\
& & - \frac{d \gamma }{d S}V(x')[2{\cal H}_Q (x) - 2 \kappa G_{klij}(x)\Pi^{ij}(x)\Phi^{kl}(x)-  
h^{-\frac{1}{2}}\Pi_{\phi}(x)\Phi_{\phi}(x)] \nonumber \\
& & + \frac{d \gamma }{d S}V(x)[2{\cal H}_Q (x')-  
2 \kappa G_{klij}(x')\Pi^{ij}(x')\Phi^{kl}(x') - h^{-\frac{1}{2}}\Pi_{\phi}(x')\Phi_{\phi}(x')] \nonumber \\
& & \approx 0 
\end{eqnarray}
O segundo lado da \'ultima express\~ao \'e fracamente zero j\'a que \'e uma 
combina\c c\~ao dos v\'{\i}nculos e
das rela\c c\~oes guia
(\ref{ch}) e (\ref{cf}). 
Nos exemplos acima foram obtidas as `constantes de estrutura' da 
\'algebra que carateriza a `pr\'e-4-geometria' gerada por $H_Q$, isto \'e, a 
estrutura de "espuma" introduzida nos trabalhos  
de J. A. Wheeler 
 \cite{whe,whe2}.

\newpage

\baselineskip=2.0 \normalbaselineskip
\chapter{Discuss\~ao e Conclus\~oes}
Nesta tese estudamos a consist\^encia e consequ\^encias da aplica\c c\~ao
da interpreta\c c\~ao de Bohm de Broglie, tanto em teorias 
qu\^anticas de campos no espa\c co-tempo plano como em gravita\c c\~ao 
qu\^antica. No primeiro caso, mostramos o resultado j\'a conhecido, utilizando um 
novo m\'etodo, que a descri\c c\~ao bohmiana  quebra a
invari\^ancia relativ\'{\i}stica de processos individuais  mas mantem a 
invari\^ancia relativ\'{\i}stica  estat\'{\i}stica, ou seja, dos processos 
observ\'aveis. Para a gravita\c c\~ao qu\^antica, chega-se a um resultado an\'alogo,
embora n\~ao se possa afirmar nada do ponto de vista estat\'{\i}stico.

A interpreta\c c\~ao de Bohm-de Broglie da cosmologia  qu\^antica can\^onica
fornece uma imagem geometrodin\^amica onde a evolu\c c\~ao 
qu\^antica bohmiana das 3-geometrias \'e sempre consistente e pode formar, 
dependendo do funcional de onda
do Universo, ou uma 4-geometria n\~ao degenerada que se for n\~ao trivial tem 
que  ser euclideana, ou uma 4-geometria
degenerada, indicando a presen\c ca de campos de vetores especiais, e a quebra
da estrutura de espa\c co-tempo como uma entidade \'unica (numa ampla 
classe de posibilidades).
Ent\~ao, em geral, e sempre que o potencial qu\^antico seja n\~ao local,
o espa\c co-tempo \'e quebrado. As 3-geometrias evolu\'{\i}das sob 
a influ\^encia do potencial qu\^antico em geral n\~ao v\~ao se superpor para formar 
uma 4-geometria n\~ao degenerada, um espa\c co-tempo com a estrutura causal da 
relatividade. Isto  ja foi antecipado  h\'a algum tempo \cite{whe2}, mas aqui vemos
 a realiza\c c\~ao destas id\'eias concretamente.

As estruturas mais gerais  formadas pela evolu\c c\~ao bohmiana
s\~ao 4-geometrias degeneradas com estruturas causais alternativas.
Obtivemos estes resultados  tomando um campo escalar de m\'ateria com 
acoplamento m\'{\i}nimo como a fonte  da gravita\c c\~ao, mas pode-se generalizar
este resultado para qualquer fonte material com acoplamento sem derivadas 
com a m\'etrica, como
 os campos de Yang-Mills.

Como foi mostrado nesta tese, uma 4-geometria n\~ao degenerada
pode ser obtida somente se o potencial qu\^antico tem a forma 
espec\'{\i}fica (\ref{q4}). Neste caso, o \'unico efeito qu\^antico relevante ser\'a
a mudan\c ca da assinatura do espa\c co-tempo, fato que aponta  em dire\c c\~ao
\`as id\'eias de Hawking. Podemos resumir dizendo que os espa\c cos-tempos qu\^anticos
 n\~ao triviais devem 
ser euclideanos.

No caso de 4-geometrias degeneradas, mostramos que qualquer solu\c c\~ao real
da equa\c c\~ao de Wheeler-DeWitt produz uma estrutura que \'e uma 
idealiza\c c\~ao do limite  de  gravita\c c\~ao forte da TRG. Este tipo de 
geometria, que \'e degenerada, ja tem sido estudada \cite{san1}.
Como esta situa\c c\~ao \'e bem geral (\'e valida para qualquer 
solu\c c\~ao real da equa\c c\~ao de Wheeler-DeWitt, que \'e real), 
ela merece 
maior aten\c c\~ao. Pode ser que estas 4-geometrias degeneradas sejam 
a correta descri\c c\~ao da geometrodin\^amica qu\^antica do universo primordial.
Tambem seria interessante investigar se essas estruturas tem um 
limite cl\'assico produzindo a 4-geometria usual da cosmologia cl\'assica.

Para potenciais qu\^anticos n\~ao locais, mostramos que a evolu\c c\~ao
qu\^antica resulta tamb\'em consistente quando as trajet\'orias s\~ao restritas 
\`as trajet\'orias bohmianas satisfazendo as rela\c c\~oes guia (\ref{grh}) e
(\ref{grf}). Este \'e um ponto que  n\~ao foi tomado com a devida aten\c c\~ao
 na literaura.

Seria interessante tamb\'em, examinar  a liga\c c\~ao entre o limite cl\'assico
e as condi\c c\~oes  para infla\c c\~ao e/ou homogeneidade 
e isotropia no universo. Por exemplo, omitindo o campo escalar, o limite 
cl\'assico dos exemplos estudados no cap.6 nas se\c c\~oes  6.1.2, 6.2.1 e 
6.2.2 implica que a 
hipersuperf\'{\i}cie cl\'assica inicial deva ter uma curvatura escalar constante, 
o que \'e mais pr\'oximo de uma hipersuperf\'{\i}cie inicial maximalmente sim\'etrica.

Como j\'a foi discutido, na interpreta\c c\~ao de Bohm-de Broglie  \'e possivel
examinar qual  tipo de estrutura \'e formada na geometrodin\^amica qu\^antica 
utilizando a rela\c c\~ao dos colchetes de Poisson (\ref{algebra1}), e 
as rela\c c\~oes guia (\ref{hdot}) e (\ref{fdot}). Assumindo a exist\^encia 
de 3-geometrias, configura\c c\~oes de campo, e os respectivos 
momentos, independentemente de quaisquer observa\c c\~oes, a
interpreta\c c\~ao de Bohm-de Broglie  permite-nos utilizar 
ferramentas cl\'assicas, como o formalismo hamiltoniano, para 
entender a estrutura 
da geometria qu\^antica. 
Se esta informa\c c\~ao resulta de alguma utilidade, n\~ao sabemos. J\'a 
na experiencia das duas fendas na mec\^anica qu\^antica n\~ao relativ\'{\i}stica,
a interpreta\c c\~ao de Bohm-de Broglie  permite dizer por qual fenda passou
a part\'{\i}cula: se ela chega \`a  metade superior da tela ent\~ao veio
da fenda superior, e vice-versa. A interpreta\c c\~ao de v\'arios mundos n\~ao
d\'a esta informa\c c\~ao. Mas esta informa\c c\~ao n\~ao tem utilidade: n\~ao
podemos verific\'a-la nem utiliz\'a-la em outros experimentos.
A situa\c c\~ao pode ser an\'aloga na cosmologia qu\^antica can\^onica.
A interpreta\c c\~ao de Bohm-de Broglie  produz muita informa\c c\~ao 
sobre a geometrodin\^amica qu\^antica que n\~ao pode ser obtida da 
interpreta\c c\~ao de v\'arios mundos, mas esta informa\c c\~ao pode ser 
in\'util. De qualquer modo, apenas com mais investiga\c c\~ao poderemos responder a
esta quest\~ao, as ferramentras est\~ao a disposi\c c\~ao.

Gostar\'{\i}amos de ressaltar que estes resultados foram obtidos 
sem que tenhamos assumido nenhum ordenamento nem regulariza\c c\~ao particular 
da equa\c c\~ao de Wheeler-DeWitt. Al\'em disso, n\~ao utlilizamos 
nehuma interpreta\c c\~ao
probabil\'{\i}stica das solu\c c\~oes. Portanto s\~ao resultados completamente 
gerais. Por\'em, gostar\'{\i}amos de comentar alguns pontos sobre o problema da 
probabilidade em cosmologia qu\^antica. A equa\c c\~ao de Wheeler-DeWitt
para  universos fechados n\~ao produz nehuma interpreta\c c\~ao probabil\'{\i}stica
 para suas solu\c c\~oes 
por causa de sua natureza hiperb\'olica. Mas, foi sugerido em v\'arias oportunidades
\cite{kow,banks,pad,kie,Halliwell90} que no regime semicl\'assico \'e possivel construir uma
medida de probabilidade  com as solu\c c\~oes  da equa\c c\~ao de Wheeler-DeWitt.
Ent\~ao, no caso de interpreta\c c\~oes onde as probabilidades s\~ao essencias, o
problema de encontrar um espa\c co de Hilbert  para as solu\c c\~oes 
da equa\c c\~ao de Wheeler-DeWitt resulta crucial se algu\'em deseja obter 
alguma informa\c c\~ao acerca do regime puramente qu\^antico.
Claro que as probabilidades s\~ao tambem \'uteis na interpreta\c c\~ao 
de Bohm-de Broglie. Quando integramos as rela\c c\~oes guia  (\ref{hdot}) e
 (\ref{fdot}), as condi\c c\~oes  iniciais s\~ao arbitr\'arias, e seria bom ter
 alguma distribui\c c\~ao de probabilidade das mesmas. Mas, como foi visto e 
 discutido nesta tese, 
 podemos extrair uma grande quantidade de informa\c c\~ao do regime de 
gravita\c c\~ao qu\^antica sem aproxima\c c\~ao semi-cl\'assica utilizando  a 
interpreta\c c\~ao 
de Bohm-de Broglie, sem ter que recorrer a nehuma no\c c\~ao de probabilidade.
Nesta interpreta\c c\~ao, as probabilidades n\~ao s\~ao essenciais.

Seria tambem muito importante examinar a interpreta\c c\~ao de Bohm-de Broglie 
para outros sitemas gravitacionais, como buracos negros. J\'a foram feitos estudos
neste sentido mas para modelos com simetria esf\'erica no vazio \cite{japa27}, onde
temos um numero finito de graus de liberdade. Seria interessante estudar modelos
mais gerais. Por\'em estes casos s\~ao qualitativamente differentes dos modelos 
cosmol\'ogicos-qu\^antico fechados. N\~ao h\'a problema pensar que existam 
observadores fora de um conjunto de buracos negros. \'E mec\^anica qu\^antica de 
um sitema aberto, que tem menos problemas conceituais de interpreta\c c\~ao.

As conclus\~oes desta tese est\~ao, claro, limitadas por varias  
suposi\c c\~oes fortes que fizemos tacitamente, a saber, que existem 
3-geometrias cont\'{\i}nuas
no regime qu\^antico (os efeitos qu\^anticos poderiam destrui-las tambem) ou 
a validade do processo de quantiza\c c\~ao usual da TRG, 
sem considerar outros formalismos como, por exemplo, a  Teoria de  Cordas.
N\~ao obstante, ainda que nosso formalismo n\~ao fosse o apropriado, \'e impressionante verificar
at\'e  onde se pode chegar com a interpreta\c c\~ao de Bohm-de Broglie,
mesmo no presente estado incompleto da gravita\c c\~ao qu\^antica can\^onica.
Provavelmente deve ser dificil, 
ou tal vez imposs\'{\i}vel,
chegar \`as 
conclus\~oes detalhadas que apresentamos nesta tese usando outras 
interpreta\c c\~oes.

Por outro lado, se esta imagem mais fina da  interpreta\c c\~ao de Bohm-de Broglie
da cosmologia qu\^antica puder fornecer informa\c c\~ao \'util na forma de 
 efeitos observ\'aveis, ent\~ao  teremos uma forma de decidir entre 
 interpreta\c c\~oes, algo que ser\'a com certeza muito importante n\~ao s\'o para 
a cosmologia qu\^antica como tamb\'em para a pr\'opria teoria qu\^antica.

\appendix

\chapter{C\'alculo de $\{ {\cal H}(x),{\cal H}(y)\}$ no caso da teoria de 
campos parametrizada}

Apresentamos aqui o c\'alculo de $\{ {\cal H}(x),{\cal H}(y)\}$ no caso da teoria de 
campos parametrizada. O super-hamiltoniano \'e dado por (\ref{shc})
\begin{equation}
{\cal H}=\frac{1}{\nu}(\Pi_{\alpha}\nu^{\alpha} + \frac{1}{2}\pi_{\phi}^{2} + 
\frac{1}{2} \nu^2 (h^{ij}\phi_{,i}\phi_{,j} +  U(\phi))) \, ,
\end{equation}
que escrevemos, para simplificar, como

\begin{equation}
{\cal H}=\nu^{-1}{\sc h} \, ,
\end{equation}
onde definimos
\begin{equation}
{\sc h}\equiv\Pi_{\alpha}\nu^{\alpha} + \frac{1}{2}\pi_{\phi}^{2} + 
\frac{1}{2} \nu^2 (h^{ij}\phi_{,i}\phi_{,j} +  U(\phi)) \, .
\end{equation}
Ent\~ao temos
\begin{eqnarray}
\{ {\cal H}(x),{\cal H}(y)\}=\frac{1}{\nu(x)}\frac{1}{\nu(y)}\{{\sc h}(x),{\sc h}(y)\}+
\frac{1}{\nu(x)}{\sc h}(y)\{{\sc h}(x), \frac{1}{\nu(y)}\} + \nonumber \\
 \frac{1}{\nu(y)}{\sc h}(x)\{\frac{1}{\nu(x)},{\sc h}(y)\} \, .
\end{eqnarray}
Utilizando o fato de que 
$\frac{\delta \nu_{\alpha}(x)}{\delta X^{\beta}(y)}=-\frac{\delta \nu_{\beta}(x)}{\delta X^{\alpha}(y)}$, 
que vem de (\ref{nu}), e as propriedades b\'asicas da $\delta(x,y)$, \'e 
poss\'{\i}vel ver que cada um dos \'ultimos dois 
par\^enteses do lado direito desta equa\c c\~ao se anulam id\'enticamente. Usando este mesmo argumento e o fato
de que o potencial $U(\phi)$ n\~ao contem derivadas da m\'etrica, a \'ultima equa\c c\~ao fica:

\begin{eqnarray}
\{ {\cal H}(x),{\cal H}(y)\}=-\frac{1}{\nu^2(y)}\Pi_{\beta}(y)
\nu^{\alpha}(y) \frac{\partial \nu^{\beta}(y)}{\partial X^{\alpha}_{i}(x)}
\frac{\partial}{\partial y^i}\delta(y,x) \nonumber \\
+\frac{1}{\nu^2(x)}\Pi_{\alpha}(x)\nu^{\beta}(x)\frac{\partial \nu^{\alpha}(x)}{\partial X^{\beta}_{i}(y)}
\frac{\partial}{\partial x^i}\delta(x,y) - \nonumber \\
h^{ij}(y)\Pi_{\phi}(y)\frac{\partial \phi}{\partial y^j}
\frac{\partial}{\partial y^i}\delta(y,x)+ h^{ij}(x)\Pi_{\phi}(x)\frac{\partial \phi}{\partial x^j}
\frac{\partial}{\partial x^i}\delta(x,y) \, .
\end{eqnarray}
Finalmente \'e poss\'ivel provar que o primeiro termo do lado direito \'e igual a

\begin{eqnarray}
-h^{ij}(y)\Pi_{\alpha}(y)\frac{\partial X^{\alpha}}{\partial y^j}\frac{\partial}{\partial y^i}\delta(y,x)
\end{eqnarray}
e o segundo \'e igual a

\begin{eqnarray}
+h^{ij}(x)\Pi_{\alpha}(x)\frac{\partial X^{\alpha}}{\partial x^j}\frac{\partial}{\partial x^i}\delta(x,y)
\end{eqnarray}
onde novamente usamos a Eq. (\ref{nu}).
Ent\~ao temos
\begin{eqnarray}
\{ {\cal H}(x),{\cal H}(y)\}=h^{ij}(x)\biggr(\Pi_{\alpha}(x)\frac{\partial X^{\alpha}}{\partial x^j}+
\Pi_{\phi}(x)\frac{\partial \phi}{\partial x^j}\biggl)\frac{\partial}{\partial x^i}\delta(x,y) \nonumber \\
-h^{ij}(y)\biggr(\Pi_{\alpha}(y)\frac{\partial X^{\alpha}}{\partial y^j}+
\Pi_{\phi}(y)\frac{\partial \phi}{\partial y^j}\biggl)\frac{\partial}{\partial y^i}\delta(y,x) \, ,
\end{eqnarray}
isto \'e,
\begin{eqnarray}
\{ {\cal H}(x),{\cal H}(y)\}={\cal H}^i(x)\frac{\partial}{\partial x^i}\delta(x,y)-
{\cal H}^i(y)\frac{\partial}{\partial y^i}\delta(y,x) \, .
\end{eqnarray}

\baselineskip=2.0 \normalbaselineskip

\chapter{A quebra da invari\^ancia de Lorentz}

Aqui vamos estudar um exemplo onde a invari\^ancia  de Lorentz \'e quebrada 
a n\'{\i}vel de eventos individuais, segundo a interpreta\c c\~ao de Bohm-de Broglie 
da teoria de campos em espa\c co-tempo plano, estudada no cap\'{\i}tulo 3. Para mostrar
isto vamos primeiramente desparametrizar a teoria, isto \'e, voltar \`as coordenadas
de Minkowski. Partimos da Eq.(\ref{supHpsi})

\begin{equation}
\label{supHpsi2}
\frac{1}{\nu}\biggr(-i\hbar \nu^{\alpha}\frac{\delta \Psi}{\delta X^{\alpha}(x) } -
 (\hbar)^2\frac{1}{2}\frac{\delta^2 \Psi}{\delta \phi(x)^2} + 
 \frac{1}{2} \nu^2 (h^{ij}(x)\phi(x)_{,i}\phi(x)_{,j} +  U(\phi(x)))\Psi\biggl) = 0 \, .
\end{equation}
A funcional $\Psi$ n\~ao depende do rotulamento da 
hipersuperf\'{\i}cie, j\'a que, como foi visto, \'e invariante perante 
transforma\c c\~oes espaciais de coordenadas. Ent\~ao podemos escolher as 
coordenadas $x$ na hipersuperficie como sendo as de Minkowsi $x^i=X^i$.
As hipersuperf\'{\i}cies ser\~ao  descritas em forma desparametrizada 
por $X^{0}=X^{0}(X^i)$ e o funcional de onda $\Psi(\phi(X), X^\alpha)$ se escreve 
$\Psi=\Psi(\phi(X), X^{0}(X^i))$. Podemos escolher uma familia de hipersuperficies 
plana uniparam\'etrica, de modo que $ X^{0}(X^i)=T$ com $-\infty < T < \infty$ e  
$\frac{\partial T}{\partial X^i}=0$. O funcional $\Psi=\Psi(\phi(X), X^{0}(X^i))$ sob esta 
fam\'{\i}lia de hipersuperf\'{\i}cies \'e agora um funcional de $\phi(X)$ que depende de 
$T$ como um par\^ametro $\Psi=\Psi(\phi(X), T)$. Temos ent\~ao para a derivada temporal

\begin{eqnarray}
\frac{\partial \Psi}{\partial T}=\int d^3X \frac{\delta \Psi}{\delta X^{0}(X)}\frac{\partial X^{0}(X)}{\partial T} \nonumber \\
=\int dX^3 \frac{\delta \Psi}{\delta X^{0}(X)}
\end{eqnarray}
Com estas defini\c c\~oes, $\nu_{\alpha}$, dado na Eq. (\ref{nu}), tem  componentes 
$\nu_0=-1, \nu_i=0$. ($\nu^0=1$ ). Substituindo tudo isto na equa\c c\~ao 
(\ref{supHpsi2}) temos:

\begin{equation}
\label{esc}
i \hbar \frac{\partial \Psi (\phi ,t)}{\partial T} = 
\int d^3X \frac{1}{2}\biggr[-\hbar ^2  
\frac{\delta^2}{\delta \phi^2} +
(\nabla \phi)^2+U(\phi)\biggl]  \Psi (\phi ,T) \, ,
\end{equation}
que \'e a equa\c c\~ao funcional de Schr\"odinger para o campo escalar $\phi$ no potencial $U(\phi)$.
Esta equa\c c\~ao tem problemas de regulariza\c c\~ao, j\'a que o operador de deriva\c c\~ao 
funcional est\'a aplicado no mesmo ponto. Acima ela est\'a escrita em forma n\~ao regularizada. 
Ela deve, portanto, ser regularizada e em seguida ser escolhido um ordenamento
que deixe a teoria livre de anomalias. As solu\c c\~oes desta equa\c c\~ao 
regularizada ser\~ao interpretadas segundo Bohm-de Broglie.
Escrevendo  o funcional de onda na forma polar $\Psi = A \exp (iS/\hbar)$, e substituindo
na Eq. (\ref{esc}), obtemos:

\begin{equation}
\label{chja}
\frac{\partial S}{\partial T} + \frac{1}{2}\int d^3X  
\biggr[ \biggr(\frac{\delta S}{\delta \phi}\biggl)^2 +
(\nabla \phi)^2+ U(\phi)\biggl] +  Q(\phi,T) = 0 ,
\end{equation}
                
\begin{equation}
\label{cp}
\frac{\partial A^2}{\partial t}+\int d^3 X \frac{\delta}{\delta \phi}
\biggr(A^2 \frac{\delta S}{\delta \phi}\biggl) = 0 ,
\end{equation}

onde 

\begin{equation}
\label{pqc}
Q(\phi) = -\hbar ^2 \frac{1}{2A} \int d^3 X \frac{\delta^2 A}
{\delta \phi^2} \, ,
\end{equation}
\'e o correspondente potencial qu\^antico, que depende da regulariza\c c\~ao e 
ordenamento escolhido na (\ref{esc}). Acima ele est\'a escrito em forma n\~ao regularizada.
A primeira equa\c c\~ao Eq.(\ref{chja}) \'e interpretada como uma  equa\c c\~ao de 
Hamilton-Jacobi 
que governa a evolu\c c\~ao  de certa configura\c c\~ao inicial do campo no 
tempo,
a qual vai ser diferente da cl\'assica devido a presen\c ca do potencial 
qu\^antico.
A Eq. (\ref{cp}) \'e uma lei de conserva\c c\~ao que justifica a suposi\c c\~ao 
de que, a tempo $T$, $A^2 D\phi$ \'e
a probabilidade  de que o campo $\phi$ esteja num elemento de `volume'  $D\phi$ 
ao redor da 
configura\c c\~ao $\phi(X)$ para todo $X$. A nota\c c\~ao $D\phi$ significa o 
produto 
infinito $\Pi_X d\phi$ dos elementos de volume do campo $d\phi$ para 
cada valor de $X$.
O funcional pode-se supor normalizado: 

\begin{equation}
\int|\Psi|^2 D\phi=1
\end{equation}
 A evolu\c c\~ao qu\^antica pode ser obtida integrando-se a  rela\c c\~ao guia de Bohm, dada 
 agora por

\begin{equation}\label{rgcampo}
\Pi _{\phi} = \frac{\partial \phi}{\partial T} = 
\frac{\delta S [\phi(X),T]}{\delta \phi(X)} |_{\phi(X)=\phi(X,T)}
\end{equation}
uma vez dada a configura\c c\~ao inicial $\phi_0(X)$.
A equa\c c\~ao de movimento da coordenada $\phi$ pode se obtida tomando a 
derivada funcional da 
equa\c c\~ao de Hamilton-Jacobi modificada (\ref{chja}). Temos

\begin{equation}
\frac{\partial}{\partial T}\biggr(\frac{\delta S}{\delta \phi(X)}\biggl) + \frac{1}{2}\int d^3Y  
\biggr[2\frac{\delta S}{\delta \phi (Y)} \frac{\delta}{\delta \phi(X)}\biggr(\frac{\delta S}{\delta \phi}\biggl) +
2(\nabla \phi)\frac{\delta (\nabla \phi(Y))}{\delta \phi(X)}+\frac{\delta U(\phi)}{\delta \phi(X)}\biggl] +  \frac{\delta Q(\phi,T)}{\delta \phi(X)} = 0 ,
\end{equation}
e usando que $\frac{\partial \phi}{\partial T} = 
\frac{\delta S [\phi(X),T]}{\delta \phi(X)} $ e denotando $\dot{\phi}\equiv \frac{\partial \phi}{\partial T}$ segue que

\begin{equation}
\frac{\partial \dot{\phi} }{\partial T}+\int d^3Y \dot{\phi}(Y)\frac{\delta \dot{\phi}(Y)}{\delta \phi(X)}+ 
\int d^3Y \nabla \phi .\nabla \delta(X,Y)+\frac{1}{2}\int d^3Y\frac{\delta U(\phi)}{\delta \phi(X)}+\frac{\delta Q(\phi,T)}{\delta \phi(X)} = 0
\end{equation}
O segundo termo do lado esquerdo \'e nulo ja que $\frac{\delta \dot{\phi}}{\delta \phi}=
\frac{\delta \Pi_{\phi}}{\delta \phi}=0$, ent\~ao, integrando por partes e desprezando 
um termo de fronteira

\begin{equation}
\frac{\partial^2 \phi}{\partial T^2}- \nabla^2 \phi +
\frac{1}{2}\int d^3Y\frac{\delta U(\phi)}{\delta \phi(X)} + \frac{\delta Q(\phi,T)}{\delta \phi(X)}=0,
\end{equation}
que podemos escrever

\begin{equation}
\label{eob}
\Box \phi(X,T)+ \frac{1}{2}\int d^3Y \frac{\delta U(\phi)}{\delta \phi(X)}=-\frac{\delta Q[\phi(X),T]}{\delta \phi(X)}|_{\phi(X) = \phi(X,T)}\, .
\end{equation}
onde $\Box\equiv-\partial_{\mu}\partial^{\mu}$.
No caso de um campo escalar livre temos $U(\phi)=m^2\phi^2$, e esta \'ultima equa\c c\~ao se reduz a

\begin{equation}
\label{eob2}
\Box \phi(X,T)+ m^2\phi(X,T) =-\frac{\delta Q[\phi(X),T]}{\delta \phi(X)}|_{\phi(X) = \phi(X,T)}
\end{equation}
Esta equa\c c\~ao, \'e a vers\~ao qu\^antica da equa\c c\~ao de onda 
cl\'assica:
\begin{equation}
\label{eoc}
\Box \phi(X,T)+ m^2\phi(X,T) = 0 \, .  
\end{equation}
A "for\c ca qu\^antica" que aparece no lado  direito de (\ref{eob}) \'e responsavel  por
 todos os efeitos qu\^anticos
da teoria.

Vamos mostrar que j\'a o estado de v\'acuo do 
campo escalar livre produz um potencial qu\^antico que quebra  a 
invariancia Lorentz dos campos.
A solu\c c\~ao para o estado de v\'acuo da equa\c c\~ao (\ref{esc}) no caso 
livre, est\'a dada por (\cite{hatfield} cap. 10):

\begin{equation}
\label{vacio}
\Psi_0[\phi,T]=e^{-\frac{iE_{0}T}{\hbar}} \eta e^{-\int d^3X d^3Y \phi(X)g(X,Y)\phi(Y)} 
\end{equation}
onde 

\begin{equation}
g(X,Y)=\frac{1}{2}\int \frac{d^3k}{(2\pi)^3} \omega_k e^{i k.(X-Y)}
\end{equation}
e $\omega_k=\hbar \sqrt{k^2+m^2}$ 

Calculando a amplitude de (\ref{vacio}) e usando (\ref{pqc}), o 
potencial qu\^antico fica

\begin{equation}
Q=-\frac{1}{2}\int d^3X  ( \int d^3Y \frac{d^3k}{(2\pi)^3} \omega_k \cos\{k.(X-Y)\} \phi(Y))^2 + \frac{1}{2}\int d^3 X \int d^3k \omega_k .
\end{equation}
O ultimo termo \'e  a energia do v\'acuo.
Derivando funcionalmente esta express\~ao com  respeito a $\phi$, tendo em conta que a energia do v\'acuo 
n\~ao depende funcionalmente de $\phi$ e usando que (\cite{hatfield} cap.10)

\begin{equation}
\int d^3X g(Z,X)g(X,Y) =\frac{1}{4}(-\nabla^2+m^2)\delta(Z,Y),
\end{equation}
segue que

\begin{equation}
\frac{\delta Q}{\delta \phi(X)}= -(-\nabla^2+m^2)\phi(X)
\end{equation}
Substituindo na equa\c c\~ao do campo $\phi$ (\ref{eob2}) temos

\begin{equation}
\label{eobp}
\Box \phi(X,T) + m^2\phi(X,T) =(-\nabla^2 + m^2)\phi(X)|_{\phi(X) = \phi(X,T)}
\end{equation}
a qual n\~ao \'e uma equa\c c\~ao invariante de Lorentz.

\baselineskip=2.0 \normalbaselineskip

\chapter{Potencial qu\^antico n\~ao local para um 
espa\c co-tempo esfericamente sim\'etrico}

Apresentamos a seguir o exemplo de   um potencial qu\^antico 
n\~ao local 
do tipo estudado nos cen\'arios cosmol\'ogicos no cap.6.
Se trata de um modelo de midi-superespa\c co esfericamente sim\'etrico com um 
campo eletromagn\'etico. Escolhemos um certo ordenamento na equa\c c\~ao 
de Wheeler-DeWitt, que n\~ao est\'a regularizada, s\'o para ter uma 
solu\c c\~ao exata
j\'a conhecida na literatura\footnote{De fato, o modelo estudado aqui 
pode se reduzir a um mini-superespa\c co, com um n\'umero finito de graus de
liberdade\cite{kuchar}. N\~ao obstante, nosso objetivo neste ap\^endice
\'e mostrar que n\~ao \'e muito dificil que um potencial do tipo estudado
no cap. 5 e que quebra a \'algebra de Dirac, apare\c ca em cosmologia qu\^antica 
can\^onica. Eles podem ser obtidos, por exemplo, em equa\c c\~oes 
de Wheeler-DeWitt com 
simetria esferica, como iremos ver.}.

Come\c cemos com a descomposi\c c\~ao ADM (Arnowit-Deser-Misner) sob a 
variedade $R\times R \times S^2$ da
 m\'etrica de um espa\c co-tempo esfericamente sim\'etrico com um campo 
eletromagn\'etico:

\begin{equation} ds^2=-N^2dt^2+\Lambda^2(dr+N^{r}dt)^2+R^2d\Omega^2 \, , 
\end{equation}
onde $N$, $N^{r}$ sao as fun\c c\~oes lapso e deslocamento, respectivamente, 
ambas dependendo de $r$ e $t$), e $d\Omega$  indica o elemento de linha 
sobre $S^2$. O potencial eletromagn\'etico est\'a descrito  pela 1-forma
esfericamente sim\'etrica:

\begin{equation} dA=\Gamma(r,t)dr+\Phi(r,t)dt \, .
\end{equation}
A a\c c\~ao ADM no midisuperespa\c co, depois de integrar sobre a 2-esfera, 
fica ($c\equiv 1$)

\begin{eqnarray}S=\int dt \int dr \frac{1}{2N} \biggr\{ \frac{1}{G}\biggr[N^2\Lambda-\Lambda \dot{R}^2
+2\frac{N^2 R \Lambda' R'}{\Lambda^2}-2\frac{N^2 R R''}{\Lambda}- \nonumber \\ \Lambda (N^{r})^2 R'^2 
+2N^{r}\dot{R}(\Lambda R)'-2 R N^{r} R'(\Lambda N^{r})'+ 2 R \Lambda (N^{r})' \dot{R} \nonumber \\ 
+ 2 R N^{r} \dot{\Lambda} R'-2 R \dot{\Lambda} \dot{R}- 
N^2 \frac{R'^2}{\Lambda}\biggl] + 
\frac{R^2}{\Lambda}(\dot{\Gamma}-\Phi')^2 \biggl\} ,
\end{eqnarray}
onde a linha indica a derivada com respeito a $r$.
Variando a a\c c\~ao  com rela\c c\~ao a $N$ e $N^{r}$ obtemos os v\'{\i}nculos
super-hamiltoniano
e   super-momento \cite{kuchar,louko}

\begin{equation} {\it {\cal H}}\equiv \frac{G}{2}\frac{\Lambda P_{\Lambda}^2}{R^2}-
G\frac{P_{\Lambda}P_{R}}{R}
+\frac{V_g}{G}+ \frac{\Lambda P_{\Gamma}^2}{2R^2}\approx 0 , 
\end{equation}

e

\begin{equation}\label{mc} {\it {\cal H}_{r}}\equiv P_{R}R'-\Lambda P_{\Lambda}'\approx 0 , 
\end{equation}
onde

\begin{equation}\frac{V_g}{G}=\frac{RR''}{\Lambda}-
\frac{RR'\Lambda'}{\Lambda^2}+\frac{R'^2}{2\Lambda}-\frac{\Lambda}{2}
\end{equation}
Variar a a\c c\~ao  com rela\c c\~ao ao multiplicador de Lagrange $\phi$
conduz a

\begin{equation}
\label{E} 
P_{\Gamma}' \approx 0 \, .
\end{equation}
Assumimos condi\c c\~oes de contorno para todos os campos de modo que todas
as integrais resultem bem definidas, e de maneira que a m\'etrica do 
espa\c co-tempo cl\'assico seja n\~ao-degenerada \cite{louko}.

Quantizaremos segundo o formalismo de Dirac. Todos os v\'{\i}nculos atuam 
sobre 
funcionais $\Psi[\Lambda(r),R(r),\Gamma(r)]$. O v\'{\i}nculo eletromagn\'etico
(\ref{E}) \'e resolvido\cite{brotz} por 
$ \Psi= f(\int_{\infty}^{\infty}\Gamma dr)\psi[\Lambda(r),R(r)]$
onde $f$ \'e uma fun\c c\~ao  diferenci\'avel.
Do v\'{\i}nculo super-hamiltoniano, obtemos a equa\c c\~ao de  Wheeler-DeWitt
que, escrita com um certo ordenamento, fica:

\begin{equation}
\biggr(-\frac{G\hbar^2 \Lambda}{2R^2}F\frac{\delta}{\delta\Lambda}F^{-1}
\frac{\delta}{\delta\Lambda}
+\frac{G\hbar^2}{R}F \frac{\delta}{\delta R} F^{-1} \frac{\delta}{\delta \Lambda}+\frac{V_g}{G}-
 \frac{\hbar^2 \Lambda \delta^2}{2R^2\delta \Gamma^2}\biggl)\Psi=0 
\end{equation}
onde

\begin{equation} F\equiv R \sqrt{\biggr(\frac{R'}{\Lambda}{\biggl)}^2+\frac{2m}{R}-\frac{q^2}{R^2}-1} 
\end{equation}
Nossa solu\c c\~ao tem a forma \cite{brotz}

\begin{equation} 
\Psi=e^{\frac{iq}{\hbar}\int_{-\infty}^{\infty}\Gamma dr}
\psi[\Lambda(r),R(r)] , 
\end{equation}
onde foram separados os graus de liberdade eletromagn\'eticos e 
gravitacionais, e $\psi$ verifica uma equa\c c\~ao de  Wheeler-DeWitt
reduzida

\begin{equation}
\label{wd} \biggr(-\frac{G\hbar^2 \Lambda}{2R^2}F\frac{\delta}{\delta\Lambda}F^{-1}
\frac{\delta}{\delta\Lambda}
+\frac{G\hbar^2}{R}F \frac{\delta}{\delta R} F^{-1} \frac{\delta}{\delta \Lambda}+\frac{V_g}{G}+
\frac{\Lambda q^2}{2 R^2}\biggl)\psi=0 .
\end{equation}
Escolhemos este ordenamento particular pois neste caso
 \'e conhecida na literatura uma 
solu\c c\~ao exata \cite{brotz}, a saber

\begin{equation}
\label{psi1} \psi_e=\exp{\frac{iS_{0}}{\hbar}} , 
\end{equation}
onde

\begin{equation}
\label{so} 
S_{0}= G^{-1}\int_{-\infty}^{\infty}dr \biggr\{\Lambda F -
 \frac{1}{2} RR'\ln{\frac{\frac{R'}{\Lambda}+\frac{F}{R}}{\frac{R'}{\Lambda}
 -\frac{F}{R}}}\biggl\} 
\end{equation}
\'E simples  verificar que, sendo a equa\c c\~ao de 
Wheeler-DeWitt  real, ent\~ao o complexo conjugado do funcional (\ref{psi1})
tambem \'e uma outra solu\c c\~ao exata. Por tanto , temos duas solu\c c\~oes
exatas independentes desta equa\c c\~ao e, por causa da sua 
linearidade, qualquer superposi\c c\~ao linear ser\'a uma outra solu\c c\~ao:

\begin{equation}
\label{psi} \psi=a\exp{\biggr(\frac{iS_{0}}{\hbar}\biggl)}+
b\exp{\biggr(\frac{-iS_{0}}{\hbar}\biggl)} .
\end{equation}
Escrevendo-a em forma polar

\begin{equation}
\psi=A\exp{\frac{iS}{\hbar}} , 
\end{equation}
e substituindo-a  na equa\c c\~ao de Wheeler-DeWitt (\ref{wd}), obtemos
duas equa\c c\~oes. Uma delas \'e

\begin{equation}
\label{cwdw}
\frac{G\Lambda}{2R^2}(\frac{\delta S}{\delta\Lambda})^2-\frac{G}{R}\frac{\delta S}{\delta\Lambda}
 \frac{\delta S}{\delta R} + V + {\cal Q}=0 ,
\end{equation}
onde  $V$ indica o potencial cl\'assico 
\begin{equation}
V \equiv \frac{V_g}{G}+
\frac{\Lambda q^2}{2 R^2} , 
\end{equation}
e  ${\cal Q}$ \'e o  potential  qu\^antico

\begin{equation}\label{Pq}
{\cal Q}=\frac{G\hbar^2}{A R}\biggr(-\frac{\Lambda \delta^2 A}{2R\delta \Lambda^2} +\frac{\delta^2 A}{\delta R \delta \Lambda}+
\biggr(-\frac{1}{F}\frac{\delta F}{\delta R} +\frac{\Lambda}{2 R F}\frac{\delta F}{\delta \Lambda}\biggl)\frac{\delta A}{\delta\Lambda}\biggl) .
\end{equation}
Para o funcional (\ref{psi}), o  potential  qu\^antico (\ref{Pq}) \'e

\begin{equation}
{\cal Q}= \gamma  V ,
\end{equation}
onde  $V$ \'e o potencial cl\'assico, e o fator $\gamma$ \'e dado por

\begin{equation}
\label{pq} \gamma=-4\biggr\{ \biggr(\frac{ab}{A^2}\biggl)^2 \sin^2{\biggr(\frac{2 S_{0}}{\hbar}\biggl)} +
\frac{ab}{A^2}\cos{\biggr(\frac{2 S_{0}}{\hbar}\biggl)} \biggl\} .
\end{equation}
J\'a que a fase $S$ do funcional  vem dado em termos de 
 $S_{0}$ por 
$S=\frac{\hbar}{2i}\ln(\frac{\psi}{\psi^*})$, onde a  $\psi$ \'e uma 
fun\c c\~ao de  $S_0$ , ent\~ao 

\begin{equation}
\gamma=\gamma(S) .
\end{equation}
Portanto temos um exemplo concreto de um potencial qu\^antico n\~ao local
 do tipo estudado no cap\'{\i}tulo 5. A \'algebra dos v\'{\i}nculos
 n\~ao \'e a de Dirac mas fecha com diferentes `constantes de estrutura',
 quebrando a 4-geometria do espa\c co tempo. As hipersuperf\'{\i}cies evoluem 
 com o hamiltoniano
 qu\^antico de Bohm-de Broglie consistentemente, mas a estrutura gerada 
 nessa evolu\c c\~ao 
 n\~ao \'e uma 4-geometria Riemanniana e por tanto a causalidade Einsteniana vai ser modificada
 de acordo com a nova \'algebra dos v\'{\i}nculos.  Explicitamente, temos que 
 o par\^enteses  de Poisson para o super-hamiltoniano qu\^antico \'e dado por:

\begin{eqnarray}
\{{\cal H}_Q,{\bar {\cal H}_Q }\}=(1+\gamma)\{ {\cal H},{\bar {\cal H}}\} 
-2\frac{\delta \gamma}{\delta S}\frac{\bar V_{G}}{G}{\cal H}_Q + \frac{\delta \gamma}{\delta S}\frac{\bar V_{G}}{G}
(\frac{G \Lambda}{R^2}P_{\Lambda} \phi_{\Lambda}-\frac{G}{R}P_{\Lambda} \phi_{R}-\frac{G}{R}P_{R} \phi_{\Lambda})   \nonumber \\
+2\frac{\delta \gamma}{\delta S}\frac{V_{G}}{G}{\bar {\cal H}_Q} + \frac{\delta \gamma}{\delta S}\frac{V_{G}}{G} \nonumber 
(\frac{\bar G}{\bar \Lambda}{{\bar R}^2}{\bar P}_{\Lambda} {\bar \phi}_{\bar \Lambda}-
\frac{\bar G}{\bar R}{\bar P}_{\Lambda}{\bar \phi}_{R}-\frac{\bar G}{\bar R}{\bar P}_{R} {\bar \phi}_{\Lambda})\, ,
\end{eqnarray}
que est\'a de acordo com (\ref{algnao}).
Aqui usamos a  barra para denotar as quantidades no ponto $r'$, por ex.
${\bar {\cal H}_Q} \equiv {\cal H}_Q (r')$, $ {\cal H} \equiv {\cal H}(r)$etc.
Vemos que o lado direito desta u\'ltima equa\c c\~ao \'e fracamente zero 
por causa do v\'{\i}nculo 
super-hamiltoniano qu\^antico ${\cal H}_Q \approx 0$, das rela\c c\~oes de 
Bohm $\phi_{\Lambda} \approx 0, \phi_{R}\approx 0$ e do v\'{\i}nculo  
super-momento, j\'a que para os 
par\^enteses de Poisson que aparecem no
 primeiro termo da soma \'e

\begin{eqnarray}
\{ {\cal H},{\bar {\cal H}}\}= {\it {\cal H}^{r}}\frac{\partial}{\partial r} \delta(r-r')-{\bar{\it {\cal H}^{r}}}\frac{\partial}{\partial r'}\delta(r'-r) \approx 0 \, .
\end{eqnarray}

\baselineskip=2.0 \normalbaselineskip

\end{document}